\documentclass[aps, prd, reprint,10pt, notitlepage, a4paper,floats, floaqutfix,amsmath, amssymb, amsfonts,superscriptaddress,showpacs, showkeys,nofootinbib,longbibliography]{revtex4-2}

\pdfoutput=1

\usepackage{epsfig}
\usepackage{graphics}
\usepackage{graphicx}
\usepackage{amsmath,amsfonts,amssymb,amsthm}
\usepackage[usenames,dvipsnames]{xcolor}
\usepackage{wasysym}
\usepackage{times}
\usepackage{mathptmx}
\usepackage{gensymb}
\usepackage{appendix}
\usepackage{listings}
\usepackage{url}
\usepackage[normalem]{ulem}
\usepackage{alltt}
\usepackage[colorlinks]{hyperref}
\usepackage{cleveref}
\usepackage{longtable}
\usepackage{enumitem}
\setlist{nosep}
\usepackage{color}
\usepackage{calc}
\usepackage{tensor}
\usepackage{bm}
\usepackage{times}
\usepackage{multirow}
\usepackage[varg]{txfonts}
\usepackage{float}
\usepackage{dcolumn}
\usepackage[nolist,nohyperlinks]{acronym}
\usepackage{xspace}
\usepackage[english]{babel}
\usepackage[abs]{overpic}
\usepackage{pict2e}
\usepackage[caption=false]{subfig} % prevent loading of caption package: http://tex.stackexchange.com/questions/22388/subfigures-with-revtex
\allowdisplaybreaks[1]
\usepackage[utf8]{inputenc}
\usepackage{gensymb}
\usepackage{bm}
\usepackage{stackengine}
\usepackage{boldline,multirow}
\usepackage{braket}
\usepackage{longtable}
\usepackage{makecell}

\usepackage{tabularx}
\usepackage{rotating}
\usepackage{booktabs}

\graphicspath{{plots/}}

\newcommand{\AEI}{Max Planck Institute for Gravitational Physics (Albert Einstein Institute), Am M\"uhlenberg 1, Potsdam 14476, Germany}
\newcommand{\Maryland}{Department of Physics, University of Maryland, College Park, MD 20742, USA}

\definecolor{dodgerblue}{HTML}{0099E6}
\definecolor{viennared}{HTML}{F23814}

\hypersetup{citecolor=dodgerblue, linkcolor=viennared, urlcolor=lapislazuli}

\def\seobv5{\texttt{SEOBNRv5PHM}}
\def\seobesgb{\texttt{SEOBNRv5PHM$\_$ESGB}}

\def\sqrtalpha{\sqrt{\alpha_{\rm{GB}}}}

\DeclareMathOperator{\Order}{\mathcal{O}}
\def\mF{\mathcal{F}}

\newcommand{\ConstraintCombinedIMR}{0.32}
\newcommand{\ConstraintCombinedIMRMerger}{0.31}
\newcommand{\ConstraintCombinedInspiral}{0.30}

\newcommand{\ConstraintAIMR}{1.81}
\newcommand{\ConstraintAIMRMerger}{1.77}
\newcommand{\ConstraintAInspiral}{2.15}
\newcommand{\BayesAIMR}{-0.52}
\newcommand{\BayesAIMRMerger}{-4.1}
\newcommand{\ConstraintBIMR}{0.50}
\newcommand{\ConstraintBIMRMerger}{0.48}
\newcommand{\ConstraintBInspiral}{0.41}
\newcommand{\BayesBIMR}{-0.02}
\newcommand{\BayesBIMRMerger}{-3.0}
\newcommand{\ConstraintCIMR}{0.31}
\newcommand{\BayesCIMR}{-0.94}

\newcommand{\ConstraintCombinedIMREll}{1.72}
\newcommand{\ConstraintCombinedIMRMergerEll}{1.65}
\newcommand{\ConstraintCombinedInspiralEll}{1.60}
\newcommand{\ConstraintAIMREll}{9.65}
\newcommand{\ConstraintAIMRMergerEll}{9.42}
\newcommand{\ConstraintAInspiralEll}{11.42}
\newcommand{\ConstraintBIMREll}{2.66}
\newcommand{\ConstraintBIMRMergerEll}{2.55}
\newcommand{\ConstraintBInspiralEll}{2.19}
\newcommand{\ConstraintCIMREll}{1.66}

\definecolor{lapislazuli}{rgb}{0.15, 0.38, 0.61}

\begin{document}

\title{
Inspiral-merger-ringdown 
	 waveforms in Einstein-scalar-Gauss-Bonnet \\gravity within the 
effective-one-body formalism
}
%%%%%%%%%%%%%%%%%%%%%%%%%%%%%%%%%%%%%%%%%%%%%%%%%%%% Title page %%%%%%%%%%%%%%%%%%%%%%%%%%%%%%%%%%%%%%%%%%%%%%%%%%%%

\author{Félix-Louis Julié}
\email{felix-louis.julie@aei.mpg.de}
\affiliation{\AEI}

\author{Lorenzo Pompili}
\email{lorenzo.pompili@aei.mpg.de}
\affiliation{\AEI}

\author{Alessandra Buonanno}
\email{alessandra.buonanno@aei.mpg.de}
\affiliation{\AEI}
\affiliation{\Maryland}

\date{\today}

%%%%%%%%%%%%%%%%%%%%%%%%%%%%%%%%%%%%%%%%%%%%%%%%%%% Abstract %%%%%%%%%%%%%%%%%%%%%%%%%%%%%%%%%%%%%%%%%%%%%%%%%%%%%%
\begin{abstract}
  Gravitational waves (GWs) provide a unique opportunity to test
  General Relativity (GR) in the highly dynamical, strong-field
  regime. So far, the majority of the tests of GR with GW signals have
  been carried out following parametrized, theory-independent
  approaches. An alternative avenue consists in developing
  inspiral-merger-ringdown (IMR) waveform models in specific beyond-GR
  theories of gravity, by combining analytical and
  numerical-relativity results. In this work, we provide the first
  example of a full IMR waveform model in a beyond-GR theory, focusing
  on Einstein-scalar-Gauss-Bonnet (ESGB) gravity.
  This theory has
  attracted particular attention due to its rich phenomenology for
  binary black-hole (BH) mergers, thanks to the presence of non-trivial
  scalar fields. Starting from the state-of-the-art, effective-one-body
  (EOB) multipolar waveform model for spin-precessing binary BHs 
\texttt{SEOBNRv5PHM}, we include theory-specific corrections
  to the EOB Hamiltonian, the metric and scalar energy fluxes, the GW
  modes, the quasi-normal-mode (QNM) spectrum and the mass and spin of the
  remnant BH. 
  We also propose a way to marginalize over the
  uncertainty in the merger morphology with additional nuisance
  parameters.
   Interestingly, we observe that changes in the frequency of the ringdown waveform due to the final mass and spin 
  corrections are significantly larger than those due to ESGB corrections to the QNM spectrum.
  By performing Bayesian parameter estimation for the GW
  events GW190412, GW190814 and GW230529$\_$181500, we place constraints on
  the fundamental coupling of the theory 
  ($\sqrt{\alpha_{\mathrm{GB}}} \lesssim \ConstraintCombinedIMRMerger~\mathrm{km}$ at $90\%$ confidence). 
  The bound could be improved by one order of magnitude by observing a single ``golden'' binary system with next-generation ground-based GW detectors.
  Our model can be used to improve
  constraints on modifications of GR with upcoming GW observations, and
  to provide forecasts for future GW detectors on the ground and in space.
\end{abstract}

\maketitle

%%%%%%%%%%%%%%%%%%%%%%%%%%%%%%%%%%%%%%%%%%%%%%%%%%%%%%%

\section{Introduction}

The observations of gravitational waves (GWs) from coalescing binary
systems~\cite{LIGOScientific:2016aoc, LIGOScientific:2018mvr,
  LIGOScientific:2021usb, KAGRA:2021vkt, 
  Nitz:2021zwj, Mehta:2023zlk} with the LIGO and Virgo
detectors~\cite{LIGOScientific:2014pky, VIRGO:2014yos} have for the
first time allowed us to test Einstein's theory of General Relativity
(GR) in the highly nonlinear and dynamical
regime~\cite{LIGOScientific:2016lio, LIGOScientific:2018dkp,
  LIGOScientific:2019fpa, LIGOScientific:2020tif,
  LIGOScientific:2021sio}. Near-future upgrades to the LIGO, Virgo and
KAGRA~\cite{KAGRA:2020tym} (LVK) interferometers, and upcoming detectors on
the ground, Einstein Telescope (ET) and Cosmic
Explorer (CE)~\cite{Punturo:2010zz, Evans:2021gyd}, and in space
(LISA)~\cite{LISA:2017pwj}, will significantly increase the number of
detected sources, some of which will be observed with signal-to-noise
ratios (SNR) reaching thousands, allowing for tests of GR with
exquisite precision (see, e.g., Ref.~\cite{Perkins:2020tra}).

One can adopt either a theory-independent or theory-specific approach
to testing GR. Theory-independent tests modify GR waveform models by
introducing some generic, parameterized deviations, which are
constrained by comparing the waveform templates with an observed GW
signal (a non-exhaustive list includes Refs.~\cite{Blanchet:1994ez,
  Arun:2006hn, Yunes:2009ke, Li:2011cg, Agathos:2013upa,
  Barausse:2016eii, Brito:2018rfr, Cardoso:2019mqo, McManus:2019ulj,
  Maselli:2019mjd, Carullo:2019flw, Ghosh:2021mrv, Mehta:2022pcn}), or
look for consistency between the signal, or portions of the signal,
and the data, without invoking any parametrization of
the deviations~\cite{Hughes:2004vw, Ghosh:2016qgn, Ghosh:2017gfp, Madekar:2024zdj}.
Theory-specific tests (see, e.g., Refs.~\cite{Will:2014kxa, Berti:2015itd, Yunes:2016jcc, Silva:2022srr, Maselli:2023khq}) use instead
waveforms predicted in a particular alternative theory of gravity and
aim at estimating its underlying physical parameters. So
far, the majority of tests of GR with GWs have been carried
out following a theory-independent approach.  While theory-independent
tests can, in principle, constrain a wide range of alternative
theories, parameterized deviations are not unique, and degeneracies
among different deviation parameters complicate constraining multiple
of them at the same time unless specific combinations are
selected~\cite{Datta:2022izc}. Moreover, there is no guarantee that
parameterized deviations can represent the potentially infinite
landscape of beyond-GR theories (see Ref.~\cite{Xie:2024ubm} for an attempt to address 
this issue).
Furthermore, although such tests are valuable in identifying possible deviations from GR, they do not necessarily inform us about the new physics at play. Thus, it is relevant to develop, both analytically (see, e.g.,
Refs.~\cite{Damour:1992we,Damour:1993hw,Pani:2009wy,Yagi:2011xp,Mirshekari:2013vb,Lang:2013fna,Lang:2014osa,Blazquez-Salcedo:2016enn,Sennett:2016klh,Julie:2017pkb,Julie:2017ucp,Julie:2017rpw,Cardenas:2017chu,Julie:2018lfp,Bernard:2018hta,Bernard:2018ivi,Bernard:2019yfz,Khalil:2018aaj,Julie:2019sab,Shiralilou:2021mfl,Pierini:2021jxd,Bernard:2022noq,Julie:2022huo,Pierini:2022eim,Jain:2022nxs,Julie:2022qux,Jain:2023fvt,Creci:2023cfx,Bernard:2023eul,Julie:2023ncq})
and numerically (see, e.g.,
Refs.~\cite{Healy:2011ef,Barausse:2012da,Berti:2013gfa,Shibata:2013pra,Palenzuela:2013hsa,Okounkova:2017yby,Cayuso:2017iqc,Witek:2018dmd,Okounkova:2019dfo,Okounkova:2019zjf,Okounkova:2020rqw,Julie:2020vov,Witek:2020uzz,Silva:2020omi,East:2020hgw,East:2021bqk,Figueras:2021abd,Lara:2021piy,Corman:2022xqg,AresteSalo:2022hua,Doneva:2022byd,Elley:2022ept,Franchini:2022ukz,Hegade:2022hlf,Ma:2023sok,Cayuso:2023xbc,AresteSalo:2023mmd,Doneva:2023oww,Brady:2023dgu,Lara:2024rwa,Corman:2024cdr,Corman:2024vlk,Nee:2024bur}),
waveform models in specific beyond-GR theories of gravity.

One avenue to conduct theory-specific tests is to develop accurate inspiral-merger-ringdown (IMR) waveform models in alternative theories of gravity, combining analytical and numerical-relativity (NR) results, as done in GR~\cite{Schafer:2018kuf,Blanchet:2013haa,Buonanno:1998gg,Barack:2018yvs,Pretorius:2005gq, Campanelli:2005dd, Baker:2005vv}.
In this work, we provide the first example of a full IMR waveform model featuring analytical beyond-GR corrections. We focus on Einstein-scalar-Gauss-Bonnet (ESGB) theories, which supplement GR with a massless scalar field coupled to the Gauss-Bonnet scalar $\mathcal G=R^{\mu\nu\rho\sigma}R_{\mu\nu\rho\sigma} - 4R^{\mu\nu}R_{\mu\nu} + R^2$, with a coupling of the form $\ell_{\rm GB}^2f(\varphi)\mathcal G$ in the action.
These theories have attracted particular attention, thanks to their rich phenomenology for black holes (BHs) with scalar ``hair''~\cite{Mignemi:1992nt,Torii:1996yi,Yunes:2011we,Sotiriou:2013qea,Sotiriou:2014pfa,Kanti:1995vq,Pani:2009wy,Pani:2011gy,Ayzenberg:2014aka,Maselli:2015tta,Kleihaus:2015aje,Antoniou:2017acq,Cunha:2019dwb,Julie:2019sab,Julie:2022huo}.
In particular, for certain functions $f(\varphi)$, BHs can undergo spontaneous~\cite{Doneva:2017bvd,Silva:2017uqg,Minamitsuji:2018xde,Silva:2018qhn,Macedo:2019sem,Minamitsuji:2019iwp,Dima:2020yac}
or dynamical~\cite{Silva:2020omi,Doneva:2022byd,Elley:2022ept, Doneva:2022ewd, Julie:2023ncq} (de)scalarization phenomena analogous to those predicted for neutron stars (NSs) in scalar-tensor (ST) theories of gravity~\cite{Damour:1993hw, Barausse:2012da, Shibata:2013pra, Palenzuela:2013hsa}.
Different methods have been proposed to include dynamical scalarization in waveform models~\cite{Sennett:2016rwa,Sennett:2017lcx,Khalil:2019wyy,Khalil:2022sii}.
Recent progress in ESGB gravity was made to obtain inspiral waveforms within the post-Newtonian (PN) and effective-one-body (EOB) formalisms~\cite{Julie:2019sab,Julie:2022qux,Shiralilou:2021mfl, Julie:2022huo,vanGemeren:2023rhh},
to calculate quasi-normal-mode (QNM) frequencies of rotating BHs at second order in a small-spin expansion~\cite{Blazquez-Salcedo:2016enn,Pierini:2021jxd,Pierini:2022eim} and very recently for large spins~\cite{Chung:2024ira, Chung:2024vaf}, 
and to numerically simulate binary black holes (BBHs)~\cite{Okounkova:2017yby,Witek:2018dmd,Okounkova:2019dfo,Julie:2020vov,Witek:2020uzz,Okounkova:2020rqw,East:2020hgw,East:2021bqk,Figueras:2021abd,Corman:2022xqg,AresteSalo:2022hua,Brady:2023dgu,Doneva:2023oww,AresteSalo:2023mmd,Lara:2024rwa,Corman:2024cdr}, binary neutron stars (BNSs)~\cite{East:2022rqi,Kuan:2023trn}
and neutron star-black hole (NSBH) binaries~\cite{Corman:2024vlk}.

The most commonly used methods to build complete IMR waveform models for compact binaries in GR are the NR surrogate~\cite{Blackman:2015pia,Varma:2018mmi,Varma:2019csw}, phenomenological~\cite{Pan:2007nw,Ajith:2007qp,Hannam:2013oca,Pratten:2020ceb,Estelles:2021gvs,Thompson:2023ase} and EOB~\cite{Buonanno:1998gg,Buonanno:2000ef,Damour:2000we,Damour:2001tu,Buonanno:2005xu,Damour:2008gu} frameworks. EOB waveform models combine and resum information from several analytical approximation methods (PN, post-Minkowskian, gravitational self-force) with a physically motivated ansatz for the merger, and BH perturbation theory for the ringdown, and are made highly accurate through calibration with NR simulations. There are currently two state-of-the-art families of EOB waveform models: \texttt{SEOBNR} (see, e.g., Refs.~\cite{Pompili:2023tna,RamosBuadesv5,VandeMeentv5,Khalilv5,Ramos-Buades:2021adz}) and \texttt{TEOBResumS} (see, e.g., Refs.~\cite{Nagar:2018zoe,Nagar:2019wds,Nagar:2020pcj,Gamba:2021ydi,Nagar:2023zxh}). Here, we focus on the former. 

In this work, we start from the state-of-the-art EOB waveform model \texttt{SEOBNRv5PHM}~\cite{RamosBuadesv5}, which includes effects of higher harmonics and spin precession, and we include ESGB corrections to the EOB Hamiltonian, the GW and scalar-energy fluxes, the spherical-harmonic GW modes, the QNM spectrum, and the mass and spin of the remnant BH.
To account for the uncertainty in merger features that may only be determined when further NR simulations are available in ESGB gravity, we use parametrized deviations inspired by the \texttt{pSEOBNR} pipeline~\cite{Brito:2018rfr, Ghosh:2021mrv, Maggio:2022hre}, interpreted as nuisance parameters. 
We implement our waveform model within the \texttt{pySEOBNR} Python package~\cite{Mihaylovv5}, a code recently developed to construct, calibrate and validate new EOB waveform models with efficiency and flexibility. We refer to our new model with the name \seobesgb.

In a second step, we use \seobesgb~to provide constraints on the coupling length $\ell_{\rm GB}$ of the dilatonic theory $f(\varphi)=\exp(2\varphi)/4$ by analyzing the
GW events GW190412, GW190814 and GW230529$\_$181500 (to which we will refer as GW230529 for brevity) through Bayesian inference, and compare
our results to the literature.
We improve previous analyses~\cite{Wang:2021jfc,Perkins:2021mhb,Lyu:2022gdr} by
including ESGB corrections to the waveform for the entire coalescence,
and as a consequence, by analyzing the complete signal rather than
restricting to its inspiral phase. 
Our results for GW190814 are in agreement with Refs.~\cite{Wang:2021jfc,Lyu:2022gdr}, with a $90 \%$ bound on their coupling length $\sqrt{\alpha_{\mathrm{GB}}} \equiv \ell_{\rm GB}/(4\pi^{1/4})\lesssim 0.4~\mathrm{km}$. For GW230529, which currently places the best constraint on this theory, we obtain $\sqrt{\alpha_{\mathrm{GB}}}\lesssim \ConstraintCIMR \rm{km}$.
This is consistent with Refs.~\cite{Sanger:2024axs, Gao:2024rel}, which obtain a $90 \%$ bound on $\sqrt{\alpha_{\mathrm{GB}}}$ ranging from $0.26~\mathrm{km}$ to $0.35~\mathrm{km}$ depending on the waveform approximant employed.
On the other hand, we obtain from GW190412 the bound
$\sqrt{\alpha_{\mathrm{GB}}} \lesssim \ConstraintAIMR ~\mathrm{km}$, roughly a factor $\sim 2$ smaller than previous analyses~\cite{Wang:2021jfc}. This is
consistent with the larger mass of the source of GW190412, resulting
in a higher relative SNR in the late-inspiral stage, for which we are
including additional corrections. 
Stacking the posteriors of the three events, we obtain a combined bound 
$\sqrt{\alpha_{\mathrm{GB}}} \lesssim \ConstraintCombinedInspiral - \ConstraintCombinedIMR~\mathrm{km}$, depending on whether we include the merger-ringdown portion of the signal or not. 
Our waveform templates
also allow for Bayesian--model-selection studies, to address whether
the GR hypothesis is favored over a specific
modified gravity theory in describing GW data~\cite{DelPozzo:2011pg}. We compute Bayes
factors between the GR and ESGB hypotheses, finding the three analyzed events to be consistent with GR.

The paper is organized as follows.
In Sec.~\ref{sec:ESGBreminders} we introduce ESGB theories and their PN description.
In Sec.~\ref{sec:esgb_hamiltonian}, we review the structure of the EOB Hamiltonian in GR, and include its ESGB corrections at 3PN.
In Sec.~\ref{sec:esgb_flux} we present the ESGB metric and scalar energy fluxes on circular orbits and at relative 2PN and 2.5PN orders. 
In Section~\ref{sec:IMRwaveforms}, we recall the structure of the IMR spherical-harmonic waveform modes in GR, and obtain their ESGB corrections.
In Sec.~\ref{sec:results} we present our results on waveform morphology and Bayesian parameter estimation.
In Section~\ref{sec:conclusions}, we summarize our main conclusions and discuss future work. Various technical details are relegated to the
appendices.
In Appendix~\ref{App:EinsteinVsJordan} we present a dictionary relating quantities in the Einstein and Jordan frames.
In Appendix~\ref{sec:ST_ESGB_HeffReminder} we recall the ESGB corrections to the EOB metric.
In Appendix~\ref{App:metricModes} we gather factors entering the metric modes.
In Appendix~\ref{App:scalarFlux} we list the coefficients of the scalar flux. Finally, in Appendix~\ref{App:scalarModes} we provide the spherical-harmonic scalar modes at relative 1.5PN order. 
Throughout this paper we use geometrical units ($c = G =1$).

\section{Einstein-scalar-Gauss-Bonnet reminder\label{sec:ESGBreminders}}

\subsection{ESGB and ST gravity}

We study the class of theories described by the Einstein-frame action~\cite{Julie:2019sab, Damour:1992we}
\begin{align}
I  = \frac{1}{16\pi}  \int\! d^{4}x & \sqrt{-g} \big(R - 2g^{\mu\nu} \partial_{\mu}\varphi \partial_{\nu}\varphi +\ell_{\rm GB}^2 f(\varphi)\mathcal G\big)\nonumber\\
&+I_{\rm m}[\Psi,\mathcal A^2(\varphi)g_{\mu\nu}]\,,
\label{eq:action}
\end{align}
where $R$ is the Ricci scalar, $g = {\rm det} \, g_{\mu\nu}$ is the metric determinant, and
$\mathcal G=R^{\mu\nu\rho\sigma}R_{\mu\nu\rho\sigma} - 4R^{\mu\nu}R_{\mu\nu} + R^2$ is the Gauss-Bonnet scalar, where $R^{\mu}{}_{\nu\rho\sigma}$ and $R_{\mu\nu}$ are the Riemann and Ricci
tensors, respectively.
The integral $\int d^4x\sqrt{-g}\,\mathcal G$ over a four-dimensional spacetime is a boundary term~\cite{Myers:1987yn}.
As for matter fields $\Psi$, they couple minimally to the Jordan metric $\tilde g_{\mu\nu}=\mathcal A^2(\varphi)g_{\mu\nu}$.
The length $\ell_{\rm GB}$ and the dimensionless functions $\mathcal A$ and $f$ specify the theory.
We recover ST theories in the limit $\ell_{\rm GB}=0$ or if $f$ is constant,
and GR if moreover $\mathcal A$ (and $\varphi$) is constant.
\eject

To model compact bodies, we use the phenomenological treatment initiated in Refs.~\cite{1975ApJ196L59E,Damour:1992we} in ST theories,
and describe them as point particles:
\begin{align}
I_{\rm m}\to I_{\rm m}^{\rm pp}[g_{\mu\nu},\varphi, \{x_A^\mu\}]=-\sum_A\int m_A(\varphi)\, d s_A\,,
\label{eq:actionSkel}
\end{align}
 where $d s_A=\sqrt{-g_{\mu\nu}dx_A^\mu dx_A^\nu}$ and $x_A^\mu[s_A]$ is the worldline of particle $A$. 
The constant GR mass becomes a function $m_A(\varphi)$ that depends on the internal structure of body $A$ and on the value of the scalar field at $x_A^\mu(s_A)$.

In this paper, we will refer to the theory with action \eqref{eq:action} as ESGB gravity, but note that ST theories are included as a particular subcase.

\begin{table*}
\centering
\begin{tabular}{| @{\quad} c @{\quad} | @{\quad} l @{\quad} |}
  \hline & \\
 0PN & $G_{AB}=1+\alpha_A^0\alpha_B^0 $\\ & \\
 1PN & $\bar\gamma_{AB}=\frac{-2\alpha_A^0\alpha_B^0}{1+\alpha_A^0\alpha_B^0}$\qquad $\bar\beta_A=\frac{\beta_A^0(\alpha_B^0)^2}{2(1+\alpha_A^0\alpha_B^0)^2}$ \\ & \\
   2PN & $\delta_A=\frac{(\alpha_A^0)^2}{(1+\alpha_A^0\alpha_B^0)^2}$\qquad $\epsilon_A=\frac{{\beta'}_A^0(\alpha_B^0)^3}{(1+\alpha_A^0\alpha_B^0)^3}$\qquad $\zeta_{AB}=\frac{{\beta}_A^0{\beta}_B^0\alpha_A^0\alpha_B^0}{(1+\alpha_A^0\alpha_B^0)^3}$ \\ & \\
 3PN & $\kappa_A=\frac{(\alpha_B^0)^4{\beta''_A}^0}{8(1+\alpha_A^0\alpha_B^0)^4}$\qquad $\psi_A=\frac{\alpha_A^0\alpha_B^0\beta_A^0}{(1+\alpha_A^0\alpha_B^0)^3}$\qquad  $\xi_A=\frac{(\alpha_A^0)^2\alpha_B^0\beta_A^0{\beta'}_B^0}{(1+\alpha_A^0\alpha_B^0)^4}$\qquad  $\omega_A=\frac{(\alpha_A^0)^2\beta_A^0(\beta_B^0)^2}{(1+\alpha_A^0\alpha_B^0)^4}$  \\ & \\
 & $k_{\rm tail}=-\frac{2}{3}\frac{\alpha_-^2\,\nu}{(1+\alpha_A^0\alpha_B^0)^2}$\qquad $k_{\rm ESGB}=-\frac{\ell_{\rm GB}^2 f'(\varphi_0)}{2M^2}\frac{3\alpha_++m_-\alpha_-}{(1+\alpha_A^0\alpha_B^0)^4}$\\ & \\
  \hline & \\
(Anti) symmetry &  $\alpha_\pm=\alpha^0_A\pm \alpha^0_B$ \qquad $X_\pm=X_A\pm X_B$\quad with\quad $X_A=\{\bar\beta_A, \delta_A, \epsilon_A, \kappa_A, \psi_A, \xi_A, \omega_A\}$\\ & \\
  \hline & \\
Mean value & $ \langle\bar\beta\rangle=\frac{m_A^0\bar\beta_B+m_B^0 \bar\beta_A}{M}$\\ & \\
     & $\langle\delta\rangle=\frac{m_A^0\bar\delta_A+m_B^0 \delta_B}{M}\qquad \langle\epsilon\rangle=\frac{m_A^0\epsilon_B+m_B^0 \epsilon_A}{M}$\\ & \\
     & $\langle\kappa\rangle=\frac{m_A^0\kappa_B+m_B^0 \kappa_A}{M}\qquad \langle\psi\rangle=\frac{m_A^0\psi_B+m_B^0 \psi_A}{M} \qquad 
\langle \xi\rangle=\frac{m_A^0 \xi_A+m_B^0 \xi_B}{M} \qquad \langle \omega\rangle=\frac{m_A^0 \omega_A+m_B^0 \omega_B}{M}$\\ & \\
  \hline
\end{tabular}
\caption{Notations of this paper, following Refs.~\cite{Damour:1992we,Damour:1995kt,Julie:2017pkb,Julie:2017ucp,Julie:2019sab,Julie:2022qux}.
The quantities in the first five lines are ordered by the PN level at which they enter the conservative dynamics.
Note the factor of 2 compared to Refs.~\cite{Sennett:2016klh,Bernard:2018hta,Bernard:2018ivi} in our (anti) symmetrization.}
\label{table:PNcombinations}
\end{table*}

\subsection{Post-Newtonian framework}

We focus on the dynamics of compact binary systems on bound orbits. When the relative orbital velocity is small and the gravitational field is weak, the motion can be studied in the PN framework.\footnote{We denote by $n$PN the $\mathcal O(v^{2n})\sim\mathcal O(M/r)^n$ corrections to the Newtonian dynamics, with $v$ the system's relative orbital velocity, $M$ the total mass, and $r$ the orbital radius.}
To this aim, the field equations of the theory \eqref{eq:action} with the substitution \eqref{eq:actionSkel} are solved iteratively around a flat metric $g_{\mu\nu}=\eta_{\mu\nu}+\delta g_{\mu\nu}$, and a constant scalar field background $\varphi=\varphi_0+\delta\varphi$, where $\varphi_0$ is imposed by the binary's cosmological environment.

From now on, a superscript $0$ denotes a quantity evaluated at $\varphi=\varphi_0$.
We consider a binary system with $m^0_A \geq m^0_B$, and define its total mass, reduced mass, and the dimensionless ratios
\begin{equation}
\begin{gathered}
	M\equiv m^0_A + m^0_B, \qquad \mu \equiv \frac{m^0_Am^0_B}{M}, \qquad \nu \equiv \frac{\mu}{M},\\
	m_{-} \equiv\frac{m^0_A - m^0_B}{M},  \qquad q \equiv \frac{m^0_A}{m^0_B}, \qquad \hat \ell_{\rm GB} \equiv \frac{\ell_{\rm GB}}{\mu}.
\end{gathered}\label{eq:massRatios}%
\end{equation}
The functions $m_A(\varphi)$ and  $m_B(\varphi)$, describing bodies $A$ and $B$, can then be further expanded at 3PN by introducing the following quantities:
\begin{subequations}
\begin{align}
\alpha_A^0&=\frac{d\ln m_A}{d\varphi}(\varphi_0),\label{eq:sensiAlpha}\\
\beta_A^0&=\frac{d\alpha_A}{d\varphi}(\varphi_0),\label{eq:sensiBeta}\\
{\beta'}_A^0&=\frac{d\beta_A}{d\varphi}(\varphi_0),\\
{\beta''}_A^0&=\frac{d{\beta'}_A}{d\varphi}(\varphi_0),\label{eq:sensiBetapp}
\end{align}\label{eq:sensis}%
\end{subequations}
and their counterparts for body $B$.
We also introduce the notation $f'(\varphi_0)=(df/d\varphi)_{\varphi_0}$.

The ESGB corrections included in this work will depend only on the theory-dependent product $\ell_{\rm GB}^2f'(\varphi_0)$ and on ten body-dependent parameters: the masses of bodies $A$ and $B$ and their logarithmic derivatives~\eqref{eq:sensis} at infinity (i.e., at $\varphi=\varphi_0$), via the combinations gathered in Table~\ref{table:PNcombinations}.
We recover ST theories in the limit $\ell_{\rm GB}^2 f'(\varphi_0)=0$, and GR when moreover $m_A(\varphi)$ and $m_B(\varphi)$ are constants: then, Eqs.~\eqref{eq:sensis} and their $B$-counterparts are zero, so that only $G_{AB}=1$ is nonzero in Table~\ref{table:PNcombinations}.

Note that $k_{\rm tail}$ differs from its original definition in Ref.~\cite{Julie:2022qux} by a factor $\mathcal A_0^2$.
This factor originates from a typographical error in Ref.~\cite{Bernard:2018ivi}, which the author communicated to us, after it was pointed out in Ref.~\cite{Julie:2022qux} [cf. Sec. II.B there] that the terms in Ref.~\cite{Bernard:2018ivi} depending on $\mathcal A_0$ or $\alpha_0=(d\ln\mathcal A/\varphi)_{\varphi_0}$ (in the Einstein frame) should be revised. In practice, the factor $1/\phi_0$ in Eq.~(A3) of Ref.~\cite{Bernard:2018ivi} must be replaced by $1$.
The terms depending on $\alpha_0$ in Ref.~\cite{Bernard:2018ivi} still require revision, and they enter the present paper via $\tilde \alpha=(1+\alpha_A^0\alpha_B^0)/(1+\alpha_0^2)$ in Eq.~\eqref{eq:da4}.

\subsection{The example of dilatonic ESGB gravity}
\label{subsec:dilatonic}

The quantities~\eqref{eq:sensis} can be calculated once the theory and the bodies are specified.
In ST theories, they were derived numerically for NSs in Refs.~\cite{Damour:1992we,Zaglauer:1992bp,Damour:1993hw,Damour:1993hw,Zhao:2022vig}.
They were also calculated in ESGB models for BHs, including spontaneously and dynamically scalarized ones, in Refs.~\cite{Julie:2019sab,Julie:2022huo,Julie:2023ncq}.

For the sake of illustration, we will apply our results explicitly to the dilatonic ESGB theory\footnote{We translate between our conventions and those of Refs.~\cite{Wang:2021jfc, Lyu:2022gdr} by setting $\varphi=\sqrt{4\pi}\phi$ and $\ell_{\rm GB}= \sqrt{\alpha_{\rm GB}}\,4\pi^{1/4}$.\label{footnote:YagiTransl}}
\begin{align}
f(\varphi)=\frac{1}{4}\exp(2\varphi),\quad \mathcal A(\varphi)=1,\label{eq:dilatonicTheory}
\end{align}
for which the action \eqref{eq:action} becomes invariant under the simultaneous redefinitions $\varphi\to\varphi+\Delta\varphi$ and $\ell_{\rm GB}\to \ell_{\rm GB}\exp(-\Delta\varphi)$, where $\Delta\varphi$ is a constant.
In this theory, the quantities~\eqref{eq:sensis} were derived analytically and numerically for nonspinning BHs in Refs.~\cite{Julie:2019sab,Julie:2022huo}.
Given a BH $A$, they are functions of two parameters only: $\varphi_0$, which we set here to zero without loss of generality, using the symmetry of the theory; and the dimensionless ratio $\ell_{\rm GB}/m_A^0\lesssim 0.831$, where the bound ensures the regularity of the horizon~\cite{Julie:2022huo}.

In this paper, we will describe BBHs and NSBH binaries as follows.
For BBHs, we use the analytical, nonperturbative $(5,5)$-Padé approximant of $\alpha_A^0$ (and $\alpha_B^0$) discussed in Sec. III.B of Ref.~\cite{Julie:2022huo}, which exhibits remarkable agreement with its numerical counterpart.
We then infer $\beta_A^0$, ${\beta'}_A^0$ and ${\beta''}_A^0$ (and their counterparts for BH $B$) using Eqs.~\eqref{eq:sensiBeta}--\eqref{eq:sensiBetapp}.
Since $m^0_A \geq m^0_B$, it is sufficient to impose the horizon regularity bound on BH $B$.

By contrast, Eqs.~\eqref{eq:sensis} were so far not calculated for NSs in the dilatonic ESGB theory.
However, we can describe NSBH binaries at leading order in the small-$\ell_{\rm GB}$ approximation.
We indeed have, for a BH $A$~\cite{Julie:2019sab,Julie:2022huo},
\eject
\begin{align}
\left.\alpha_A^0\right|_{\rm BH}=-\frac{1}{4}\left (\frac{\ell_{\rm GB}}{m_A^0}\right)^2+\mathcal O\left (\frac{\ell_{\rm GB}}{m_A^0}\right )^4.\label{eq:sensiLeadingEll}
\end{align}
At this order, the theory reduces to the shift-symmetric ESGB model~\cite{Julie:2019sab}, in which NSs are known to carry no scalar monopole~\cite{Yagi:2011xp,Yagi:2015oca}.
Since Eq.~\eqref{eq:sensiAlpha} is proportional to the latter~\cite{Damour:1992we,Julie:2017rpw}, we conclude that for a NS $B$~\cite{Lyu:2022gdr},
\begin{align}
\left.\alpha_B^0\right|_{\rm NS}=\mathcal O\left (\frac{\ell_{\rm GB}}{m_B^0}\right )^4.\label{eq:sensiLeadingEllns}
\end{align}
In the small-$\ell_{\rm GB}$ approximation, the contributions from $\beta_A^0$, ${\beta'}_A^0$ and ${\beta''}_A^0$ (and their counterparts for NS $B$) are subleading and can be discarded, because they only enter the dynamics via the combinations gathered in Table \ref{table:PNcombinations}, which are already quadratic in $\alpha_A^0$ or $\alpha_B^0$.

The ST two-body Lagrangian was derived at 1PN by Damour and Esposito-Farèse~\cite{Damour:1992we}, at 2PN by Mirshekari and Will~\cite{Mirshekari:2013vb} and at 3PN by Bernard~\cite{Bernard:2018ivi}. It was then generalized by Julié and Berti, who derived its ESGB corrections in Ref.~\cite{Julie:2019sab}.
The resulting conservative dynamics was 
included within an EOB Hamiltonian at 2PN by Julié and Deruelle~\cite{Julie:2017pkb,Julie:2017ucp,Julie:2022qux}, and at 3PN by Julié et al.~\cite{Julie:2022qux}; see also Jain et al.~\cite{Jain:2022nxs,Jain:2023fvt} in the ST limit.
We include the main results as corrections to the \seobv5 Hamiltonian in Sec.~\ref{subSec:EOBHamiltonian}, and extend those works to the radiative sector in Secs.~\ref{sec:EOMs}, \ref{sec:esgb_flux} and~\ref{sec:IMRwaveforms}.

\section{Effective-one-body Hamiltonian and equations of motion}
\label{sec:esgb_hamiltonian}

\subsection{The EOB Hamiltonian\label{subSec:EOBHamiltonian}}

In the EOB formalism~\cite{Buonanno:1998gg,Buonanno:2000ef,Damour:2000we,Damour:2001tu,Buonanno:2005xu}, the two-body Hamiltonian in the center-of-mass frame is mapped, through canonical transformations, to the effective Hamiltonian $H_{\rm eff}$ of a test mass in a deformed BH spacetime, the deformation parameter being 
the symmetric mass-ratio $\nu$, as~\cite{Buonanno:1998gg}
\begin{equation}
    \hat{H}_{\mathrm{EOB}}=\frac{H_{\mathrm{EOB}}}{\mu}=\frac{1}{\nu} \sqrt{1+2 \nu\left(\hat{H}_{\mathrm{eff}}-1\right)}\,, 
\end{equation}
where $\hat{H}_{\rm eff} = H_{\rm eff}/\mu$. For nonspinning (noS) binaries, the motion is planar and we can use polar coordinates $(r,\phi,p_r,p_\phi)$.
We introduce the following reduced variables:
\begin{align}
\hat r=\frac{r}{M},\quad \hat p_r=\frac{p_r}{\mu},\quad \hat p_\phi=\frac{p_\phi}{M\mu},\quad \hat t=\frac{t}{M}.
\end{align}
The effective Hamiltonian then takes the form
\begin{equation}
\hat {H}_{\mathrm{eff}}^{\mathrm{noS}}=\frac{{H}_{\mathrm{eff}}^{\mathrm{noS}}}{\mu}=\sqrt{\hat p_{r_*}^2+A_{\mathrm{noS}}(\hat r)\left[1+\frac{\hat p_\phi^2}{\hat r^2}+Q_{\mathrm{noS}}\left(\hat r, \hat p_{r_*}\right)\right]}\,,\label{def:effectiveHamiltonian}
\end{equation}
where we use the tortoise-coordinate $\hat p_{r_*}$ instead of $\hat p_r$ to facilitate the numerical integration of the equations of motion close to merger.
Indeed, $\hat p_r$ diverges when approaching the zero of $A_{\mathrm{noS}}(\hat r)$, while $\hat p_{r_*}$ is finite~\cite{Damour:2007xr,Pan:2009wj}. For nonspinning binaries, the tortoise-coordinate $\hat r_*$ is defined by
\begin{equation}
\frac{d \hat r_*}{d\hat r}=\frac{1}{\xi_{\mathrm{noS}}(\hat r)}\,, \quad \xi_{\mathrm{noS}}(\hat r) \equiv A_{\mathrm{noS}}(\hat r) \sqrt{\bar{D}_{\mathrm{noS}}(\hat r)}\,,
\end{equation}
and the conjugate momentum $\hat p_{r_*}$ is
\begin{equation}
\label{prstar}
\hat p_{r_*}=\hat p_r\, \xi_{\mathrm{noS}}(\hat r)\,.
\end{equation}

In the general relativistic \seobv5 Hamiltonian~\cite{Khalilv5}, the potentials $A^{\rm GR}_\text{noS}$ and $\bar{D}^{\rm GR}_\text{noS}$ are based on the 5PN results of Refs.~\cite{Bini:2019nra,Bini:2020wpo}, which are complete modulo two unknown
constants at 5PN order, denoted $ a_6^{\nu^2}$ and $\bar d_5^{\nu^2}$ there.
In \seobv5, one sets $\bar d_5^{\nu^2}=0$ and replaces the 5PN coefficient in $A^{\rm GR}_\text{noS}$, with the exception of the part proportional to $\ln(\hat r)$, by a function $\nu\,a_6(\nu)$ calibrated to NR simulations~\cite{Pompili:2023tna}.
The potential $Q^{\rm GR}_\text{noS}$ uses the full 5.5PN expression of Refs.~\cite{Bini:2020wpo,Bini:2020hmy}, once $\hat p_{r}$ is traded for $\hat p_{r_*}$
by inverting Eq.~(\ref{prstar}) at the same order.
This potential is known to $\Order(\hat p_{r}^8)$, thus accounting for nonlocal-in-time ``tail" effects (entering at 4PN in GR) to eighth order in the binary's orbital eccentricity.
The complete potentials are given explicitly in Ref.~\cite{Khalilv5}

To include the ESGB effects up to 3PN order in the \seobv5 Hamiltonian, we follow the prescription of Ref.~\cite{Julie:2022qux}:\\

\begin{enumerate}
\item Formally replace $\hat r^{-1}$ by $u$, defined as
\begin{align}
u\equiv \frac{G_{AB}M}{r},\label{def:uESGB}
\end{align}
at all PN orders in the potentials $A^{\rm GR}_\text{noS}$, $\bar{D}^{\rm GR}_\text{noS}$ and $Q_\text{noS}$, where $G_{AB}$ is defined in Table~\ref{table:PNcombinations}.\\
\item Supplement the resulting potentials, which are for now Taylor-expanded (Tay) with respect to $u$, with the corrections found in Refs.~\cite{Julie:2017pkb,Julie:2022qux}: 
\begin{subequations}
\begin{align}
A^\text{Tay}_\text{noS}(u)&=A^{\rm GR}_\text{noS}(u)+2(\langle\bar\beta\rangle-\bar\gamma_{AB})u^{2}+\delta \bar a_{3}u^{3}\nonumber\\
&+[\delta  \bar a_4+ \bar a_{4,\ln}\ln u]u^{4},\\
\bar D^\text{Tay}_\text{noS}(u)&=\bar D^\text{\rm GR}_\text{noS}(u)
-2\bar\gamma_{AB} u+\delta \bar d_{2}u^{2}
+\delta \bar d_3 u^{3},\\
Q^\text{Tay}_\text{noS}(u,\hat p_{r_*})&=Q^{\rm GR}_\text{noS}(u,\hat p_{r_*})+\delta  \bar q_{1}\hat p_{r_*}^4u^{2}+ \bar q_{2} \hat p_{r_*}^6u,\label{eq:finalEOBpotentialsQ}%
\end{align}\label{eq:finalEOBpotentials}%
\end{subequations}
where $\langle\bar\beta\rangle$ and $\bar\gamma_{AB}$ are defined in Table~\ref{table:PNcombinations}, and where the explicit expressions of $(\delta \bar a_{3},\, \delta \bar a_{4},\, \bar a_{4,\ln})$,
 $(\delta \bar d_{2},\, \delta \bar d_3)$ and $(\delta\bar q_{1},\, \bar q_{2})$ are given in Appendix~\ref{sec:ST_ESGB_HeffReminder}.\\
\end{enumerate}
We note that the potential $Q$ in Ref.~\cite{Julie:2022qux} is a function of $\hat p_r$, and it is linked to our potentials through $Q=A_\text{noS}Q_\text{noS}$.
Since it was calculated at leading PN order, our $(\delta\bar q_{1},\, \bar q_{2})$ are the same as those presented in Ref.~\cite{Julie:2022qux}.
The ESGB corrections in Eq.~\eqref{eq:finalEOBpotentialsQ} are known to $\mathcal O(\hat p_r^6)$, which amounts to including the ESGB ``tail" effects to sixth order in the eccentricity.
These effects already appear at 3PN, due to (dipolar) scalar radiation that is absent in GR.

To improve the accuracy of the model against NR waveforms in the GR limit, we resum $A_\text{noS}^\text{Tay}(u)$ using a (1,5)-Pad\'e approximant with respect to $u$~\cite{Damour:2000we,Buonanno:2007pf,Damour:2007yf,Nagar:2018zoe,Pompili:2023tna}
\begin{equation}
\label{eq:A_esgb}
A_\text{noS}(u) = P^1_5[A_\text{noS}^\text{Tay}(u)],
\end{equation}
which features a simple zero (by construction), consistently with the Schwarzschild metric recovered in the GR, test-mass limit.
We also resum $\bar{D}_\text{noS}(u)$ as a (2,3)-Pad\'e approximant~\cite{Pompili:2023tna,Nagar:2021xnh}
\begin{equation}
\label{eq:D_esgb}
\bar{D}_\text{noS}(u) = P^2_3[\bar{D}_\text{noS}^\text{Tay}(u)],
\end{equation}
which remains numerically close to the Schwarzschild value 1 in the range $u\lesssim 1$. 
Note that the ESGB corrections in the $A_\text{noS}$ and $\bar{D}_\text{noS}$ potentials are added before performing the Pad\'e resummations (which are the same as in GR) so that the potentials remain qualitatively consistent with their GR counterparts as $\ell_{\rm GB}$ is increased.

\begin{figure}
	\includegraphics[width=\linewidth]{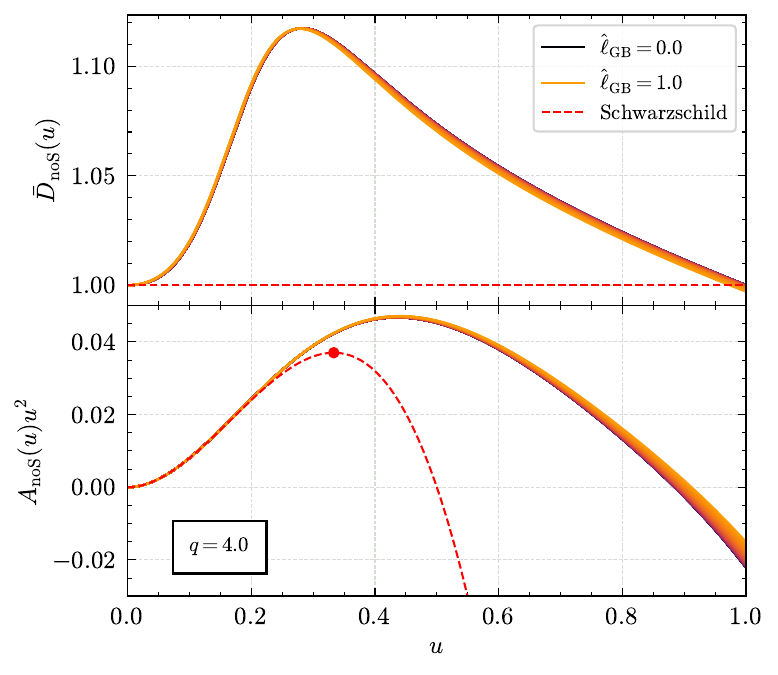}\\
	\caption{
    Pad\'e-resummed EOB potentials $A_{\rm noS}(u) u^2$ and $\bar{D}_{\rm noS}(u)$ for a nonspinning BBH with $q=4$.
    We include ESGB corrections for the dilatonic theory $f(\varphi)=\exp(2\varphi)/4$ and $\mathcal A(\varphi)=1$, and vary the Gauss-Bonnet coupling $\hat \ell_{\mathrm{GB}}$ between $0$ and $1$.
    We add the ESGB corrections before performing the Pad\'e  resummations, so that the potentials remain qualitatively consistent with their GR counterparts as $\hat \ell_{\mathrm{GB}}$ increases.
    The peak of the $A_{\rm noS}(u)u^2$ function corresponds to the value of $u=G_{AB}M/r$ at the effective light-ring, located at $u=1/3$ in the Schwarzschild metric (red dot).} 
	\label{fig:potentials}
\end{figure}

For generic nonzero spins, the \seobv5 Hamiltonian takes inspiration from that of a test mass in a deformed Kerr background~\cite{Damour:2001tu,Balmelli:2015zsa,Khalil:2020mmr,Khalilv5}. 
It includes spin-orbit (SO) information up to next-to-next-to-leading order (NNLO), spin-spin (SS) information to NNLO, as well as cubic- and quartic-in-spin terms at leading order (LO), corresponding to the 4PN order for {generic, precessing spins. For efficiency, some spin-precession effects are analytically averaged out over circular orbits~\cite{Khalilv5}.
The tortoise coordinate is also extended as
\begin{equation}
  \xi(\hat r)=\frac{\sqrt{\bar{D}_{\mathrm{noS}}}\left(A_{\mathrm{noS}}+\chi_{+}^2 / \hat r^2\right)}{1+\chi_{+}^2 / \hat r^2},\label{def:newXi}
\end{equation}
with $\chi_{+} = (m_A^0/M) \chi_A + (m_B^0/M) \chi_B$, where $\chi_A$ and $\chi_B$ are the dimensionless spins of the bodies {projected along the orbital angular momentum at Newtonian order.
More details and full expressions can be found in Refs.~\cite{Khalilv5,Pompili:2023tna,RamosBuadesv5}.
We do not add any ESGB correction in the spin contributions to the Hamiltonian, because they are currently unknown.

The ESGB corrections to the EOB potentials in Eqs.~\eqref{eq:A_esgb} and ~\eqref{eq:D_esgb} are valid for arbitrary choices of the functions $f(\varphi)$ and $\mathcal A(\varphi)$ entering the action~\eqref{eq:action}.
They can be specified to a particular theory and compact binary once the corresponding values of the quantities~\eqref{eq:sensis} are given.
As an explicit illustration, we show in Fig.~\ref{fig:potentials} the potentials $A_\text{noS}$ and $\bar{D}_\text{noS}$ for a nonspinning BBH with $q=4$ in the dilatonic ESGB case~\eqref{eq:dilatonicTheory}, varying the rescaled Gauss-Bonnet coupling [cf. Eq.~\eqref{eq:massRatios}] in the range $0<\hat \ell_{\mathrm{GB}}<1$.
The value $\hat \ell_{\mathrm{GB}}=1$ is just below that at which the BH $B$ becomes a naked singularity.
Adding back the units, this corresponds to a coupling $\sqrtalpha \simeq 0.044\,\mathrm{km} ~(M/M_{\odot})$, in the notation of Ref.~\cite{Lyu:2022gdr}.
We note that the peak of $A_{\rm noS}(u)\,u^2$ corresponds to the value of $u$ at the effective light-ring.
While the differences in $A_{\rm noS}(u)$ may seem small, especially below $u \simeq 0.4$, we recall that $\Delta A_{\rm noS}\gtrsim 10^{-4}$ can still lead to significantly different conservative dynamics~\cite{Nagar:2021xnh}, as is the case here.

 \subsection{The EOB equations of motion and radiation-reaction force\label{sec:EOMs}}

For nonspinning binaries or in the spin-aligned case,
the \seobv5 equations of motion are given by Eqs.~(15) of Ref.~\cite{Pompili:2023tna}. They read:
\begin{align}
\label{eq:eom_prst}
\begin{aligned}
\dot{\hat r} &= \xi \frac{\partial \hat H_{\rm EOB}}{\partial \hat p_{r_*}}, \quad
&\dot{\hat p}_{r_*} &= -\xi \frac{\partial \hat H_{\rm EOB}}{\partial \hat r} + \hat \mF_r, \\
\dot{\phi} &= \frac{\partial \hat H_{\rm EOB}}{\partial \hat p_\phi}, \quad
&\dot{\hat p}_\phi &= \hat \mF_\phi,
\end{aligned}
\end{align}
where the dot indicates a derivative with respect to $\hat t$,  while $\hat \mF_r = \mF_r/\mu$ and $\hat \mF_\phi = \mF_\phi/\mu$.
The radiation reaction (RR) force is computed as \cite{Buonanno:2005xu}
\begin{equation}
\mathcal{F}_\phi=-\frac{\mF}{\Omega}, \quad \mathcal{F}_r=\mathcal{F}_\phi \frac{\hat p_{r_*}}{\hat p_\phi},
\end{equation}
where $M\Omega= \dot{\phi}$ is the reduced orbital frequency, and $\mF$ is the energy flux radiated by the binary on quasi-circular orbits.
The metric flux is obtained by summing over the contributions of the factorized metric modes $h_{\ell m}^{\mathrm{F}}$~\cite{Damour:2007xr,Damour:2007yf,Damour:2008gu,Pan:2010hz}, which we will define in Sec.~\ref{sec:waveform-modes}, that is
\begin{equation}
\label{RRforce}
\mF^{\mathrm{metric}} \equiv \frac{\Omega^2}{8 \pi} \sum_{\ell=2}^8 \sum_{m=1}^{\ell} m^2\left|R h_{\ell m}^{\mathrm{F}}\right|^2,
\end{equation}
where $R$ is the distance to the source.
In ESGB theories, we have to account for an additional contribution to the flux from the scalar field,
\begin{equation}
    \mF = \mF^{\mathrm{metric}} + \mF^{\mathrm{scalar}},
\end{equation}
which we will discuss in Sec.~\ref{sec:scalar_flux}.
In the following, we express $\mF$ in terms of $\Omega$,
and insert the result back into the system~\eqref{eq:eom_prst}.

In the spin-precessing case, the \texttt{SEOBNRv5PHM} model uses the same equations of motion as in Eqs.~(\ref{eq:eom_prst}) in the co-precessing frame, in which the GW radiation resembles that of an aligned-spin binary~\cite{Buonanno:2002fy, Schmidt:2010it, Boyle:2011gg, OShaughnessy:2011pmr, Schmidt:2012rh}, together with separate PN-expanded evolution equations for the spins and angular momentum inferred from $\hat H_{\rm EOB}$~\cite{Khalilv5, RamosBuadesv5}.
For this work, we employ the same description of the precession dynamics as in the \texttt{SEOBNRv5PHM} model and refer the reader to Ref.~\cite{Khalilv5}  and Sec. II of Ref.~\cite{RamosBuadesv5} for more details.
We point out that in the future, the spin and angular momentum evolution equations should be generalized to account for their ESGB corrections.

\section{Gravitational radiation in ESGB gravity}
\label{sec:esgb_flux}

In this section, we adopt the conventions of Refs.~\cite{Faye:2012we,Henry:2022dzx}, and define the metric perturbation:
\begin{align}
h^{\mu\nu}=\sqrt{-g}g^{\mu\nu}-\eta^{\mu\nu},\label{def:metricPerturb}
\end{align}
whose indices are manipulated with the Minkowski metric.
Let us introduce a Cartesian orthonormal basis $(\mathbf e_X,\mathbf e_Y,\mathbf e_Z)$.
In the center-of-mass frame of a spin-aligned binary, the relative motion is planar and we set the source  position as $ r^i=r\left(\cos \phi,\,\sin \phi,\,0\right)$. 
We locate the observation point with spherical coordinates, such that
 $R^i=R\left(\sin\Theta\,\cos\Phi,\,\sin\Theta\,\sin\Phi,\,\cos\Theta\right)$.
When $R\gg r$, the metric waveform is the transverse, traceless part of the metric perturbation,
\begin{align}
h_{ij}^{\rm TT}=\left(P^k_i P^l_j-\frac{1}{2}P^{kl}P_{ij}\right)\,h_{kl}+\mathcal O\left(\frac{1}{R^2}\right)\,,\label{def:TTmetricWaveform}
\end{align}
where $P^{ij}=\delta^{ij}-N^iN^j$ and $N^i=R^i/R$.
Finally, we introduce the polarization vectors $P^i=-\partial_\Phi N^i/\sin\Theta$ and $Q^i=\partial_\Theta N^i$ and define:
 \begin{subequations}
 \begin{align}
 h_+&=\frac{1}{2}\left(P^iP^j-Q^iQ^j\right)h_{ij}^{\rm TT},\\
 h_\times&=\frac{1}{2}\left(P^iQ^j+Q^iP^j\right)h_{ij}^{\rm TT},
 \end{align}\label{eq:PlusCrossPolarizations}%
 \end{subequations}
which can in turn be decomposed in spin-weighted spherical harmonics of weight $-2$,
\begin{align}
h_+-i h_\times=\sum_{\ell= 2}^\infty\sum_{m=-\ell}^\ell {}_{-2} Y_{\ell m} (\Theta, \Phi) \,h_{\ell m}\,.\label{eq:sphericalHamonicsMetricDecomposition}
\end{align}

\subsection{Spherical-harmonic metric modes at 2PN}
\label{sec:metric_flux}

In ST theories, 
the metric waveform~\eqref{def:TTmetricWaveform}
was derived for nonspinning binaries at 0PN by Damour and Esposito-Farèse~\cite{Damour:1992we} and at 2PN by Lang~\cite{Lang:2013fna,Lang:2014osa} relatively to the quadrupolar order.
The corresponding metric modes  $h_{\ell m}$ were then inferred by Sennett et al. in Ref.~\cite{Sennett:2016klh} for circular orbits. The corrections beyond ST
from the Gauss-Bonnet coupling $\ell_{\rm GB}$ were calculated at leading PN order by Shiralilou et al.~\cite{Shiralilou:2020gah,Shiralilou:2021mfl}, but they enter at 3PN.
This means that the metric modes are complete only at 2PN in ST and ESGB gravity, an order at which their expressions are the same.
Rather, they differ by the specific values taken by the quantities \eqref{eq:sensis}.

Yet, the results in Ref.~\cite{Sennett:2016klh} were presented using another, Jordan-frame formulation of ST theories based on a set of Brans-Dicke-inspired parameters.
To recover the conventions of the present paper, we must proceed as follows:\\

\begin{enumerate}
\item \label{point1:MetricDecomposition}  Translate the parameters in Ref.~\cite{Sennett:2016klh} in terms of the quantities \eqref{eq:sensis}. The conversion is detailed in Appendix~\ref{App:EinsteinVsJordan}.\\
\item \label{point2:MetricDecomposition} Notice that Ref.~\cite{Sennett:2016klh} uses a coordinate system $\{ {x'}^\mu\}$ such that the Jordan metric $\tilde g_{\mu\nu}=\mathcal A^2(\varphi)g_{\mu\nu}$ is Minkowski at infinity, $\tilde g'_{\mu\nu}\to\eta_{\mu\nu}$.
By contrast, we use here coordinates $\{x^\mu\}$ such that $g_{\mu\nu}\to\eta_{\mu\nu}$.
The coordinate systems are thus related by
\begin{align}
{x'}^\mu=\mathcal A_0 x^\mu,
\end{align}
with $\mathcal A_0=\mathcal A(\varphi_0)$.
This means that $R'$, $t'$, and the orbital velocity $\omega'$ entering Ref.~\cite{Sennett:2016klh} must be rescaled as $R'=\mathcal A_0 R$, $t'=\mathcal A_0 t$, and $\omega'=\Omega/\mathcal A_0 $.\\
\item Observe that Ref.~\cite{Sennett:2016klh} studies the perturbation of $(\phi/\phi_0)\,\tilde g_{\mu\nu}'$ where $\phi=\mathcal A(\varphi)^{-2}$ is the scalar field in the Jordan frame.
This conformal metric identifies to our $g_{\mu\nu}$ given the coordinate change above.
The perturbation~\eqref{def:metricPerturb}
also differs from that of Ref.~\cite{Sennett:2016klh} by an overall minus sign.
However, their polarization vectors are $\tilde P^i=Q^i$ and $\tilde Q^i=-P^i$, so that our $h_+$ and $h_\times$ are the same.\\
\end{enumerate}

In harmonic coordinates $(t,R^i)$ such that $\partial_\mu h^{\mu\nu}=0$, the metric waveform exhibits a logarithmic correction to its $\sim 1/R$ decay, due to tail effects at 1.5PN~\cite{Lang:2013fna,Lang:2014osa}. To eliminate it, we can introduce the radiative time
\begin{align}
T=t-2M\ln\left(\frac{R}{R_0}\right),\label{def:radiativeTime}
\end{align}
where $R_0$ is an arbitrary length.
The ESGB metric modes then read, at 2PN and in the radiative coordinate system $(T,R^i)$:
\begin{align}
h_{\ell m}(T,R)=\frac{2M\nu x}{R}\sqrt{\frac{16\pi}{5}}\hat H_{\ell m}e^{-im\psi},\label{def:metricModes}
\end{align}
where
\begin{align}
x=(G_{AB}M\Omega)^{2/3},\label{def:x}
\end{align}
with, recall, $M \Omega= \dot{\phi}$ [cf. below Eq.~\eqref{eq:eom_prst}], and where 
\begin{align}
\psi&=\phi-\Phi+\frac{\pi}{2}-\frac{2x^{3/2}}{1+\alpha_A^0\alpha_B^0}\left(\gamma_{\rm E}-\frac{11}{12}+\ln\left(4\Omega R_0\right)\right),\label{def:effectivePhase}
\end{align}
where $\gamma_E$ is Euler's constant.
The effective phase \eqref{def:effectivePhase} was first introduced in GR to factor out $\ln(\Omega)$-dependent tail corrections to the modes~\cite{Blanchet:1995ez}, and it is now dressed-up by the scalar field effects.
As for the dimensionless quantities $\hat H_{\ell m}$, we gather their explicit expressions in Appendix \ref{App:metricModes}.
Note that the orbit being planar, we have by symmetry~\cite{Faye:2012we}
\begin{align}
h_{\ell,-m}=(-1)^\ell (h_{\ell m})^*,
\end{align}
where $*$ denotes complex conjugation. Thus, it is sufficient to present the modes with $m\geq 0$.

In Sec.~\ref{sec:waveform-modes}, we shall resum the modes \eqref{def:metricModes} by rewriting them in a factorized form $h_{\ell m}^{\mathrm{F}}$, allowing in turn to compute the metric flux~\cite{Damour:1992we,Deruelle:2018ltn} (cf. also Refs.~\cite{Khalil:2018aaj,Julie:2018lfp,Shiralilou:2021mfl}),
\begin{align}
\mathcal F^{\rm metric}=\frac{R^2}{32\pi}\int_{R\to\infty}\!\!\!\!\!\! d^2\Omega\, \left(\frac{\partial h_{ij}^{\rm TT}}{\partial U}\right)^2\,,\label{def:metricFlux}
\end{align}
where $U=T-R$.
This flux formula returns Eq.~\eqref{RRforce} on circular orbits, given the orthogonality of the spherical harmonics $\int d^2\Omega\, ({}_{-2} Y_{\ell m}) ({}_{-2} Y_{\ell' m'})^*=\delta_{\ell \ell'}\delta_{m m'}$, the contribution of each mode $m<0$ being the same as that of its counterpart $-m$.

\subsection{Scalar flux at relative 2.5PN order}
\label{sec:scalar_flux}

In ST theories, the scalar waveform
\begin{align}
\delta\varphi=\varphi-\varphi_0+\mathcal O\left(\frac{1}{R^2}\right)\label{def:scalarWaveform}
\end{align}
was derived for nonspinning binaries at 2PN by Lang~\cite{Lang:2014osa} and at 2.5PN by Bernard et al.~\cite{Bernard:2022noq} relatively to the (scalar) dipolar order.
The decomposition of $\delta\varphi$ on spherical harmonics was performed in Ref.~\cite{Bernard:2022noq}, but the associated scalar modes are presented there through 1.5PN only.
In this paper, we thus start from the scalar flux itself, which was derived at 2.5PN in Ref.~\cite{Bernard:2022noq}.
As for the corrections beyond ST from the Gauss-Bonnet coupling $\ell_{\rm GB}$ to the scalar flux, they were calculated at leading PN order by Shiralilou et al.~\cite{Shiralilou:2020gah,Shiralilou:2021mfl}, but they enter at 3PN, an order at which even the ST subcase is unknown.

Once again, the results in Refs.~\cite{Lang:2014osa,Bernard:2022noq} were presented using the Jordan-frame formulation of the theory.
We recover our conventions as follows:\\

\begin{enumerate}
\item Repeat steps \ref{point1:MetricDecomposition} and \ref{point2:MetricDecomposition} from Sec.~\ref{sec:metric_flux}.\\
\item Notice that Refs.~\cite{Lang:2014osa,Bernard:2022noq} study the scalar perturbation $\Psi=\phi/\phi_0-1$, where $\phi=\mathcal A(\varphi)^{-2}$ is the scalar field in the Jordan frame.
Our $\delta\varphi$ is thus related to $\Psi$ through $\delta\varphi=-\Psi/(2\alpha_0)$, where $\alpha_0$ is defined in Appendix~\ref{App:EinsteinVsJordan}.\\
\item Since $ t'=\mathcal A_0 t$ and the total flux is identified to the mechanical energy loss $\mathcal F=-dH/dt$ of the system, we must rescale the scalar fluxes of Refs.~\cite{Lang:2014osa,Bernard:2022noq} as $\mathcal F^{\rm scalar}=\mathcal A_0^2\,\tilde{\mathcal F}^{\rm scalar}$.\\
\end{enumerate}

As a result of this procedure, the scalar flux formula in Refs.~\cite{Lang:2014osa,Bernard:2022noq} becomes
\begin{align}
\mathcal F^{\rm scalar}=\frac{R^2}{4\pi}\int d^2\Omega\left(\frac{\partial \delta\varphi}{\partial U}\right)^2,\label{def:scalarFlux}
\end{align}
which coincides with our Einstein-frame definition, cf. also Refs.~\cite{Damour:1992we,Julie:2018lfp,Khalil:2018aaj}.
For circular orbits, the ESGB scalar flux then reads:
\begin{align}
    \label{eq:scalar_flux}
    \mathcal F^{\rm scalar}=\frac{\nu^2x^5}{\left(1+\alpha_A^0\alpha_B^0\right)^2}
&\Big[x^{-1}f_{-1\rm{PN}}+f_{0\rm{PN}}+x^{1/2}f_{0.5\rm{PN}}\nonumber\\
&+xf_{1\rm{PN}}+x^{3/2}f_{1.5\rm{PN}}+\cdots\Big],
\end{align}
where $x$ was defined in Eq.~\eqref{def:x}, and where the lengthy expressions of the coefficients $f_{i\rm{PN}}$ are relegated to Appendix \ref{App:scalarFlux}.
To make the comparison with the metric flux transparent, the labels $i$PN are here relative to the quadrupolar order.

We point out that $f_{-1\rm{PN}}$ is proportional to $(\alpha_A - \alpha_B)^2$. Thus, when restricted to the dipolar order, the scalar flux vanishes for symmetric binaries.
In Fig.~\ref{fig:flux-1}, we show the behavior of the scalar flux~(\ref{eq:scalar_flux}) truncated at successive PN orders, as a function of $x$.
As an illustration, we consider a BBH with $q=4$ and $\hat \ell_{\mathrm{GB}} =1$ in the dilatonic ESGB theory~\eqref{eq:dilatonicTheory}. The curves terminate at merger, taken as the peak of the $(2,2)$ mode amplitude when all PN orders of the scalar flux contribute to the binary's evolution. We also show the corresponding number of GW cycles.
The ratio between the scalar flux truncated at different PN orders and the leading order -1PN term highlights the poor convergence of the low PN order contributions in the late inspiral, which seems, however, improved when including higher-order PN corrections.
We leave the comparison of potential resummation methods for the scalar flux to future work, when more NR waveforms become available in ESGB gravity.
\begin{figure}
\includegraphics[width=\linewidth]{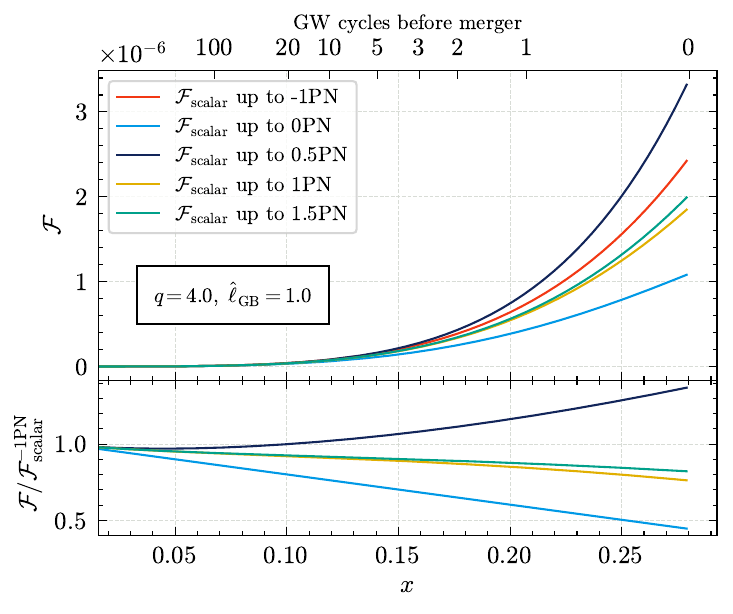}\\
	\caption{\textit{Top panel:} The PN-expanded scalar flux given in Eq.~(\ref{eq:scalar_flux}), truncated at successive PN orders. We consider a BBH with $q=4$ and $\hat \ell_{\mathrm{GB}} =1$ in the dilatonic theory $f(\varphi)=\exp(2\varphi)/4$ and $\mathcal A(\varphi)=1$. The curves terminate at merger, taken as the peak of the $(2,2)$ mode when all PN orders of the scalar flux contribute to the binary's evolution, and the corresponding number of GW cycles. \textit{Bottom panel:} Ratio between the scalar flux truncated at different PN orders and the leading -1PN term.}
	\label{fig:flux-1}
\end{figure}

For completeness, we present in Appendix~\ref{App:scalarModes} the decomposition of the scalar waveform in spherical harmonics, once translated in our Einstein-frame conventions. As explained below Eq.~\eqref{def:scalarWaveform},
we do not infer the scalar flux from these modes, because the result would be one PN order lower than in Eq.~\eqref{eq:scalar_flux}.

\section{EOB inspiral-merger-ringdown metric waveform in ESGB gravity\label{sec:IMRwaveforms}}

In the EOB framework, the metric modes defined in Eq.~(\ref{eq:sphericalHamonicsMetricDecomposition}) are decomposed into inspiral-plunge and merger-ringdown modes. The \texttt{SEOBNRv5PHM} model includes the modes $(\ell, |m|)=$  (2,2), (2,1), (3,3), (3,2), (4,4), (4,3) and (5,5) in the co-precessing frame.
We have: 
\begin{equation}
	\label{eq:h_match}
	h_{\ell m}(\hat t)= \begin{cases}h_{\ell m}^{\text {insp-plunge }}(\hat t), & \hat t<\hat t_{\text {match }}^{\ell m} \\ h_{\ell m}^{\text {merger-RD }}(\hat t), & \hat t>\hat t_{\text {match }}^ {\ell m }\end{cases},
\end{equation}
where $\hat t_{\text {match }}^{\ell m}$ is the peak time of the $(2,2)$-mode amplitude ($\hat t_{\text {peak }}^{22}$) for all the modes except $(5,5)$, for which it is taken as the peak time of the $(2,2)$ mode minus $10 M$~\cite{Pompili:2023tna}.
From now on, we shall refer to $\hat t_{\text {peak }}^{22}$ as the merger time. Note that the other modes with $\ell\leq 8$ are known in their factorized form (cf. below), but they are only used to compute the fluxes until merger. 

In GR, $\hat t_{\text {peak}}^{22}$ is suitably chosen to agree with the corresponding time in NR simulations. Specifically, in the \seobv5 model, one defines
\begin{equation}
\label{eq:tISCO}
\hat t_{\text {peak }}^{22}=\hat t_{\mathrm{ISCO}}+\Delta \hat t_{\mathrm{NR}}
\end{equation}
where $\hat t_{\rm{ISCO}}$ is the time at which $u = u_{\rm{ISCO}}$ reaches the inverse radius of the Kerr innermost stable circular orbit (ISCO)~\cite{Bardeen:1972fi} with the mass and spin of the remnant BH, as predicted by NR fitting formulas~\cite{Jimenez-Forteza:2016oae, Hofmann:2016yih}.
The calibration parameter $\Delta \hat t_{\rm NR}$ is determined via comparisons to NR simulations~\cite{Pompili:2023tna}.
In this work, we also use Eq.~\eqref{eq:tISCO}, formally replacing $u$ by its ESGB expression~\eqref{def:uESGB}, to determine the merger-ringdown attachment time. We note that $\hat t_{\mathrm{ISCO}}$ is affected by ESGB corrections, as they give rise to differences in the orbital dynamics, but we don't include additional corrections to $\Delta \hat t_{\rm NR}$, which should be determined from NR simulations in ESGB gravity.

\subsection{Inspiral-plunge $h_{\ell m}$ modes}
\label{sec:waveform-modes}

The inspiral-plunge EOB modes can be written as
\begin{equation}
	\label{eq:hellm_insp}
	h_{\ell m}^{\text {insp-plunge }}=h_{\ell m}^{\mathrm{F}} N_{\ell m}\,,
\end{equation}
where $h_{\ell m}^{\mathrm{F}}$ is a factorized, resummed form of the PN-expanded metric modes for aligned-spin binaries on circular orbits 
\cite{Damour:2007xr,Damour:2008gu,Pan:2010hz},
while $N_{\ell m}$ is the non-quasi-circular (NQC) correction~\cite{Damour:2002vi} which does not enter the flux~\eqref{RRforce} in the \seobv5 model. More precisely, the factorized metric modes are defined as
\begin{equation}
\label{hlmFactorized}
	h_{\ell m}^{\mathrm{F}}=h_{\ell m}^{\text{N}} \hat{S}_{\mathrm{eff}} T_{\ell m} f_{\ell m} e^{i \delta_{t m}}.
\end{equation}

In GR, the first factor $h_{\ell m}^{\text{N,GR}}$ is the mode at leading PN order (called Newtonian).
Its explicit expression can be found in Refs.~\cite{Damour:2008gu,Pan:2010hz} for all $(\ell,m)$, with an overall minus sign compared to our conventions.
 The second factor $\hat{S}_\text{eff}^{\rm GR}$ takes inspiration from the Regge-Wheeler-Zerilli equation sourced by
 a test mass $\mu$, on circular orbits
 of the Schwarzschild spacetime of mass $M$~\cite{Damour:2008gu}.
Depending on the parity of $\ell+m$, it is either identified to the (reduced) effective energy of the EOB framework, or to the orbital angular momentum, normalized by its Newtonian value on circular orbits:
\begin{equation}
\hat{S}_\text{eff}^{\rm GR} = \left\{
        \begin{array}{ll}
            \hat H_\text{eff}^{\rm GR}, & \quad \ell + m \text{ even}, \\\\
          \sqrt{x_{\rm GR}} {\hat p_\phi}, & \quad \ell + m \text{ odd}.
        \end{array}
    \right.
\end{equation}
Here, $x_{\rm GR}=(M\Omega)^{2/3}$ and we recall that $ M\Omega=\dot{\phi}$ was introduced below Eq.~\eqref{eq:eom_prst}.
The third factor $T_{\ell m}^{\rm GR}$ resums leading logarithms entering the tail contributions~\cite{Blanchet:1997jj}:
\begin{align}
T_{\ell m}^{\rm GR}=\frac{\Gamma\left(\ell+1-2i\hat k_{\rm GR}\right)}{\Gamma(\ell+1)}e^{\pi \hat k_{\rm GR}}e^{2i\,\hat k_{\rm GR}\ln(2m\Omega R_0)},
\end{align}
where $\Gamma(.)$ is the Euler gamma function, while $\hat k_{\rm GR}=m\, H_{\rm EOB}^{\rm GR}\Omega$ is proportional to the binary's resummed EOB energy.
The constant $R_0$ is the same as that entering Eq.~\eqref{def:effectivePhase}. In GR, it is usually set to $R_0=2M/\!\sqrt{e}$, where $e\simeq 2.718$ is Euler's number, to reproduce the correct test-particle limit waveforms~\cite{Pan:2010hz}.
The remaining factor depends on an amplitude $f_{\ell m}^{\rm GR}$ and a phase $\delta_{\ell m}^{\rm GR}$, whose PN expansions in $x_{\rm GR}$ can be determined by identifying order-by-order the mode $h^{\rm GR}_{\ell m}$ to its resummation $h_{\ell m}^\text{F,GR}$ on circular orbits.

The variable $x_{\rm GR}$ can be traded for $y_{\rm GR}=\big(H_{\rm EOB}^{\rm GR}\Omega\big)^{2/3}$ in $\delta_{\ell m}^{\rm GR}$ to gather relativistic corrections sourced by the binary's Arnowitt-Deser-Misner (ADM) mass~\cite{Damour:2008gu}.
For $m$ even, or for nonspinning binaries, $f_{\ell m}^{\rm GR}$ is further resummed as $f_{\ell m}^{\rm GR} = \big(\rho_{\ell m}^{\rm GR}\big)^\ell$ to reduce the magnitude of the 1PN coefficient, which grows linearly with $\ell$.

By contrast, for $m$ odd, the spins deserve a particular treatment.
Indeed, the procedure above yields spin contributions to $\delta_{\ell m}^{\rm GR}$ inversely proportional to $m_-$ [cf. Eq.~\eqref{eq:massRatios}], thus making $h_{\ell m}^\text{F,GR}$ ill-defined in the equal-mass case.
This issue traces back to $h_{\ell m}^{\rm GR}$ being zero when $m_-=0$ at Newtonian order, while some of its spin contributions, entering at subleading PN orders, are not.
We thus adopt the treatment of Refs.~\cite{Pan:2010hz,Taracchini:2012ig,Taracchini:2013rva}, and redefine the amplitude and phase as follows:
\begin{align}
\label{frholm}
\bar f_{\ell m}^{\rm GR} = \left\{
        \begin{array}{ll}
           (\rho_{\ell m}^{\rm GR})^\ell, & \quad m \text{ even}, \\\\
           (\rho_{\ell m}^\text{noS,GR})^\ell + \bar f_{\ell m}^\text{S,GR}, & \quad m \text{ odd},
        \end{array}
    \right.
\end{align}
\begin{align}
\label{drholm}
\bar \delta_{\ell m}^{\rm GR} = \left\{
        \begin{array}{ll}
          \delta_{\ell m}^{\rm GR}, & \quad m \text{ even}, \\\\
        \delta_{\ell m}^{\rm noS,GR}, & \quad m \text{ odd},
        \end{array}
    \right.
\end{align}
where $\rho_{\ell m}^\text{noS,GR}$ and $ \delta_{\ell m}^{\rm noS,GR}$ are inferred from the nonspinning limit of $f_{\ell m}^{\rm GR}$ and $\delta_{\ell m}^{\rm GR}$,
while the additive constant $\bar f_{\ell m}^{\rm S,GR}$ reabsorbs their spin contributions, by expanding the (complex) exponential in Eq.~\eqref{hlmFactorized}.
The amplitude and phase being replaced by their ``barred'' versions, Eq.~\eqref{hlmFactorized} has a non-ambiguous equal-mass limit, because the terms inversely proportional to $m_-$ only enter $\bar f_{\ell m}^{\rm GR}$ (and not $\bar \delta_{\ell m}^{\rm GR}$), which is multiplied by $h_{\ell m}^{\text{N,GR}}\propto m_-$.

The \seobv5 model also incorporates information from second-order self-force (2GSF) ~\cite{Warburton:2021kwk, Wardell:2021fyy} in the GW mode amplitudes and radiation-reaction force, by introducing new calibration parameters in the $\rho_{\ell m}^{\rm GR}$ coefficients and matching them to the 2GSF energy-flux multipolar data for quasicircular nonspinning BBHs~\cite{VandeMeentv5}.
The explicit expressions of $\bar f_{\ell m}^{\rm GR}$ and $\bar \delta_{\ell m}^{\rm GR}$ used in the \seobv5 model are provided in Appendix B of Ref.~\cite{Pompili:2023tna}.

The steps described above can now be generalized to ESGB theories, starting from the 2PN metric modes discussed in Sec.~\ref{sec:metric_flux} [cf. also Appendix~\ref{App:metricModes}].
To do so, we must Taylor-expand $\hat H_{\rm eff}$ and $\hat H_{\rm EOB}$ [defined in Sec.~\ref{subSec:EOBHamiltonian}] at 1PN, as well as $\hat p_\phi$ at 2PN on circular orbits.
Now, observe that when $m$ is odd, the modes $h_{\ell m}$ defined in Eq.~\eqref{def:metricModes} and Appendix~\ref{App:metricModes} are imaginary and proportional to $m_-$ at Newtonian order.
However, some real ESGB corrections, entering at subleading PN orders, are not.
The situation is reminiscent of that with spins, and thus, we shall treat ESGB effects in a similar fashion.

More explicitly, we include the ESGB corrections to the metric modes up to 2PN within \seobv5 as follows:\\

\begin{enumerate}
\item In the first three factors $h_{\ell m}^{\text{N,GR}}$, $\hat{S}_\text{eff}^{\rm GR}$ and $T_{\ell m}^{\rm GR}$ of Eq.~\eqref{hlmFactorized}, formally replace the effective and EOB energies by their ESGB expressions $\hat H_\text{eff}$ and $\hat H_\text{EOB}$ [cf. Sec.~\ref{subSec:EOBHamiltonian}], and $x_{\rm GR}$ by $x$ [cf. Eq.~\eqref{def:x}].\\

\item For $\ell+m$ odd, normalize $\hat{S}_\text{eff}^{\rm GR}(x,\hat p_\phi)$ by its Newtonian value on circular orbits in ESGB gravity, that is
\begin{align}
\hat S_{\rm eff}=\frac{\sqrt{x}\hat p_\phi}{1+\alpha_A^0\alpha_B^0},\qquad \ell + m \text{ odd}.
\end{align}

\item In $\rho_{\ell m}^{\rm noS,GR}$ and $\delta_{\ell m}^{\rm noS,GR}$ and in the nonspinning parts of $\rho_{\ell m}^{\rm GR}$ and $\delta_{\ell m}^{\rm GR}$, formally replace $x_{\rm GR}$ by  $x$ [cf. Eq.~\eqref{def:x}], and $y_{\rm GR}$ by
\begin{align}
y=\big(H_\text{EOB}\Omega\big)^{2/3}.
\end{align}

\item Deform the resulting amplitude and phase by ESGB corrections:
\begin{align}
\bar f_{\ell m} = \left\{
        \begin{array}{ll}
           (\rho_{\ell m}^{\rm GR}+\delta \rho_{\ell m})^\ell, & \quad m \text{ even}, \\\\
           (\rho_{\ell m}^\text{noS,GR})^\ell +\bar  f_{\ell m}^\text{S,GR}+\delta \bar  f_{\ell m}, & \quad m \text{ odd},
        \end{array}
    \right.
\end{align}
\begin{align}
\bar  \delta_{\ell m} = \left\{
        \begin{array}{ll}
          \delta_{\ell m}^{\rm GR}+\delta\delta_{\ell m}, & \quad m \text{ even}, \\\\
        \delta_{\ell m}^{\rm noS,GR}, & \quad m \text{ odd}.
        \end{array}
    \right.
\end{align}
\end{enumerate}
At 2PN, the modes with $\ell\leq 6$ contribute, and we find that the nonzero ESGB corrections read:
\begin{subequations}
\begin{align}
    \delta\rho_{22}&=-\frac{1}{3} x\left(\bar{\gamma }_{AB}+2\langle\bar\beta\rangle\right)+x^2\left[\frac{1}{504} \bigg(-8 \langle\bar\beta\rangle \left(14 \bar{\gamma }_{AB}+42 \bar{\beta }_+-3\right)\right.\nonumber\\
    &+77 \bar{\gamma }_{AB}^2-352 \bar{\gamma }_{AB}+84\left( \bar{\beta }_+^2-\bar{\beta }_-^2+\langle\delta\rangle+\langle\epsilon\rangle\right)-112 \langle\bar\beta\rangle^2\bigg)\nonumber\\
    &+\frac{\nu}{252}  \bigg(84\left( \zeta _{AB} \bar{\gamma }_{AB}- \delta _+\right)-42 \left(\bar{\gamma }_{AB}^2+ \epsilon _+\right)-389 \bar{\gamma }_{AB}+168 \bar{\beta }_+^2\nonumber\\
    &+378 \bar{\beta }_++252 \zeta _{AB}-232 \langle\bar\beta\rangle\bigg)\Bigg]
    +\mathcal O\left(x^{5/2}\right),\\
    \delta\rho_{32}&=-\frac{4}{9}x \left(\bar{\gamma }_{AB}+\langle\bar\beta\rangle\right)+\mathcal O\left(x^{3/2}\right),\\
    \delta\rho_{44}&=-\frac{1}{3} x \left(\bar{\gamma }_{AB}+2 \langle\bar\beta\rangle\right)+\mathcal O\left(x^{3/2}\right),\\
    \delta\rho_{42}&=-\frac{1}{3} x \left(\bar{\gamma }_{AB}+2 \langle\bar\beta\rangle\right)+\mathcal O\left(x^{3/2}\right),
\end{align}
\end{subequations}
\begin{align}
   \delta\delta_{22}= \frac{1}{24} y^{3/2} \left[ 2\left( \alpha _+^2- \alpha _-^2\right)-9\nu \alpha _-^2\right]+\mathcal O\left(y^{5/2}\right),
\end{align}
\begin{subequations}
\begin{align}
   \delta \bar f_{21}&=-\frac{1}{6} x \left(5 \bar{\gamma }_{AB}+4 \langle\bar\beta\rangle\right)-\frac{i\nu }{3}y^{3/2}\left(\alpha _-^2+\frac{ \alpha _- \alpha _+}{ m_-}\right)
   +\mathcal O\left(x^2\right), \\
    \delta \bar f_{33}&= -x \left(\bar{\gamma }_{AB}+2 \langle\bar\beta\rangle\right)\nonumber\\
    &+iy^{3/2}\left[\frac{3}{40}\left( \alpha _+^2-\alpha _-^2\right)-\frac{2\nu}{9}\left(\alpha _-^2-\frac{  \alpha _- \alpha _+}{ m_-}\right)\right]
   +\mathcal O\left(x^2\right),\\
   \delta \bar f_{31}&=-x \left(\bar{\gamma }_{AB}+2 \langle\bar\beta\rangle\right)\nonumber\\
  &+ i y^{3/2} \left[\frac{1}{40} \left(\alpha _+^2-\alpha _-^2\right)-\frac{2\nu}{3}  \left(5 \alpha _-^2-\frac{\alpha _- \alpha _+}{m_-}\right)\right]
  +\mathcal O\left(x^2\right).
\end{align}
\end{subequations}
We do not add any ESGB correction in the spin contributions to the phase and amplitude, because they are currently unknown.

The remaining $N_{\ell m}$ factor in the inspiral-plunge modes (\ref{eq:hellm_insp}) is the NQC correction and reads
\begin{equation}
	\begin{aligned}
	N_{\ell m} &=\left[1+\frac{{\hat p}_{r_*}^2}{(\hat r M \Omega)^2}\left(a_1^{h_{\ell m}}+\frac{a_2^{h_{\ell m}}}{{\hat r}}+\frac{a_3^{h_{\ell m}}}{{\hat r}^{3 / 2}}\right)\right] \\
	&\quad \times \exp \left[i\left(b_1^{h_{\ell m}} \frac{{\hat p}_{r_*}}{\hat r M \Omega}+b_2^{h_{\ell m}} \frac{{\hat p}_{r_*}^3}{\hat r M \Omega}\right)\right].
	\end{aligned}\label{def:NQCfactor}%
\end{equation}
This factor guarantees that the modes' amplitude and frequency agree with NR input values, given by NR fits, at the matching point $\hat t^{\ell m}_{\text{match}}$.
In practice, one first solves for the binary's trajectory to infer $h_{\ell m}^{\rm F}(\hat t)$. Second, one fixes the 5 constants $(a_1^{h_{\ell m}}$, $a_2^{h_{\ell m}}$, $a_3^{h_{\ell m}}$, $b_1^{h_{\ell m}}$, $b_2^{h_{\ell m}})$ by requiring that the amplitude of the EOB modes $\left|h_{\ell m}^{\text {insp-plunge }}\left(\hat t_{\text {match }}^{\ell m}\right)\right|$ and its first two derivatives, and the frequency of the EOB modes $\omega_{\ell m}^{\text {insp-plunge }}\left(\hat t_{\text {match }}^{\ell m}\right)$ and its first derivative, are the same as that of the NR modes at the matching point $\hat t_{\text {match }}^{\ell m}$~\cite{Taracchini:2013rva, Bohe:2016gbl, Pompili:2023tna}. 
In particular, the NQC correction enforces that the (2,2)-mode's derivative vanishes (i.e., it peaks) at the merger time defined in Eq.~\eqref{eq:tISCO}. Since the factorized modes include ESGB corrections, the five constants above are changed so that the inspiral-plunge waveform~\eqref{eq:hellm_insp} still matches the NR predictions in GR at $\hat t_{\ell m}^{\rm match}$. In Sec.~\ref{sec:mrd-waveform}, we shall discuss how to relax this assumption.

\begin{figure}
\includegraphics[width=\linewidth]{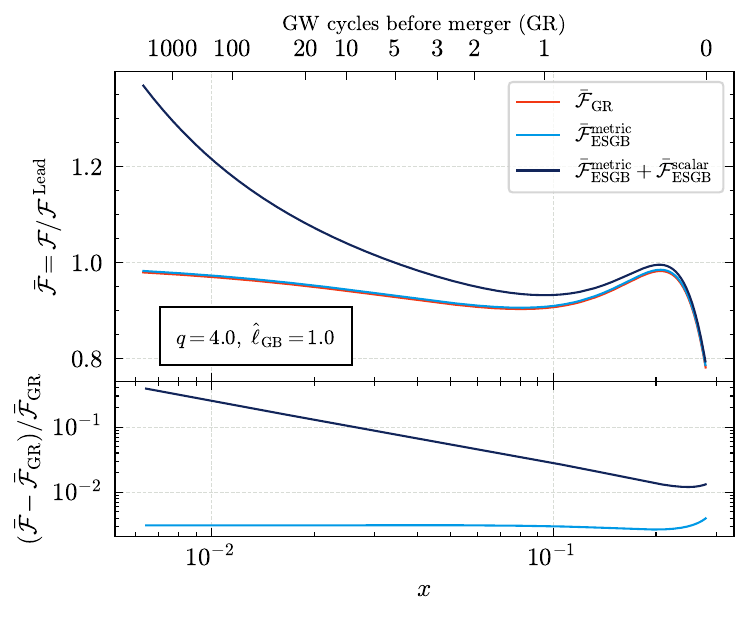}\\
	\caption{\textit{Top panel}: Fluxes in GR and ESGB gravity normalized by the leading order GR contribution, considering only the ESGB corrections to the metric, or also adding the scalar flux.
	We consider a BBH with $q=4$ and $\hat \ell_{\mathrm{GB}} =1$ in the dilatonic theory $f(\varphi)=\exp(2\varphi)/4$ and $\mathcal A(\varphi)=1$. \textit{Bottom panel}: Fractional difference between the ESGB fluxes and the GR flux. We note that the difference in the metric flux at low frequencies is constant with $x$ (e.g., at leading PN order), and is due to the redefinition of $x$ as in Eq.~\eqref{def:x}, while the difference when including the scalar flux increases at low frequencies as $x^{-1}$, due to the dipolar -1PN term.}
	\label{fig:flux-2}
\end{figure}

In Fig.~\ref{fig:flux-2}, we show the GR and ESGB fluxes normalized by the leading order GR contribution
\begin{equation}
    \bar{\mathcal{F}} = \frac{\mathcal{F}}{\frac{32}{5} \nu^2 x_{\rm GR}^5},
\end{equation} 
considering only the ESGB corrections to the metric flux, or also adding the scalar flux. As in Fig.~\ref{fig:flux-1}, we consider a binary with $q=4$ and $\hat \ell_{\mathrm{GB}} =1$ in the dilatonic theory~\eqref{eq:dilatonicTheory}.
Looking at the fractional difference between the ESGB fluxes and the GR flux, we note that the difference in the metric flux at low frequencies is constant with $x$ (e.g., at leading PN order), and is due to the replacement of $x_{\mathrm{GR}}$ with $x$ as in Eq.~\eqref{def:x}, while the difference when including the scalar flux increases at low frequencies as $x^{-1}$, due to the dipolar -1PN term.

\subsection{Merger-ringdown $h_{\ell m}$ modes}
\label{sec:mrd-waveform}

For the merger-ringdown waveform in ESGB, we assume the same phenomenological ansätze as in GR~\cite{Damour:2014yha,Bohe:2016gbl,Cotesta:2018fcv,Pompili:2023tna}, but introduce an appropriate parameterization 
for the modes' amplitude and frequency~\cite{Brito:2018rfr,Ghosh:2021mrv,Maggio:2022hre}.
We also use the QNMs derived recently in ESGB gravity~\cite{Chung:2024ira}.

More precisely, for all harmonics except $(\ell, |m|)= (3,2)$ and $(4,3)$, which exhibit post-merger oscillations due to mode-mixing~\cite{Buonanno:2006ui, Kelly:2012nd}, the merger-ringdown GR waveform employs the following ansatz~\citep{Baker:2008mj,Damour:2014sva,Bohe:2016gbl,Pompili:2023tna}:
\begin{equation}
\label{RD}
h_{\ell m}^{\textrm{merger-RD}}(\hat t) = \nu \ \tilde{\mathcal{A}}_{\ell m}(\hat t)\ e^{i \tilde{\phi}_{\ell m}(\hat t)} \ e^{-i M \sigma_{\ell m 0}(\hat t- \hat t_{\textrm{match}}^{\ell m})},
\end{equation}
where $\sigma_{\ell m0}$ is the complex frequency of the least-damped QNM (with overtone number zero) of the remnant BH.
The corresponding oscillation frequency $f_{\ell m 0}$ and damping time $\tau_{\ell m 0}$ read
\begin{subequations}
\begin{align}
f_{\ell m 0}&=\frac{1}{2 \pi} \operatorname{Re}\left(\sigma_{\ell m 0}\right),\\
\tau_{\ell m 0}&=-\frac{1}{\operatorname{Im}\left(\sigma_{\ell m 0}\right)}.
\end{align}
\end{subequations}
The functions $\tilde{\mathcal{A}}_{\ell m}(\hat t)$ and $\tilde{\phi}_{\ell m}(\hat t)$ are given by~\cite{Pompili:2023tna}:
\begin{subequations}
\begin{align}
\label{eq:ansatz_amp}
\tilde{\mathcal{A}}_{\ell m}(\hat t) &= c_{1,c}^{\ell m} \tanh[c_{1,f}^{\ell m}\ (\hat t-\hat t_{\textrm{match}}^{\ell m}) \ +\ c_{2,f}^{\ell m}] \ + \ c_{2,c}^{\ell m},\\
\label{eq:ansatz_phase}
\tilde{\phi}_{\ell m}(\hat t) &= \phi_{\textrm{match}}^{\ell m} - d_{1,c}^{\ell m} \log\left[\frac{1+d_{2,f}^{\ell m} e^{-d_{1,f}^{\ell m}(\hat t- \hat t_{\textrm{match}}^{\ell m})}}{1+d_{2,f}^{\ell m}}\right],
\end{align}\label{eq:ansatz_ringdown}%
\end{subequations}
where $ \phi_{\textrm{match}}^{\ell m}$ is the phase of the inspiral-plunge mode $(\ell, m)$ at $\hat t = \hat t_{\textrm{match}}^{\ell m}$ [cf. below Eq.~\eqref{eq:h_match}].
The coefficients $d_{1, c}^{\ell m}$ and $c_{i, c}^{\ell m}$ ($i=1,2$) are constrained by the requirement that the amplitude and phase of $h_{\ell m}(\hat t)$ are continuously differentiable at $\hat t=\hat t_{\text {match}}^{\ell m}$, while the coefficients $c_{i,f}^{\ell m}$ and $d_{i,f}^{\ell m}$ are obtained through fits to a large set of $\sim 440$ aligned-spin NR waveforms, spanning mass ratios up to 20 and spins up to 0.998, and BH perturbation-theory merger-ringdown waveforms with mass ratio 1000. 
For the $(3,2)$ and $(4,3)$ modes, the mode-mixing behavior is modeled by applying the previous construction to the spheroidal harmonics~\cite{Berti:2005gp} $(3,2,0)$ and $(4,3,0)$, which feature a monotonic amplitude and frequency evolution~\cite{KumarMehta:2019izs}. The $(3,2)$ and $(4,3)$ modes are then obtained by combining the $(3,2,0)$ and $(4,3,0)$ modes with the $(2,2)$ and $(3,3)$ modes using the appropriate mixing coefficients~\cite{Berti:2014fga}.

In the \texttt{SEOBNRv5PHM} model~\cite{Pompili:2023tna, RamosBuadesv5}, the complex QNM frequencies in GR are obtained for each $(\ell, m)$ harmonic as a function of the BH's (dimensionless) final mass and spin $(\hat M_f=M_f/M,\, \chi_f)$ using the \texttt{qnm} Python package \cite{Stein:2019mop}.
  Note that the \texttt{qnm} package provides $M_f \sigma_{\ell m 0}$, as a function of the final spin $\chi_f$, so in Eq.~\eqref{RD} we use
  \begin{equation}
    M \sigma_{\ell m 0} = \frac{1}{\hat M_f} M_f \sigma_{\ell m 0}(\chi_f).
  \end{equation}
The BH's mass and spin are in turn computed using the fitting formulas of Refs.~\cite{Jimenez-Forteza:2016oae} and \cite{Hofmann:2016yih}, respectively. 

We now describe how we account for ESGB effects in the merger-ringdown $h_{\ell m}$ modes.
For the merger, as reminded on Sec.~\ref{sec:waveform-modes}, the NQC corrections impose that the modes' amplitude and frequency agree with fits to NR waveforms in GR.
In recent years, there has been a significant effort to produce NR simulations in beyond-GR theories~\cite{Cayuso:2017iqc, Okounkova:2019zjf, Lara:2021piy, Franchini:2022ukz, Ma:2023sok, Cayuso:2023xbc}, with a few examples of NR evolutions of BBHs in ESGB gravity~\cite{Witek:2018dmd, Witek:2020uzz, Okounkova:2020rqw, Elley:2022ept, Corman:2022xqg, AresteSalo:2023mmd}. 
For example, Ref.~\cite{Corman:2022xqg} found a negligible effect on the GW amplitude at merger with varying ESGB coupling, in contrast with the large effect found in the order-reduced simulation of Ref.~\cite{Okounkova:2020rqw}, possibly due to the presence of secularly growing errors present in the latter perturbative approach (see~\cite{Corman:2024cdr} for a detailed comparison of different numerical approaches).
In any case, making parameter-space fits (calibration) of such merger properties, as done in GR, requires a large number of NR waveforms, which are not yet available.

To account for possible deviations in the waveform at merger, we take inspiration from the parametrized \texttt{SEOBNR} approach (\texttt{pSEOBNR}) \cite{Brito:2018rfr,Ghosh:2021mrv,Maggio:2022hre}, and introduce fractional deviations to the NR predictions in GR for the amplitude and frequency at $\hat t_{\text {match }}^{\ell m}$ for each mode.
These deviations parameterize our ignorance of the merger morphology with extra nuisance parameters, to be marginalized over in parameter estimation. Given this idea, we want the corrections to vanish in the GR limit, so we rescale them as
\begin{subequations}
    \begin{align}
    |h_{\ell m}^{\rm NR,GR}| & \rightarrow |h_{\ell m}^{\rm NR,GR}|\, \left(1 + {\hat{\ell}_{\mathrm{GB}}}^4 \delta A_{\ell m }\right),\label{eq:nongr_freqs_a} \\ 
    \omega _{\ell m}^{\rm NR,GR} & \rightarrow \omega _{\ell m}^{\rm NR,GR}\, \left(1 +  {\hat{\ell}_{\mathrm{GB}}}^4 \delta \omega_{\ell m }\right), 
\end{align}\label{eq:nongr_mrg}%
\end{subequations}
where the ${\hat{\ell}_{\mathrm{GB}}}^4$ dependence is inspired by the scaling of the PN corrections in the inspiral modes in the small-${\hat{\ell}_{\mathrm{GB}}}$ regime.
In turn, this rescaling changes the values taken by the constants $(a_i,b_i)$ and $(c^{\ell m}_{i,c},d^{\ell m}_{i,c})$ entering Eqs.~\eqref{def:NQCfactor} and \eqref{eq:ansatz_ringdown}, respectively.
The $\hat \ell_{\rm GB}$-dependence in Eqs.~\eqref{eq:nongr_mrg}, and how the merger morphology is modified for different modes, could be further studied using NR waveforms in ESGB gravity. Still, current NR simulations~\cite{Corman:2022xqg, AresteSalo:2023mmd} suggest that corrections at merger are small, for values of the coupling allowed by current observations.

In Sec.~\ref{sec:PE}, we quantify the impact of marginalizing over these corrections, applied for simplicity only to the dominant $(2,2)$ mode, on the analysis of GW signals, and find it to be small.

\subsection{Corrections to the quasi-normal-mode frequencies}
\label{sec:ringdown}
To incorporate ESGB corrections in the merger-ringdown waveform using the ansatz of Sec.~\ref{sec:mrd-waveform}, one should account for the following effects:\\
\begin{enumerate}
\item The complex frequency of the least-damped QNM for each $(\ell, m)$ harmonic $\sigma_{\ell m 0}$ should vary, given the BH's final mass and spin.\\
\item In several modified gravity theories, the BH metric perturbations with even and odd parity have different QNM frequencies, breaking the isospectrality of GR~\cite{Chandrasekhar:1985kt}.\\
\item The final mass and spin of the remnant BH should deviate from the GR predictions due to differences in both the conservative and dissipative binary dynamics.
For example, the final mass and spin are smaller in ESGB gravity in the NR simulations of Ref.~\cite{AresteSalo:2023mmd}.\\
\item The $c_{i,f}^{\ell m}$ and $d_{i,f}^{\ell m}$ coefficients, modeling the early ringdown, are obtained through fits of NR waveforms in GR, and should also be corrected.
Since these could only be obtained through fits of beyond-GR NR waveforms, we neglect this last correction.\\
\end{enumerate}

QNM frequencies for nonspinning ESGB BHs were first computed in Refs.~\cite{Pani:2009wy, Blazquez-Salcedo:2016enn}.
They were then extended up to the second order in a spin expansion in Refs.~\cite{Pierini:2021jxd, Pierini:2022eim}, focusing on the polar sector.
Binary coalescences typically lead to BHs with final spin $\chi_f\sim0.7$, making the accuracy of an expansion to second order in the spin limited.
While Ref.~\cite{Pierini:2022eim} suggested that its domain of validity could be extended using Pad\'e approximants, Refs.~\cite{Cano:2023qqm, Cano:2023jbk} argued that higher orders are necessary to achieve accuracy better than the statistical uncertainties of current and future GW detectors.
Various groups developed modified Teukolsky formalisms~\cite{Li:2022pcy, Hussain:2022ins, Cano:2023tmv, Li:2023ulk, Wagle:2023fwl} and a novel approach based on perturbative spectral expansions~\cite{Chung:2023wkd, Chung:2023zdq} to calculate QNM frequencies in a wide class of modified gravity theories for BHs with generic spins.
Ref.~\cite{Chung:2024ira} applied the latter formalism to ESGB gravity at leading order in $\ell_{\rm GB}/M_f$, to compute corrections to the $(\ell, m, n) = (2,2,0)$ QNM accurately for up to $\chi_f\sim0.8$, which we implement in this work.\footnote{We translate between the conventions of Ref.~\cite{Chung:2024ira} and ours by setting $\zeta^{1/4}=\ell_{\rm GB}/ (4\pi^{1/4} M_f)$.}

Ref.~\cite{Chung:2024ira} computed corrections for both the polar and axial sectors of the QNMs. However, understanding how axial and polar QNMs appear in the GW signal is a complex task outside the scope of this paper, to be answered on a theory-by-theory basis.
Following~\cite{Silva:2022srr}, we only apply corrections coming from the polar sector, which for typical spin magnitudes $\chi_f\sim0.7$ is the least damped parity, as well as the one giving rise to larger frequency corrections.

We note that Ref.~\cite{Chung:2024ira} focuses on the shift-symmetric theory, while we consider here the dilatonic case~\eqref{eq:dilatonicTheory}.
However, when analyzing real events, which also include the inspiral phase, we will obtain constraints that are within the small coupling limit.
In this case, the dilatonic theory reduces to the shift-symmetric one at leading order in $\hat\ell_{\rm GB}$ [cf. Sec.~\ref{subsec:dilatonic}], which justifies a-posteriori to use the results of Ref.~\cite{Chung:2024ira}.

\subsection{Corrections to the final mass and spin\label{sec:remnant}}

In this section, we discuss the ESGB corrections to the mass and spin of the BH remnant as a function of the coupling, for a BBH in the dilatonic ESGB case~\eqref{eq:dilatonicTheory}.
While a precise estimate of these corrections relies on NR simulations, in EOB models they can be approximated from the binary's dynamics. For simplicity, we consider spin-aligned binaries and only study the final mass and spin magnitude. We also compare our results to a self-consistent estimate following Refs.~\cite{Ghosh:2016qgn, Ghosh:2017gfp}.

\begin{table}[b!]
  \centering
  \begin{ruledtabular}
  \begin{tabular}{ c c c }
  Coefficient & $\Delta  \hat M_f^{\mathrm{ESGB}}$ & $\Delta  \hat J_f^{\mathrm{ESGB}}$ \\
  \hline
  $a_0$ & $-4.1011 \times 10^{-5}$ & $-5.1696 \times 10^{-4}$ \\
  $a_1$ & $-21.4091$ & $-21.7572$ \\
  $a_2$ & $-1.5879$ & $-1.2942$ \\
  $a_3$ & $155.6233$ & $179.7129$ \\
  $a_4$ & $0.0340$ & $-0.0853$ \\
  $a_5$ & $27.9436$ & $21.2925$ \\
  $a_6$ & $7.9436$ & $5.2301$ \\
  $a_7$ & $-48.0261$ & $-30.6130$ \\
  $a_8$ & $-230.0593$ & $-537.8520$ \\
  \end{tabular}
  \end{ruledtabular}
  \caption{Coefficients for the fits of the final mass and final spin corrections~\eqref{eq:remnant_fits}, for a BBH in the dilatonic theory $f(\varphi)=\exp(2\varphi)/4$ and $\mathcal A(\varphi)=1$.}
  \label{tab:fit}
\end{table}

A simple phenomenological estimate for the final mass and spin of a BBH coalescence can be made under the assumption that the system radiates most of its energy and angular momentum in the inspiral stage until the ISCO, while during the merger and ringdown stages the mass and angular momentum of the system are roughly conserved~\cite{Buonanno:2007sv}.
A similar procedure was applied to ESGB gravity in Refs.~\cite{Carson:2020cqb, Carson:2020ter}.
Here, we estimate the (dimensionless) final mass and orbital angular momentum from the values of $\nu \hat H_{\mathrm{EOB}}=H_{\mathrm{EOB}}/M$ and $\nu \hat p_{\phi}=p_{\phi}/M^2$ at the merger time, associated in the model with the peak of the (2,2)-mode waveform~\eqref{eq:tISCO}. We compute the quantities above in both ESGB and GR, and add the difference to the GR estimates given by the NR fitting formulas of Refs.~\cite{Jimenez-Forteza:2016oae} and \cite{Hofmann:2016yih}, used in \seobv5. This amounts to neglecting the effects of ESGB corrections to the post-merger radiation. Proceeding in an iterative manner, we evaluate the ESGB corrections to the remnant mass and spin by first inserting their GR values in Eq.~\eqref{eq:tISCO}. In a second step, in Sec.~\ref{sec:results} we build waveforms by evaluating $\hat t_{\rm ISCO}$ in Eq.~\eqref{eq:tISCO} using the corrected remnant mass and spin. Another limitation of our estimates is that the merger time uses the $\Delta \hat t_{\mathrm{NR}}$ parameter~\eqref{eq:tISCO}, which is calibrated from NR simulations in GR, and should thus also be corrected in the future.

To avoid generating an additional GR waveform each time we wish to include this correction, we provide fits as a function of the intrinsic parameters and of the coupling $\hat \ell_{\mathrm{GB}}$.
For simplicity, we explore configurations with equal-spin magnitude $\chi \in [-0.9,0.9]$ with increment $\Delta\chi= 0.18$, for several mass ratios $q\in\{1,2,4,10\}$, and coupling $\hat \ell_{\rm GB}^4\in[0,0.07]$ with increment $\Delta\hat\ell_{\rm GB}^4=0.007$. We use a polynomial ansatz in $(\nu, \chi)$ (with $\chi$ being promoted to $\chi_{+} = (m_A^0/M) \chi_A + (m_B^0/M) \chi_B$ for unequal-spin cases below), while the dependence on $\hat \ell_{\mathrm{GB}}$ is well approximated by a simple $\hat \ell_{\mathrm{GB}}^4$ scaling. We define:
\eject
\begin{subequations} \label{eq:remnant_corrections}
  \begin{align}
  \hat M_f &= \hat M_f^{\mathrm{GR}} + \Delta \hat M_f^{\mathrm{ESGB}},\\
  \hat J_f &= \hat J_f^{\mathrm{GR}} + \Delta \hat J_f^{\mathrm{ESGB}},\\
  \chi_f &= \frac{\hat J_f}{\hat M_f^2}=\chi_f^{\mathrm{GR}} + \Delta \chi_f^{\mathrm{ESGB}},
  \end{align}
\end{subequations}
where, recall, $\hat M_f=M_f/M$ and $\hat J_f=J_f/M^2$ are the dimensionless final mass and total angular momentum. $\hat M_f^{\mathrm{GR}}$, $\chi_f^{\mathrm{GR}}$ are given by the NR fitting formulas of Refs.~\cite{Jimenez-Forteza:2016oae,Hofmann:2016yih}.
We obtain the following fits for the ESGB corrections:
\begin{equation}
  \begin{aligned}
  \Delta X^{\mathrm{ESGB}} = & \, a_0 \hat{\ell}_{\mathrm{GB}}^4 \left( 1 + a_1 \nu + a_2 \chi + a_3 \nu^2 + a_4 \chi^2 \right. \\
  & \left. + a_5 \nu \chi + a_6 \nu \chi^2 + a_7 \nu^2 \chi + a_8 \nu^3 \right),
  \label{eq:remnant_fits}
  \end{aligned}
\end{equation}
where $X = (\hat{M}_f, \hat J_f)$.
The respective coefficients are listed in Table~\ref{tab:fit}.
By combining Eqs.~\eqref{eq:remnant_corrections}, we can compute the linear correction to the final spin as 
\begin{equation}
  \Delta \chi_f^{\mathrm{ESGB}} = \frac{\Delta  \hat J_f^{\mathrm{ESGB}}}{(\hat M_f^{\mathrm{GR}})^2} - 2 \chi_f^{\mathrm{GR}} \frac{\Delta  \hat M_f^{\mathrm{ESGB}}}{\hat M_f^{\mathrm{GR}}}.
\end{equation}
\begin{figure}
  \includegraphics[width=\linewidth]{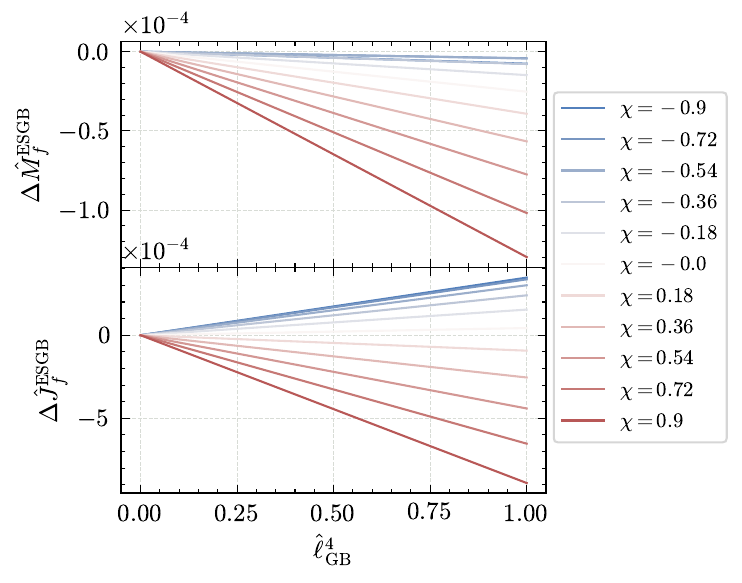}
\caption{ESGB corrections to the (dimensionless) final mass (top panel) and angular momentum (bottom panel), as given by the fits~\eqref{eq:remnant_fits}. We consider BBHs with equal-spin magnitude $\chi$ and mass ratio $q=4$, in the dilatonic theory $f(\varphi)=\exp(2\varphi)/4$ and $\mathcal A(\varphi)=1$. }
\label{fig:remnant-correction-1}
\end{figure}

Figure~\ref{fig:remnant-correction-1} shows the values of the ESGB corrections to the (dimensionless) final mass and angular momentum for a set of BBHs with equal-spin magnitude $\chi$ and mass ratio $q=4$, as a function of the ESGB coupling $\hat \ell_{\mathrm{GB}}^4$ in the dilatonic theory~\eqref{eq:dilatonicTheory}. 
These corrections are quite modest compared to those affecting the inspiral dynamics, being $\mathcal{O}(10^{-4})$ for $\hat \ell_{\mathrm{GB}}^4 \sim 1$.
We find that the corrections to the final mass are always negative, while the final angular momentum corrections can take both signs depending on the mass ratio and spins.
Note that the value of $\hat p_{\phi}$ at the ISCO can be larger when adding ESGB corrections to the EOB Hamiltonian [cf. Fig.~\ref{fig:dynamics} below].

One can also estimate the final mass and angular momentum by subtracting the energy and angular momentum radiated during the entire coalescence (including the merger-ringdown stage) from their initial values.
These can in turn be obtained by integrating the energy and angular momentum fluxes, in both the metric and the scalar channels.
The energy emitted through GWs depends on the final mass and spin $(\hat M_f, \chi_f)$, as they enter the computation of the merger-ringdown waveform modes used to compute the GW flux.
One can then numerically solve for self-consistent values of the final mass and spin~\cite{Ghosh:2016qgn, Ghosh:2017gfp}, which minimize the difference between the values set for the remnant BH and those obtained by energy and angular-momentum balance.
Using this method, we find that the corrections show a less clear dependence on intrinsic parameters, due to errors in the numerical integration and root-finding, making a parameter-space fit harder to perform. However, the median values of the corrections (across 10,000 configurations uniform in $\nu$ with $q\in[1,20]$ and uniform in $\chi_i\in[-0.99,0.99]$) still show a clear scaling with $\hat \ell_{\mathrm{GB}}^4$, with the average correction being consistent with Eqs.~\eqref{eq:remnant_fits}.

\begin{figure}
    \includegraphics[width=\linewidth]{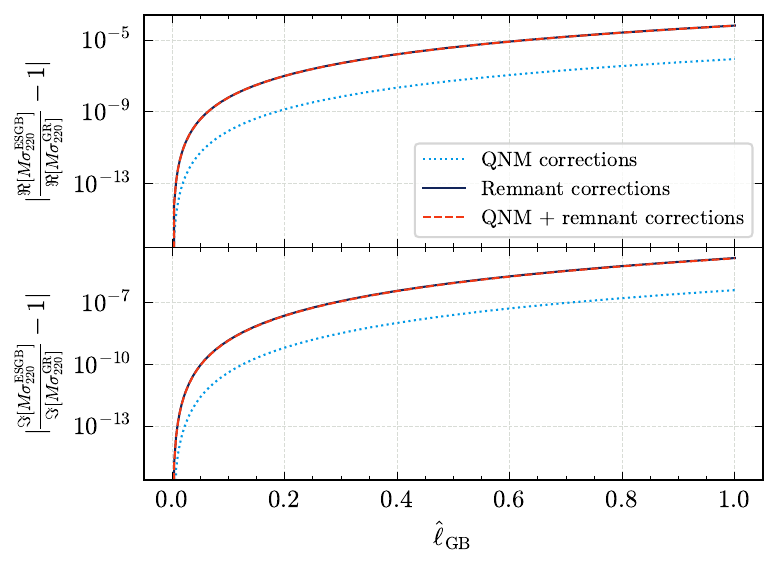}
	\caption{Fractional change between ESGB and GR in the real and imaginary part of the $(2,2)$-mode's ringdown frequency, as a function of $\hat \ell_{\mathrm{GB}}$, for a $q=4$ nonspinning BBH ($\chi_f \simeq 0.47$) in the dilatonic theory $f(\varphi)=\exp(2\varphi)/4$ and $\mathcal A(\varphi)=1$. We show the change only including corrections to the QNM spectrum, but fixing the final mass and spin to the GR values, only changing the final mass and spin using Eq.~\eqref{eq:remnant_fits}, as well as considering both corrections. 
  }
	\label{fig:qnms}
\end{figure}

\begin{figure*}[t!]
    \includegraphics[width=\columnwidth]{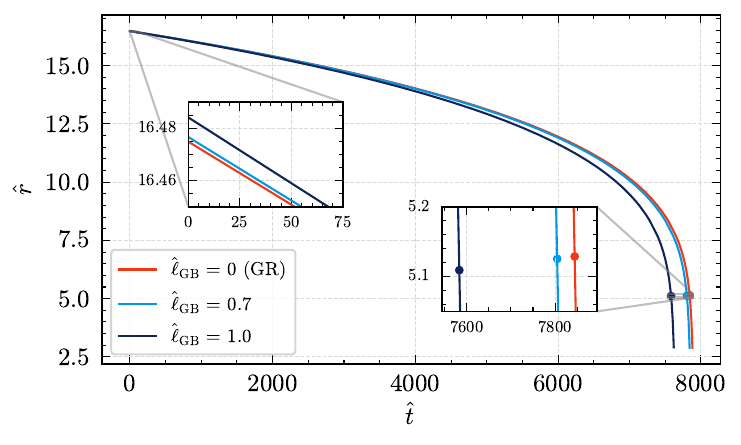}
    \includegraphics[width=\columnwidth]{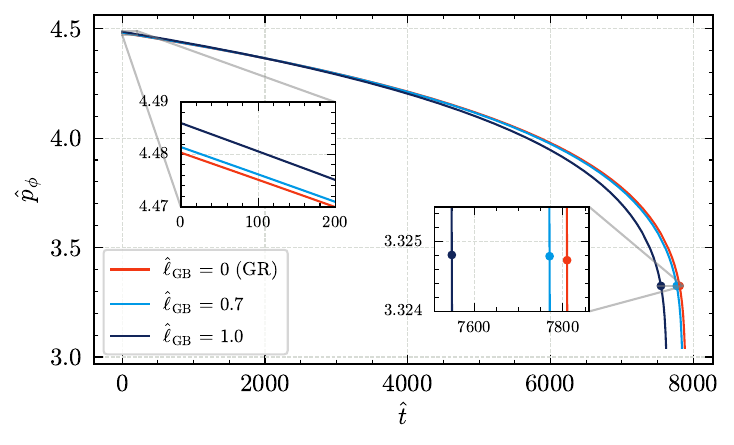}
	\caption{The orbital separation $\hat r$ (left panel) and angular momentum $\hat p_{\phi}$ (right panel) as a function of time for a nonspinning BBH with $q=4$, starting at the same dimensionless orbital frequency $M \Omega_i = 0.015$, for different values of the ESGB coupling $\hat \ell_{\mathrm{GB}} \in \{ 0, 0.7, 1.0 \}$ in the dilatonic theory $f(\varphi)=\exp(2\varphi)/4$ and $\mathcal A(\varphi)=1$. The insets highlight the differences in the dynamics at the start of the evolution, and close to the ISCO, indicated by the dots. Each curve ends at merger, taken as the peak of the (2,2) mode.
We note that: (i) The binary merges at earlier times for larger values of the ESGB coupling, since more energy is radiated away through the additional scalar field, (ii) ESGB binaries at the same initial dimensionless frequency start at a larger radius and angular momentum. This is due to the leading-order correction $1/\hat r \rightarrow G_{AB} / \hat r$ in the EOB potentials, with $G_{AB}\geq 1$ (iii) The ISCO is at a smaller separation and slightly larger angular momentum compared to GR. This is a non-trivial effect due to higher-order PN corrections in the EOB Hamiltonian.}
	\label{fig:dynamics}
\end{figure*}

In Fig.~\ref{fig:qnms}, we show the fractional change between ESGB and GR in the real and imaginary part of the $(2,2)$ mode ringdown frequency, as a function of $\hat \ell_{\mathrm{GB}}$, for a $q=4$ nonspinning BBH ($\chi_f \simeq 0.47$).
We either include corrections to the QNM spectrum, change the final mass and spin using fits from Eq.~\eqref{eq:remnant_fits}, or consider both corrections. 
As noted in previous works~\cite{Pierini:2022eim, Maselli:2023khq}, typical corrections to the QNMs are very small.
While corrections to the remnant mass and spin are also modest, they can cause changes in the frequency of the ringdown waveform from one to two orders of magnitude larger than the ESGB corrections to the QNM spectrum.
The precise estimation of such corrections would be useful for high-precision tests of GR with next-generation detectors, since prospects for BH spectroscopy in ESGB could be much more optimistic than originally thought. Reference~\cite{Maselli:2023khq} found that, even with next-generation GW detectors, constraints on ESGB gravity from ringdown observations alone are weaker than current bounds from LVK inspiral observations, when only accounting for corrections to the QNM spectrum. Our results suggest that much better constraints might be achieved by exploiting complementary information from the full IMR signal.

\section{Results}
\label{sec:results}

Given the coupling of matter fields to the metric in ESGB gravity~\eqref{eq:action}, the mirrors of a GW interferometer must follow the geodesics of the perturbed Jordan metric $\tilde g_{\mu\nu}=\mathcal A^2(\varphi)g_{\mu\nu}$ which reads, in the Solar System:
\begin{align}
\tilde g_{\mu\nu}=\mathcal A^2_\odot \left[\eta_{\mu\nu}(1+2\alpha_\odot \delta \varphi)-h_{\mu\nu}^{\rm TT}\right]+\mathcal O\left(\frac{1}{R^2}\right)\,,
\end{align}
where $R$ is the distance to the source, while $\mathcal A_\odot=\mathcal A(\varphi_\odot)$ and $\alpha_\odot=(d\ln\mathcal A/d\varphi)(\varphi_\odot)$ are the couplings of matter to the scalar field in the Solar System.
Note that for the Jordan metric to reduce to Minkowski in the absence of metric and scalar waves, we must perform the local coordinate change $d\tilde t= \mathcal A_\odot dt$ and $d\tilde x^i= \mathcal A_\odot dx^i$.

In what follows, we focus on the metric waveform $h_{\mu\nu}^{\rm TT}$, which was discussed in the previous sections.
Indeed, the contribution of the scalar waveform $\delta\varphi=\varphi-\varphi_\odot$ is cut off by the numerical factor $\alpha_\odot^2\lesssim 10^{-5}$ in the Solar System \cite{Bertotti:2003rm}.
Moreover, we illustrate our results in the dilatonic ESGB case~\eqref{eq:dilatonicTheory} for which $\mathcal A(\varphi)=1$, and thus $\mathcal A_\odot=1$ and $\alpha_\odot=0$.

\subsection{Waveform morphology}
\label{sec:morph}

In this section, we describe the impact of the ESGB corrections on the dynamics and waveforms of BBH systems, modeled with \seobesgb.
As a concrete illustration, we consider a nonspinning binary with mass ratio of $q=4$, and $\hat{\ell}_{\mathrm{GB}} \in \{0, 0.7, 1.0\}$. 
Our conclusions remain qualitatively unchanged for different mass ratios and non-zero spins.
However, in the equal-mass limit, the deviations from the GR flux are smaller, because the -1PN leading-order term in the scalar flux~\eqref{eq:scalar_flux} then vanishes. 

First, we examine in Fig.~\ref{fig:dynamics} the orbital separation $\hat r$ and angular momentum $\hat p_{\phi}$ as a function of time.
The insets highlight the differences in the dynamics at the start of the evolution, which for all cases is taken at a dimensionless orbital frequency $M \Omega_i = 0.015$, and close to the ISCO, indicated by the dots. The values of $\hat r$ and $\hat p_{\phi}$ at the ISCO are determined by setting both the first and second derivatives of the Hamiltonian with respect to $\hat r$ to zero, taking $\hat p_{r_*}=0$ (circular orbits) (i.e., $\partial H_{\mathrm{EOB}}(\hat p_{r_*}=0) / \partial \hat r=0=\partial^2 H_{\mathrm{EOB}}(\hat p_{r_*}=0) / \partial \hat r^2$). Each curve ends at merger, taken as the peak of the (2,2) mode waveform~\eqref{eq:tISCO}. Observe that, on the examples considered: \\
\begin{enumerate}
\item The binary merges at earlier times for larger values of the ESGB coupling, since more energy is radiated away due to the additional scalar field.\\
\item Given a fixed initial dimensionless frequency, binaries start at a larger radius and angular momentum for larger values of the ESGB coupling.
This is due to the replacement $1/\hat r \to  G_{AB} /\hat r$ in the EOB potentials, with $G_{AB}\geq 1$.
Note that the model uses the quasi-circular, adiabatic, initial condition formulas derived in Ref.~\cite{Buonanno:2005xu}, also used in the \seobv5 model.\\
\item The ISCO is at a smaller separation [cf. also Ref.~\cite{Julie:2022qux}] and slightly larger angular momentum compared to GR. This is a non-trivial effect due to higher-order PN corrections in the EOB Hamiltonian.\\
\end{enumerate}

In Fig.~\ref{fig:amplitude} we show, for the same set of binaries, the amplitude of the (2,2) mode waveform, to highlight how a non-zero value of the coupling changes the time-to-merger when including different beyond-GR corrections. The dash-dotted vertical lines indicate the merger, taken as the peak of the (2,2) mode, and the dotted vertical lines correspond to the time at which $\hat r=\hat r_{\rm{ISCO}}$. We set $\hat t=0$ at the GR merger time. In the top panel, we incorporate ESGB corrections in the Hamiltonian, while using the GR flux, and mostly observe that modifications to the conservative dynamics tend to delay the merger. In the middle panel, we include instead ESGB corrections in the dissipative sector and use the GR Hamiltonian. As expected, these corrections accelerate the inspiral, mostly due to the additional energy dissipation via the scalar field. In the bottom panel, we combine both contributions.
Changes to the dissipative dynamics are predominant, and the binary merges earlier than GR for a non-zero value of the ESGB coupling. The trend seen in this example is consistent with other cases we considered, and agrees qualitatively with NR simulations~\cite{Corman:2022xqg, AresteSalo:2023mmd}.
Finally, Fig.~\ref{fig:waveform} shows the real part of the (2,2) mode for the same set of binaries. The waveforms are aligned over a low-frequency interval, by shifting their time of coalescence $t_c$ and phase $\Phi$ [cf. Sec.~\ref{sec:esgb_flux}] as done in Ref.~\cite{Taracchini:2012ig}, and display a difference in the time-to-merger and a noticeable dephasing as they approach the merger.

Note that the ESGB corrections to the remnant mass and spin [entering the merger time and ringdown frequency, cf. Sec.~\ref{sec:remnant}] and to the QNM spectrum [cf. Sec.~\ref{sec:ringdown}] are included in our waveforms. However, their impact is subdominant compared to the effects above.

\begin{figure}
  \includegraphics[width=\linewidth]{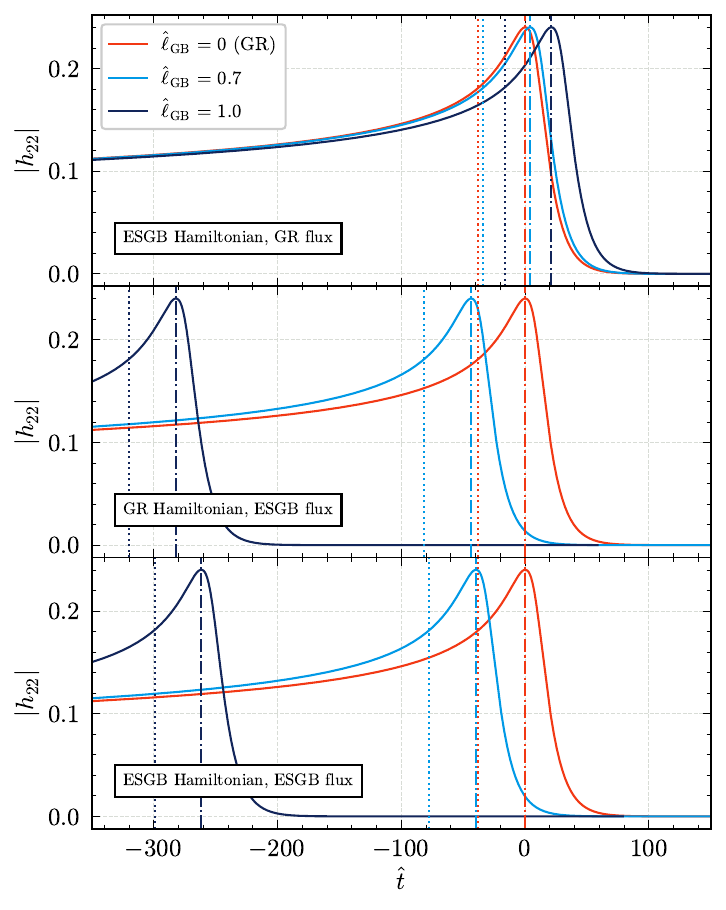}
	\caption{Amplitude of the (2,2) mode of a nonspinning BBH with $q=4$, starting at a dimensionless frequency $M\Omega_i = 0.015$, for $\hat \ell_{\mathrm{GB}} \in \{ 0, 0.7, 1.0 \}$ in the dilatonic theory $f(\varphi)=\exp(2\varphi)/4$ and $\mathcal A(\varphi)=1$.
	The dash-dotted vertical lines indicate the merger, taken as the peak of the (2,2) mode, and the dotted vertical lines correspond to the time at which $\hat r=\hat r_{\rm{ISCO}}$.
	We set $\hat t=0$ at the GR merger time. \textit{Top panel}: we include ESGB corrections in the Hamiltonian, and use the GR flux. \textit{Middle panel}: we include ESGB corrections in the flux, and use the GR Hamiltonian. \textit{Bottom panel}: we include ESGB corrections in both the Hamiltonian and the flux. We see that corrections to the conservative dynamics tend to delay the merger, while corrections to the dissipative sector accelerate the inspiral. The trend seen here is consistent with other cases we considered: when combining all contributions, the changes to the dissipative dynamics are predominant, and the binary merges earlier than in GR for a non-zero value of the coupling.}
	\label{fig:amplitude}
\end{figure}

\subsection{Quantifying measurability of deviations from GR}
\label{sec:meas}

We now quantify which deviation magnitudes are detectable by GW detectors, and, if so, estimate the SNR required for such measurements. For this purpose, we employ a common metric in GW data analysis - the mismatch. Given two waveforms $h_1(t)$ and $h_2(t)$, we introduce the overlap, defined as the noise-weighted inner product \cite{Finn:1992xs, Sathyaprakash:1991mt}
\begin{equation}
	\label{eq:overlap}
	\left(h_1 \mid h_2\right) \equiv 4 \operatorname{Re} \int_{f_l}^{f_h} \frac{\tilde{h}_1(f) \tilde{h}_2^*(f)}{S_n(f)} \mathrm{d} f ,
\end{equation}
where $\tilde h(f)$ is the Fourier transform of the time-domain signal, and $S_n(f)$ is the one-sided power spectral density (PSD) of the detector noise. The faithfulness is then defined as the overlap between the normalized waveforms, maximized over the relative time of coalescence and phase, that is 
\begin{equation}
	\label{eq:def_match}
	\left\langle h_1 \mid h_2\right\rangle=\max _{\Phi, t_c} \frac{\left(h_1\left(\Phi, t_c\right) \mid h_2\right)}{\sqrt{\left(h_1 \mid h_1\right)\left(h_2 \mid h_2\right)}}.
\end{equation}
We finally define the mismatch, or unfaithfulness, as 
\begin{equation}
    \mathcal{M} = 1 - \left\langle h_1 \mid h_2\right\rangle.
\end{equation}
In Eq.~(\ref{eq:overlap}), we fix $f_h = 2048 ~\text{Hz}$ and choose different values for $f_l$ depending on the detector. We consider the advanced LIGO (aLIGO) detector, for which we assume the design zero-detuned high-power noise PSD~\cite{Barsotti:2018} and $f_l=20\mathrm{Hz}$, and the ET detector, for which we assume the \texttt{EinsteinTelescopeP1600143} PSD implemented in \texttt{pyCBC}~\cite{alex_nitz_2023_10137381} and $f_l=5\mathrm{Hz}$. For simplicity, we only consider the dominant (2,2) mode of the waveform when computing the mismatch.

\begin{figure*}
  \includegraphics[width=\linewidth]{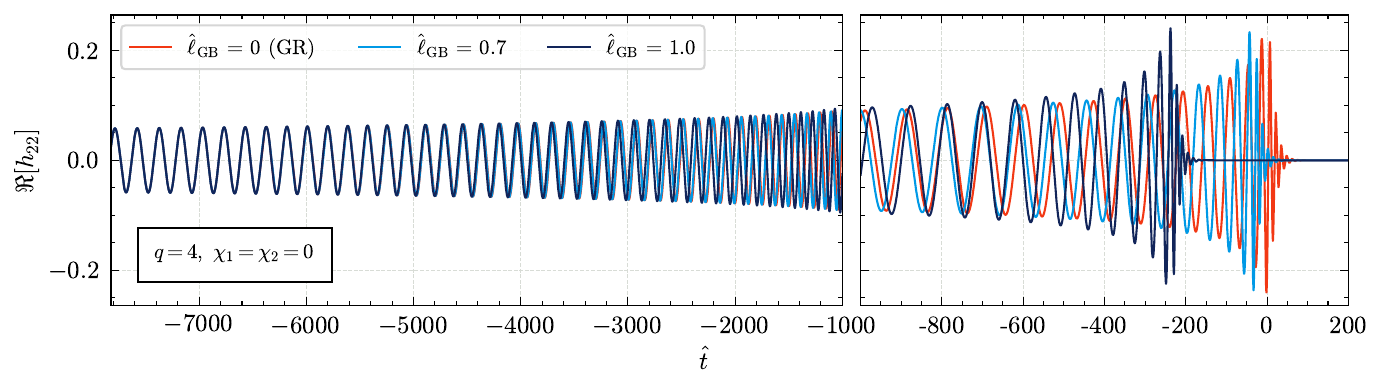}
	\caption{(2,2) mode of the \seobesgb\,waveform of a nonspinning BBH with $q=4$, starting at a dimensionless orbital frequency $\Omega_i = 0.015$, for different values of the ESGB coupling $\hat \ell_{\mathrm{GB}} \in \{ 0, 0.7, 1.0 \}$ in the dilatonic theory $f(\varphi)=\exp(2\varphi)/4$ and $\mathcal A(\varphi)=1$. We set $\hat t=0$ at the peak of the (2,2) mode of the GR waveform. In this example, the merger parameters $\delta A_{\ell m}$ and $\delta \omega_{\ell m}$ entering Eqs.~\eqref{eq:nongr_mrg} are set to zero.}
	\label{fig:waveform}
\end{figure*}

The mismatch can be used to estimate at which SNR two signals could be distinguishable, given the detector sensitivity. This is usually formulated in terms of an indistinguishability criterion~\cite{Flanagan:1997kp, Lindblom:2008cm, Chatziioannou:2017tdw, Purrer:2019jcp}, which states that if two waveforms fulfill the condition
\begin{equation}
    \label{eq:indist_criterion}
    \mathcal{M} < \frac{D}{2~\mathrm{SNR}^2},
\end{equation}
for a given PSD and SNR, then these waveforms are considered indistinguishable, and the differences in the recovered parameters are expected to be smaller than statistical errors.
The prefactor $D$ can be estimated as the number of intrinsic parameters whose measure is affected by model differences~\cite{Chatziioannou:2017tdw}, or it can be tuned by considering synthetic injections at increasing SNR~\cite{Purrer:2019jcp}.
In Fig.~\ref{fig:mismatch_aLIGO_ET} we show the mismatch of \seobesgb\, against its GR limit as a function of $\hat \ell_{\mathrm{GB}}$, for the aLIGO and ET detectors.
We consider a nonspinning binary with total mass $20 M_{\odot}$ and mass ratios $q\in\{2,4,8\}$ as a signal.
The horizontal lines correspond to the indistinguishability thresholds obtained by taking $D=9$ (our templates depend on two masses, $6$ spin components and $\hat \ell_{\mathrm{GB}}$) for some realistic SNRs ($10, 30$ for aLIGO, $100, 300$ for ET).
As expected, we find a larger mismatch for more asymmetric binaries, due to the scalar dipole radiation.
This results in a stronger bound on $\hat \ell_{\mathrm{GB}}$ if the data is consistent with GR. The minimum value of $\hat \ell_{\mathrm{GB}}$ that could be distinguished from the GR case goes from around $0.8$ ($q=2$, $\rm{SNR}=10$) to $0.3$ ($q=8$, $\rm{SNR}=30$) for aLIGO and from $0.3$ ($q=2$, $\rm{SNR}=100$) to $0.1$ ($q=8$, $\rm{SNR}=300$) for ET. The main source of improvement for ET is its larger low-frequency bandwidth, where the scalar flux dominates (see Fig.~\ref{fig:flux-2}).
Mismatch values $\lesssim 10^{-5}$ are not to be attributed to a non-zero coupling value, as differences of this order can be simply due to errors in the numerical integration of Hamilton's equations with finite tolerance.
Comparing our results to current constraints on ESGB gravity with LVK observations, the bound on GW190814 from Refs.~\cite{Wang:2021jfc, Lyu:2022gdr} (assuming the event is from a BBH), is given as a function of the coupling $\sqrt{\alpha_{\rm{GB}}} =  \ell_{\rm{GB}} / (4 \pi^{1/4})$ with $\ell_{\rm GB}/M=\nu \hat \ell_{\rm{GB}}$ by $\sqrt{\alpha_{\rm{GB}}} \lesssim 0.4 \mathrm{km}$. Taking a total mass of $\sim 26 M_{\odot}$ and a mass ratio $q=10$, this corresponds to $\hat \ell_{\rm{GB}} \lesssim 0.67$.
Our prospective bound $\hat \ell_{\rm{GB}} \lesssim 0.1$ for an ET source with $q=8$, $M = 20 M_{\odot}$ translates instead to $\sqrt{\alpha_{\rm{GB}}} \lesssim 0.055 \mathrm{km}$.

\begin{figure*}
    \includegraphics[width=\columnwidth]{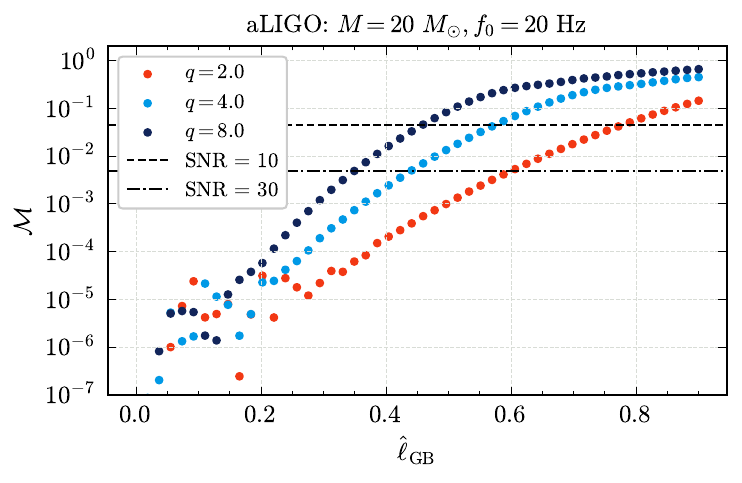}
    \includegraphics[width=\columnwidth]{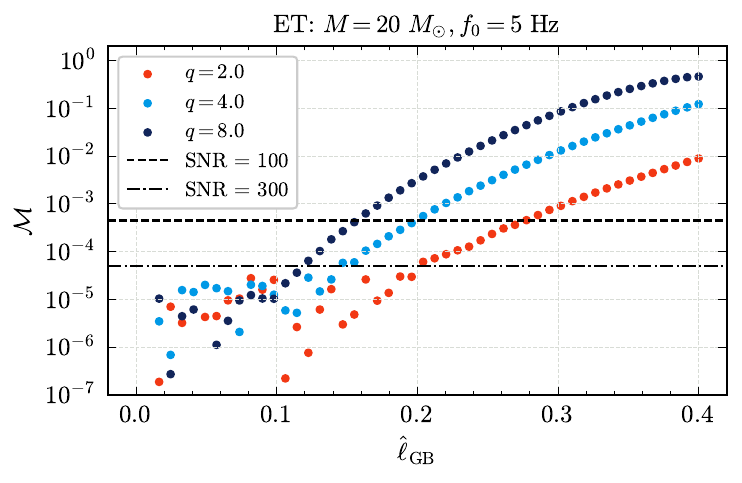}
	\caption{Mismatch of the (2,2) mode waveform of \seobesgb\, against its GR limit, as a function of the ESGB coupling $\hat \ell_{\mathrm{GB}}$ for the aLIGO (left panel) and ET (right panel) detectors. We consider a nonspinning BBH with total mass $20 M_{\odot}$ and mass ratios $q\in\{2,4,8\}$ in the dilatonic theory $f(\varphi)=\exp(2\varphi)/4$ and $\mathcal A(\varphi)=1$. The horizontal lines show the indistinguishability thresholds given by Eq.~(\ref{eq:indist_criterion}) for different SNRs.}
	\label{fig:mismatch_aLIGO_ET}
\end{figure*}

The potential constraints on $\hat \ell_{\mathrm{GB}}$ should only be regarded as an order of magnitude estimate.
Since the indistinguishability criterion offers a necessary, but not sufficient measure, if the mismatch exceeds the threshold, the differences are not necessarily measurable~\cite{Toubiana:2024car, Toubiana:2023cwr}.
Furthermore, only varying $\hat \ell_{\mathrm{GB}}$ does not account for possible correlations between the ESGB coupling and GR parameters.
For example, the main effect of a non-zero coupling being to shorten the waveform, one could expect it to be anti-correlated with the chirp mass $\mathcal{M}_c = \mu^{3/5}M^{2/5}$ of the binary~\cite{Cutler:1994ys},which affects the GW phase at leading PN order. Spin effects also affect the waveform in a qualitatively similar fashion, although they enter the GW phase starting from 1.5 PN order~\cite{Blanchet:2013haa}, and are thus expected to be less degenerate with ESGB corrections (starting from -1 PN) than the chirp mass. 
Therefore, this provides an optimistic estimate of the constraint derived from a single event.
On the other hand, these bounds could be significantly improved by combining results from multiple observations.
Since the coupling is common to all analyzed events, one can stack constraints from multiple observations by simply multiplying the marginalized likelihoods~\cite{DelPozzo:2011pg, Perkins:2021mhb}, instead of combining the results using a hierarchical approach, assuming that beyond-GR parameters in each event are drawn from a common underlying distribution~\cite{Isi:2019asy}, as typically done for parametrized deviations.

More sophisticated predictions of the measurability of the ESGB coupling could be either based on the Fisher matrix formalism or employ full Bayesian inference.
Forecasts of constraints with future detectors on ESGB gravity, as well as on other beyond-GR theories, have been obtained in Ref.~\cite{Perkins:2020tra}.
This work performed a Fisher analysis including corrections to the inspiral waveform based on the parameterized post-Einsteinian (ppE) framework~\cite{Yunes:2009ke}, mapping then the ppE results to theory-specific constraints, also including the effect of stacking multiple observations. A similar study, focused instead on the ringdown stage, was recently presented in Ref.~\cite{Maselli:2023khq}, using the parametrized spin-expansion coefficients (\texttt{ParSpec}) framework~\cite{Maselli:2019mjd}.

Similar analyses could be refined by repeating them with \seobesgb\,. Ref.~\cite{Perkins:2020tra} employed as baseline GR waveform \texttt{IMRPhenomPv2}~\cite{Hannam:2013oca}, which is a less accurate model than \texttt{SEOBNRv5PHM}.
It does not include higher modes and employs a simplified modeling of spin precession.
It also focused on estimating modifications at a single PN order at a time, while in our model, modifications at each PN order are specified as a function of the coupling parameter of the theory.
Furthermore, we include beyond-GR corrections in both the inspiral and the ringdown stages.
One could employ Bayesian parameter estimation to
provide forecasts for the next-generation GW detectors ET and CE, examining correlations among different parameters.
We leave such analyses for future work.
However, to show a practical application of \seobesgb\, to GW data analysis, we perform full Bayesian parameter estimation on some of the GW events reported by the LVK collaboration.

\subsection{Constraining the ESGB coupling}
\label{sec:PE}

As an application of \seobesgb\, to GW data analysis, we perform full Bayesian parameter estimation for some GW events reported by the LVK collaboration, also including the ESGB coupling constant $\ell_{\mathrm{GB}}$ [or equivalently $\sqrt{\alpha_{\mathrm{GB}}}$, cf. Footnote~\ref{footnote:YagiTransl}] in the analysis.
We consider three signals, GW190814~\cite{LIGOScientific:2020zkf}, GW190412~\cite{LIGOScientific:2020stg} and GW230529~\cite{LIGOScientific:2024elc}, which were recently analyzed in Refs.~\cite{Perkins:2021mhb, Wang:2021jfc, Lyu:2022gdr, Sanger:2024axs, Gao:2024rel} to provide constraints on the coupling of the dilatonic ESGB theory~\eqref{eq:dilatonicTheory}, using waveform models featuring ESGB corrections in the inspiral phase.

We expect larger corrections to the waveform for systems with smaller masses, as the leading ESGB corrections to the phase are proportional to $\hat \ell_{\rm GB}^4$~\cite{Julie:2019sab,Julie:2022qux,Lyu:2022gdr,Sanger:2024axs}, as well as for large mass ratios, due to the increased scalar dipole radiation [cf. Sec~\ref{sec:scalar_flux}].
As the sources of GW190412 and GW190814 are relatively low-mass binaries with asymmetric masses, these events are expected to offer better constraints on this theory. The source of GW190814, having a secondary with a mass around $2.6 M_{\odot}$, is consistent with being either a BBH or a NSBH. Under the hypothesis of GW190814 being a BBH, it imposes stringent bounds on ESGB gravity around $\sqrt{\alpha_{\mathrm{GB}}}\lesssim0.4~\mathrm{km}$~\cite{Wang:2021jfc, Lyu:2022gdr}.
However, this bound is not necessarily robust as it changes significantly whether one assumes the event to be an NSBH or a BBH.
Indeed, the scalar monopole vanishes for a NS in shift-symmetric ESGB gravity, so the correction can be much larger for a BBH than for a NSBH, whose hair is carried by the BH (which is more massive)~\cite{Perkins:2021mhb, Lyu:2022gdr}.

The most probable interpretation of the source of GW230529 is that of a NSBH with the primary being a BH with mass $2.5\text{--}4.5~\ensuremath{M_\odot}$. Under this hypothesis, it currently places the best constraint on ESGB gravity~\cite{Sanger:2024axs, Gao:2024rel}, with a $90 \%$ bound on $\sqrt{\alpha_{\mathrm{GB}}}$ ranging from $0.26~\mathrm{km}$ to $0.35~\mathrm{km}$ depending on the waveform approximant employed, when including ESGB corrections in the inspiral phase.
We note that for GW230529 one cannot determine from GW data alone if either component of the source is a NS or a BH, and the primary analysis from the LVK collaboration~\cite{LIGOScientific:2024elc} employed BBH waveforms, including the \texttt{SEOBNRv5PHM} model used here as a baseline GR approximant. 
Furthermore, Ref.~\cite{Sanger:2024axs} finds that, when considering parameterized deviations to the inspiral phase, NSBH and BNS results with the tides constrained to realistic values are very similar to the BBH results. Therefore, in our analysis, we also neglect tidal effects, but we take the sensitivities for the secondary to be 0, as is the case for NSs in shift-symmetric ESGB theories and in the dilatonic ESGB theory at leading order in the coupling [cf. Sec~\ref{subsec:dilatonic}]. In this work we concentrate on GW190814 and GW190412 under the BBH assumption, and on GW230529 under the NSBH assumption, for simplicity. 

\begin{figure*}[t]
  \includegraphics[width=0.666 \columnwidth]{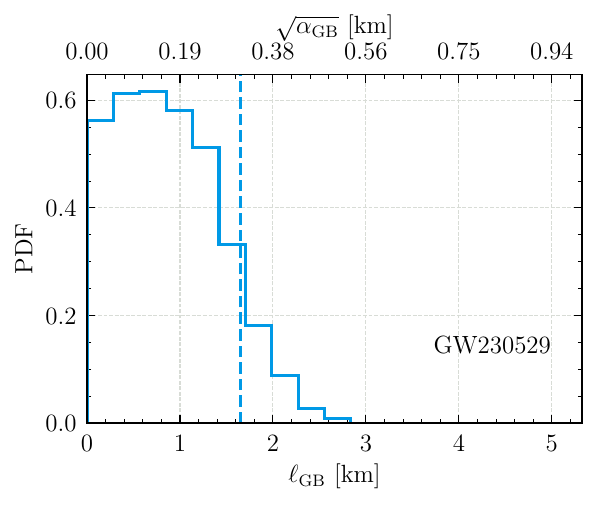}
  \includegraphics[width=0.666 \columnwidth]{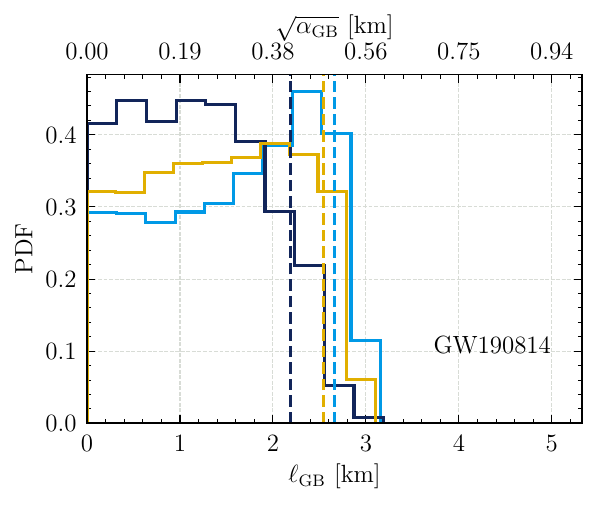}
  \includegraphics[width=0.666 \columnwidth]{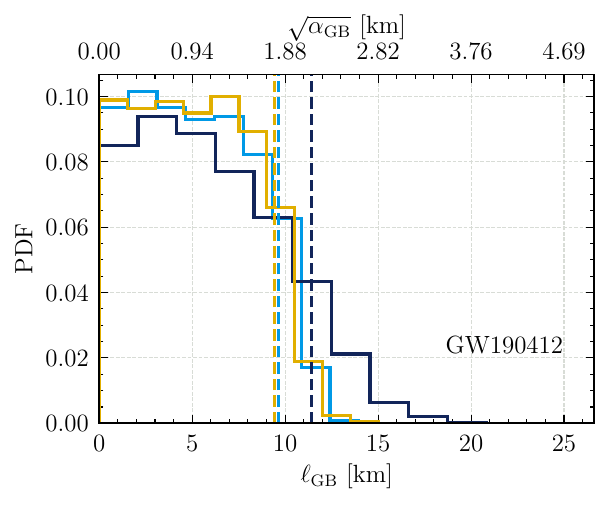}
  \caption{1-dimensional marginalized posterior of $\ell_{\rm GB}$ ($\sqrt{\alpha_{\mathrm{GB}}}$) obtained from the analysis of GW230529, GW190814 and GW190412, using the \seobesgb\, model in the dilatonic case $f(\varphi)=\exp(2\varphi)/4$ and $\mathcal A(\varphi)=1$. The vertical dashed lines indicate the $90 \%$ confidence bound. We show the posteriors for the full IMR analyses, without adding the merger corrections~\eqref{eq:nongr_mrg} (light blue) or marginalizing over them (yellow), and for inspiral-only analyses (dark blue). We perform a single analysis for GW230529, since it has negligible SNR in the merger-ringdown stage.}
  \label{fig:sqrt_alpha_GW190814}
\end{figure*}

\setlength{\extrarowheight}{8pt}
\begin{table*}
    \centering
    \begin{ruledtabular}
    \begin{tabular}{ l  c  c  c  c }
    % \hline
    Event & \makecell[cc]{GW230529} & \makecell[cc]{{GW190814}} & \makecell[cc]{GW190412} & \makecell[cc]{{Combined}}\\
    \hline
    \centering
    IMR &  $\ConstraintCIMREll~\text{km}$ ($\ConstraintCIMR~\text{km}$)   &   $\ConstraintBIMREll~\text{km}$ ($\ConstraintBIMR~\text{km}$)  & $\ConstraintAIMREll~\text{km}$ ($\ConstraintAIMR~\text{km}$)  & $\ConstraintCombinedIMREll~\text{km}$ ($\ConstraintCombinedIMR~\text{km}$) \\
    %\hline
    IMR with merger deviations &  --  & $\ConstraintBIMRMergerEll~\text{km}$ ($\ConstraintBIMRMerger~\text{km}$)  & $\ConstraintAIMRMergerEll~\text{km}$ ($\ConstraintAIMRMerger~\text{km}$)  & $\ConstraintCombinedIMRMergerEll~\text{km}$ ($\ConstraintCombinedIMRMerger~\text{km}$) \\
    %\hline
    Inspiral only &   --  & $\ConstraintBInspiralEll~\text{km}$ ($\ConstraintBInspiral~\text{km}$)  & $\ConstraintAInspiralEll~\text{km}$   ($\ConstraintAInspiral~\text{km}$)  & $\ConstraintCombinedInspiralEll~\text{km}$ ($\ConstraintCombinedInspiral~\text{km}$)\\
    % \hline
    \end{tabular}
    \end{ruledtabular}
    \caption{Summary of the $90 \%$ confidence bound on $\ell_{\rm GB}$ ($\sqrt{\alpha_{\mathrm{GB}}}$) from the analysis of GW230529, GW190814 and GW190412, using the \seobesgb\, model in the dilatonic case $f(\varphi)=\exp(2\varphi)/4$ and $\mathcal A(\varphi)=1$. We list the bounds from the full IMR analyses, without adding deviations to the merger and marginalized over parameterized merger deviations, and for inspiral-only analyses. We perform a single analysis for GW230529, since it has negligible SNR in the merger-ringdown stage.} 
    \label{tab:bounds}
\end{table*}

The goal of Bayesian parameter estimation is to infer the posterior distribution $p(\theta|d)$ for the parameters $\theta$ given the observed data $d$, using Bayes theorem
\begin{equation}
    p(\theta \mid d)=\frac{\mathcal{L}(d \mid \theta) \pi(\theta)}{Z}
\end{equation}
where $\mathcal{L}(d|\theta)$ is the likelihood of the data $d$ given the parameters $\theta$, $\pi(\theta)$ is the prior on $\theta$, and $Z \equiv \int \mathrm{d} \theta \mathcal{L}(d \mid \theta) \pi(\theta)$ is the evidence.
To determine whether a model $A$ is preferred over a model $B$, one can look at the Bayes factor, defined as the ratio of the evidence for the two different models $\mathrm{BF}^A_B=Z_A/Z_B$ (see, e.g., Ref.~\cite{Thrane:2018qnx} for a review).

We perform the analysis using {\tt parallel Bilby} \cite{Smith:2019ucc}, a highly parallelized version of the {\tt Bilby} \cite{Ashton:2018jfp, Romero-Shaw:2020owr} parameter-estimation code, and the nested sampler \texttt{dynesty} \cite{Speagle:2019ivv} using the \texttt{acceptance-walk} stepping method. We use the recommended setting for the number of accepted MCMC-chains $\mathrm{naccept}=60$, number of live points $\mathrm{nlive}=1000$, and set the remaining sampling parameters to their default values. For GW190814, we find that these settings were not sufficient to fully resolve the $\ell_{\mathrm{GB}}$ posterior distribution, and use $\mathrm{naccept}=100$, and $\mathrm{nlive}=2000$}. We employ strain data from the Gravitational Wave Open Source Catalog (GWOSC) \cite{LIGOScientific:2019lzm} and the released PSD and calibration envelopes. We use the same priors for the GR parameters as those employed in Refs.~\cite{RamosBuadesv5, LIGOScientific:2024elc}, in which these events have been analyzed using the \texttt{SEOBNRv5PHM} waveform model, and uniform priors on $\ell_{\rm GB}$, 
with the constraint $\ell_{\mathrm{GB}}/m^0_B \lesssim 0.831$ in the BBH case [cf. Sec.~\ref{subsec:dilatonic}]. In the NSBH case, we rather impose $\ell_{\mathrm{GB}}/m^0_A \lesssim 0.831$. To facilitate a more direct comparison with other results in the literature, we also perform inspiral-only analyses, taking a maximum frequency for the likelihood of $137~\mathrm{Hz}$ for GW190814 and $83~\mathrm{Hz}$ for GW190412, following Ref.~\cite{Wang:2021jfc}. These frequencies coincide with the ISCO frequency in GR, and thus, doing so amounts to cutting off the late stages of our inspiral-plunge waveforms above. We don't perform an inspiral-only analysis for GW230529 since the cutoff frequency for the inspiral stage would be around $600~\mathrm{Hz}$, and while the default analysis extends up to $1792~\mathrm{Hz}$, we don't expect significant differences due to the reduced sensitivity of the LIGO detectors in this high-frequency range.

As detailed in Sec. \ref{sec:mrd-waveform}, our model allows for parametrized deviations in the merger waveform amplitude and frequency~\eqref{eq:nongr_mrg}, which can be varied during sampling.
We perform runs assuming that they take GR values, as well as allowing them to vary for the $(2,2)$ mode.
We use uniform priors in the range $[-1,1]$ for both $\delta A_{22}$ and $\delta\omega_{22}$. 
We introduce additional merger parameters only in the IMR analysis, as the merger deviations do not significantly impact the results when only analyzing the inspiral.
For the same reason, we don't include these corrections in the analysis of GW230529, which has negligible SNR in the merger-ringdown stage.

\begin{figure}
    \includegraphics[width=\columnwidth]{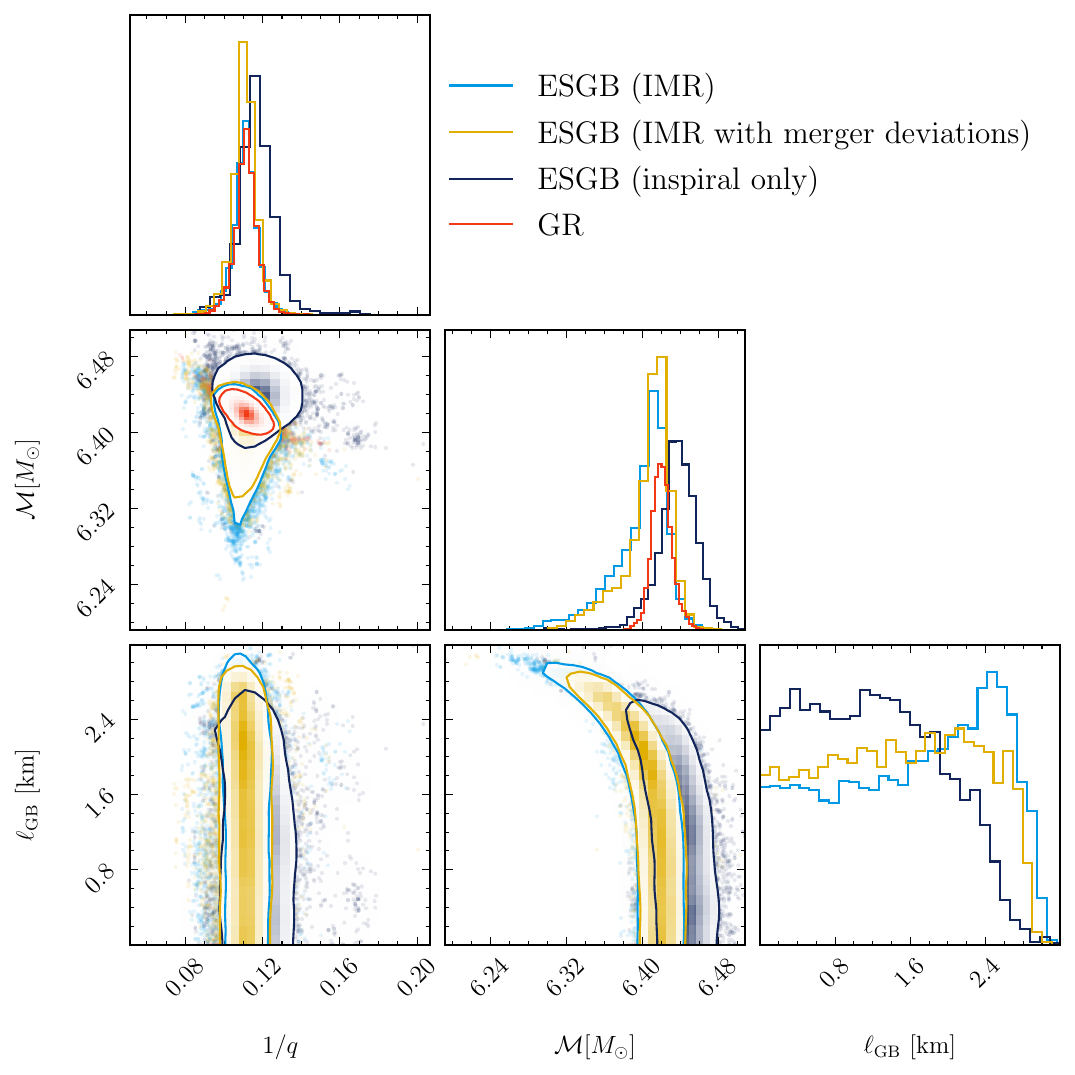}
	\caption{Posterior distribution of $\ell_{\mathrm{GB}}$, inverse mass ratio $1/q$ and chirp mass $\mathcal{M}_c$ obtained from the analysis of GW190814 using the \seobesgb\, model in the dilatonic case $f(\varphi)=\exp(2\varphi)/4$ and $\mathcal A(\varphi)=1$. For the masses, we also show the values recovered assuming GR using the \seobv5\, model. We notice a slight anti-correlation between $\ell_{\mathrm{GB}}$ and the chirp mass, as well as a wider chirp mass posterior shifted to larger values for the inspiral-only analysis. This might contribute to a marginally better constraint on $\ell_{\mathrm{GB}}$ for the inspiral-only analysis, despite the lower recovered SNR.}
	\label{fig:GW190814_chirp_mass}
\end{figure}

Figure~\ref{fig:sqrt_alpha_GW190814} presents the 1-dimensional posteriors of $\ell_{\mathrm{GB}}$ (or, equivalently, $\sqrt{\alpha_{\mathrm{GB}}}$), marginalized over all other parameters, resulting from the analysis of GW230529, GW190814 and GW190412. The vertical dashed lines indicate the $90 \%$ confidence bounds, also summarized in Table~\ref{tab:bounds}. We show the posteriors for the full IMR analyses, without adding corrections to the merger amplitude and frequency (light blue) and marginalized over parameterized merger deviations (yellow), and for inspiral-only analyses (dark blue). In all cases, the posteriors are compatible with $\ell_{\rm GB}=0$. 

For the NSBH event GW230529, we find a $90 \%$ bound on $\sqrt{\alpha_{\mathrm{GB}}}$ of $\ConstraintCIMR~\mathrm{km}$. Our bound is slightly less stringent than those of Refs.~\cite{Sanger:2024axs, Gao:2024rel} when relying on a BBH baseline model in GR, as they found $\sqrt{\alpha_{\mathrm{GB}}}\lesssim 0.28~\mathrm{km}$ and $\sqrt{\alpha_{\mathrm{GB}}}\lesssim 0.26~\mathrm{km}$, respectively.
However, our analysis differs in three ways:
(i) we include ESGB corrections at higher PN orders. The references above indeed include corrections to the conservative sector at 2PN, and to the dissipative sector at 1.5PN and 1PN relatively to the quadrupolar order, respectively;
(ii) our model resums these ESGB corrections within the EOB framework;
(iii) the waveform models of these works are not the same as ours in the GR limit, i.e. \texttt{SEOBNRv4HM$\_$ROM} \cite{Cotesta:2020qhw} and \texttt{IMRPhenomXPHM} \cite{Pratten:2020ceb} respectively. This last point alone can induce comparable bound discrepancies between these works.

For GW190814, our results are also in agreement with those of Refs.~\cite{Wang:2021jfc, Lyu:2022gdr}, with a $90 \%$ bound on $\sqrt{\alpha_{\mathrm{GB}}}$ ranging from $\ConstraintBInspiral~\mathrm{km}$ to $\ConstraintBIMRMerger~\mathrm{km}$.
We notice that, for this event, the inspiral-only analysis provides a slightly better constraint compared to the IMR analyses.
Figure~\ref{fig:GW190814_chirp_mass} shows the posterior distribution of $\ell_{\mathrm{GB}}$, inverse mass ratio $1/q$ and chirp mass $\mathcal{M}_c$ inferred from GW190814, along with the mass values recovered assuming GR. We observe: (i) a slight anti-correlation between $\ell_{\mathrm{GB}}$ and the chirp mass, whether the plunge-merger-ringdown is included or not, as expected for inspiral-dominated signals [cf. Sec.~\ref{sec:meas}]; and (ii) that neglecting the plunge-merger-ringdown signal results in a wider chirp mass posterior shifted towards larger values.
These two points might contribute to the slightly better constraint from the inspiral-only analysis.
Another factor could be that the merger-ringdown waveform is practically indistinguishable from GR for this event. Therefore, its inclusion reduces the mismatch of the waveform with its GR counterpart at fixed $\ell_{\rm GB}$.

For the BBH event GW190412, we obtain a constraint $\sqrt{\alpha_{\mathrm{GB}}}=\ConstraintAInspiral~\mathrm{km}$ from the inspiral-only analysis and a slightly tighter bound $\sqrt{\alpha_{\mathrm{GB}}}=\ConstraintAIMRMerger~\mathrm{km}$ when including the plunge-merger-ringdown portions of the signal, thanks to the additional SNR. This is roughly a factor $\sim 2$ smaller compared to previous analyses~\cite{Wang:2021jfc}. 
The larger chirp mass of the source ($\mathcal{M}_c\sim 15~M_{\odot}$, versus $\mathcal{M}_c\sim 6~M_{\odot}$ for GW190814 and $\mathcal{M}_c\sim 2~M_{\odot}$ for GW230529) results in a higher relative SNR in the late-inspiral stage, increasing the impact of higher order PN corrections and the EOB resummation.
For GW190412 and GW190814, the bounds on $\sqrt{\alpha_{\mathrm{GB}}}$ are robust with respect to uncertainties in the merger morphology, with a slight decreas} when marginalized over parameterized merger deviations.

Combining the constraints from individual events by multiplying the marginalized posteriors, we obtain a combined bound on $\sqrt{\alpha_{\mathrm{GB}}}$ ranging from $\ConstraintCombinedInspiral~\mathrm{km}$ (for the inspiral-only analyses) to $\ConstraintCombinedIMRMerger - \ConstraintCombinedIMR~\mathrm{km}$ (for the IMR analyses, with and without merger deviations respectively).
The combined result is very similar to the one obtained by analyzing GW230529 alone, and if anything slightly worsened in the IMR analyses by the fact that the $\sqrt{\alpha_{\mathrm{GB}}}$ posterior for GW190814 peaks slightly away from 0.

We find, a posteriori, that our bounds are within the small-coupling limit. This justifies using the QNM expressions from Ref.~\cite{Chung:2024ira}, as well as employing the leading-order expressions~(\ref{eq:sensiLeadingEll}) and~\eqref{eq:sensiLeadingEllns} for the quantities in Eqs.~(\ref{eq:sensis}) when analyzing GW230529 [cf. Sec.~\ref{subsec:dilatonic}]. For GW230529, we find $\ell_{\mathrm{GB}}/m_A^0 \lesssim 0.28$ at $90 \%$ confidence. For GW190412 we obtain $\ell_{\mathrm{GB}}/M_f \lesssim 0.15$, while for GW190814 $\ell_{\mathrm{GB}}/M_f \lesssim 0.065$.

Table~\ref{tab:bayes_factors} summarizes the natural log-Bayes factors between the ESGB and GR hypotheses, inferred from the full IMR analyses of the signals. Without adding deviations to the merger amplitude and frequency, we find a log Bayes factor of $\BayesCIMR$ for GW230529, of $\BayesBIMR$ for GW190814 and of $\BayesAIMR$ GW190412, indicating no substantial preference for either hypothesis~\cite{kass:bayesfactor}.
When adding parametrized merger deviations, we obtain lower Bayes factors, of $\BayesBIMRMerger$ for GW190814 and of $\BayesAIMRMerger$ GW190412.
This means that they do not improve how well our model fits the data, which is consistent with GW190412 and GW190814 being inspiral-dominated signals with low SNR in the merger-ringdown.
The decrease in the Bayes factor is then the manifestation of the Occam penalty for introducing additional parameters. Accurate modeling of the merger phase with NR simulations in ESGB gravity would be valuable to improve this analysis.

We note that the computational cost of these runs is similar to the corresponding GR analyses with \seobv5, slightly higher due to additional parameters and a marginally longer waveform evaluation time. They took around 36 hours for GW190412 (256 cores), 3 days for GW190814 (384 cores) and 5 days for GW230529 (640 cores), with each analysis requiring $\mathcal{O}(10^8)$ waveform evaluations.

Our combined result could be further improved by also including the NSBH events GW200105, GW200115, which also place competitive bounds on this theory~\cite{Lyu:2022gdr}, as well as selected BBH events. 
Since the NSBHs are low-mass binaries, leading to inspiral-dominated signals, the constraints on ESGB gravity would likely be similar to those obtained by Ref.~\cite{Lyu:2022gdr}, which found the result to be robust with respect to the inclusion of higher-order PN corrections. On the other hand, slightly improved bounds could be obtained for heavier binaries when using \seobesgb\,. 
However, the result obtained by combining several events would likely improve only marginally, as GW230529 provides a much more stringent bound compared to other events. This is a computationally expensive but still feasible analysis that we leave for future work.
Another way to improve the combined constraint we present here would be to incorporate information about the underlying astrophysical population following Ref.~\cite{Payne:2023kwj}, by simultaneously inferring the astrophysical distribution to correctly account for correlations between intrinsic and beyond-GR parameters. 

\begin{table}[t]
  \centering
  \begin{ruledtabular}
  \begin{tabular}{ l  c  c  c}
  % \hline
  Event & \makecell[cc]{GW230529} & \makecell[cc]{GW190814} & \makecell[cc]{{GW190412}} \\
  \hline
  \centering
  IMR &   $\BayesCIMR$  & $\BayesBIMR$ & $\BayesAIMR$ \\
  %\hline
  IMR with merger deviations &  ---  & $\BayesBIMRMerger$ & $\BayesAIMRMerger$ \\
  %\hline
  \end{tabular}
  \end{ruledtabular}
  \caption{Natural log Bayes factors between the dilatonic ESGB and GR hypotheses from the analysis of GW230529, GW190814 and GW190412, using the \seobesgb\, model. We consider the IMR analyses in ESGB, both with and without parameterized deviations to the waveform amplitude and frequency at merger. 
  } 
  \label{tab:bayes_factors}
\end{table}

\section{Conclusions}
\label{sec:conclusions}

Using the \texttt{pySEOBNR} code~\cite{Mihaylovv5}, we built the first IMR waveform model in a beyond-GR theory within the EOB approach, focusing on ESGB gravity for BBHs. The \seobesgb\, waveforms 
reduce to the \texttt{SEOBNRv5PHM} model~\cite{RamosBuadesv5} in the GR limit. The EOB Hamiltonian and 
fluxes in the \texttt{pySEOBNR} code can be used to build IMR waveforms for BBHs and inspiral waveforms for BNS and NSBHs in any ESGB (or ST) theory.
The EOB Hamiltonian [cf. Sec.~\ref{sec:esgb_hamiltonian}] includes 
ESGB (and ST) corrections through 3PN order~\cite{Julie:2022qux}, while the metric modes in factorized form and the scalar flux entering the RR force contain ESGB (and ST) corrections through 2PN and 2.5PN (relative to the scalar
dipolar) orders, respectively [cf. Sec.~\ref{sec:esgb_flux}]. We recall that the factorized ESGB (and ST) metric modes [cf. Sec.~\ref{sec:waveform-modes}] are a new result of this paper. 
While the ESGB inspiral waveform is theory specific, 
the merger-ringdown waveform follows phenomenological ansätze used in GR~\cite{Damour:2014yha,Bohe:2016gbl,Cotesta:2018fcv,Pompili:2023tna}, augmented with an agnostic parameterization 
for the modes' amplitude and frequency at merger, inspired by Refs.~\cite{Brito:2018rfr,Ghosh:2021mrv,Maggio:2022hre}. It also uses corrections to the QNMs frequencies and 
decay times derived in ESGB theory~\cite{Chung:2024ira} and corrections to the remnant mass and spin, which we find can 
cause changes in the frequency of the ringdown waveform from one to two orders of magnitude larger than the ESGB corrections to the QNM spectrum.
Although our model can be applied to different ESGB coupling functions, we focused 
our investigation to dilatonic ESGB gravity.

In Sec.~\ref{sec:morph}, we analyzed the morphology of the dilatonic ESGB waveforms, and contrasted them with the GR ones, finding that the ESGB corrections in the conservative and 
dissipative dynamics cause a BBH to merge in a shorter time than in GR, when starting at the same orbital frequency (see Fig.~\ref{fig:waveform}).
We provided
measurability estimates of the ESGB coupling [cf. Sec.~\ref{sec:meas}] based on the mismatch criterion for upcoming runs of the LVK collaboration and future 
ground-based detectors, such as the Einstein Telescope [cf. Fig.~\ref{fig:mismatch_aLIGO_ET}]. 
The bound from a single asymmetric ($q\sim 8$) BBH observed with ET could be one order of magnitude better than current constraints using LVK data.
Low-mass NSBHs could also be ``golden'' binary sources, as exemplified by GW230529.
This indicates that stellar-mass binaries observed with next-generation ground-based detectors are ideal candidates to test this theory, also thanks to their high expected rates~\cite{Borhanian:2022czq}, and would likely provide better constraints than extreme-mass-ratio-inspirals seen with LISA~\cite{Speri:2024qak}.

We also used \seobesgb\, to place constraints on the coupling of the dilatonic ESGB theory using GW observations 
and performed Bayesian model selection between ESGB and GR [cf. Sec.~\ref{sec:PE}].
We found that the GW events GW190412, GW190814 and GW230529 are consistent with GR, and we constrained the fundamental coupling of the theory to be $\ell_{\mathrm{GB}} \lesssim \ConstraintCombinedIMRMergerEll~\mathrm{km}$ ($\sqrt{\alpha_{\mathrm{GB}}} \lesssim \ConstraintCombinedIMRMerger~\mathrm{km}$) at 90 $\%$ confidence. Among the events we analyzed, GW230529 provides the most stringent constraint, consistently with other recent analyses~\cite{Sanger:2024axs, Gao:2024rel}.

It will be very beneficial to validate \seobesgb\, against NR simulations in ESGB gravity and to calibrate it using NR data, particularly by including NR information for the merger phase, and by guiding resummations of the PN scalar flux. Despite remarkable progress in NR evolutions of BBHs in ESGB gravity~\cite{Witek:2018dmd,Julie:2020vov,Witek:2020uzz,Okounkova:2020rqw,East:2020hgw,East:2021bqk,Corman:2022xqg,Doneva:2023oww,AresteSalo:2023mmd} and NSBH binaries~\cite{Corman:2024vlk}, the number and length of current simulations, as well as their accuracy, have so far prevented us from carrying out these studies.

We limited our investigation to BHs that have already scalarized. However, ESGB gravity also allows for dynamical scalarization~\cite{Khalil:2019wyy,Julie:2023ncq}. Interesting work has been done to model dynamical scalarization~\cite{Sennett:2016rwa,Sennett:2017lcx,Khalil:2019wyy,Khalil:2022sii}, and we plan to extend \seobesgb\, to include these nonperturbative effects, which can largely affect the waveforms toward merger, producing distinctive signatures that are absent in GR.

Furthermore, it should be straightforward to include in our waveform model ESGB corrections in the spinning sector, once they are available.
Another extension of our model would involve describing complete waveforms for BNSs and NSBHs in both ST and ESGB theories, once these models have been
finalized in GR within the \texttt{SEOBNRv5} family.

On the data-analysis side, the simple forecast we presented for upcoming runs of the LVK collaboration and 
  next-generation ground-based GW detectors can be extended to a broader region of 
the binary's parameter space by using linear-signal approximation methods, such as the Fisher matrix formalism.  
 Furthermore, they can be made more realistic by stacking many events and taking into account astrophysical
  populations~\cite{Payne:2023kwj}. Results for selected events can be
  validated by performing synthetic injection and analyzing them with
  Bayesian inference. Furthermore, our model can be used to validate
  results obtained with parametrized tests of GR~\cite{Agathos:2013upa, Mehta:2022pcn} and to test their
  effectiveness in different regions of the parameter space. It would
  also be interesting to analyze more LVK events, including the NSBHs
  GW200105, GW200115 and selected low-mass BBHs, to improve the
  combined constraints on $\ell_{\rm GB}$. However, the
  results would likely only improve marginally, as GW230529 provides a
  much more stringent bound compared to other events.

\section*{Acknowledgments}
It is a pleasure to thank Maxence Corman, Quentin Henry, Elisa Maggio, Elise S\"{a}nger, Hector Okada da Silva, and David Trestini for useful discussions. We also thank Raffi Enficiaud for his assistance with scientific computing.

The computational work for this manuscript was carried out on the \texttt{Hypatia} computer cluster at the Max Planck Institute for Gravitational Physics in Potsdam.
This work made use of the python package \texttt{pySEOBNR}~\cite{Mihaylovv5}. Stable versions of \texttt{pySEOBNR} are published through the Python Package Index (PyPI), and can be installed via ~\texttt{pip install pyseobnr}.

This research used data or software obtained from the Gravitational Wave Open Science Center (gwosc.org), a service of the LIGO Scientific Collaboration, the Virgo Collaboration, and KAGRA. This material is based upon work supported by NSF's LIGO Laboratory, a major facility fully funded by the National Science Foundation, as well as the Science and Technology Facilities Council (STFC) of the United Kingdom, the Max-Planck-Society (MPS), and the State of Niedersachsen/Germany for support of the construction of Advanced LIGO and construction and operation of the GEO600 detector. Additional support for Advanced LIGO was provided by the Australian Research Council. Virgo is funded, through the European Gravitational Observatory (EGO), by the French Centre National de Recherche Scientifique (CNRS), the Italian Istituto Nazionale di Fisica Nucleare (INFN) and the Dutch Nikhef, with contributions by institutions from Belgium, Germany, Greece, Hungary, Ireland, Japan, Monaco, Poland, Portugal, Spain. KAGRA is supported by Ministry of Education, Culture, Sports, Science and Technology (MEXT), Japan Society for the Promotion of Science (JSPS) in Japan; National Research Foundation (NRF) and Ministry of Science and ICT (MSIT) in Korea; Academia Sinica (AS) and National Science and Technology Council (NSTC) in Taiwan.

\appendix

\section{Einstein and Jordan frames\label{App:EinsteinVsJordan}}

In Refs.~\cite{Mirshekari:2013vb,Bernard:2018hta,Bernard:2018ivi}, the ST two-body Lagrangian was derived up to 3PN order by adopting the Jordan-frame formulation of the theory:
\begin{align}
I_{\rm ST} =  \int \frac{d^{4}x  \sqrt{-\tilde g}}{16 \pi}  \left(\phi \tilde R - \frac{\omega(\phi)}{\phi}(\partial\phi)^2\right)+I_{\rm m}[\Psi,\tilde g_{\mu\nu}],\label{eq:JFaction1}
\end{align}
using tildes for clarity. Here, $(\partial\phi)^2=\tilde g^{\mu\nu} \partial_{\mu}\phi \partial_{\nu}\phi$, and $\omega(\phi)$ is a function defining the theory. As for compact bodies, they were described by the point particle action
\begin{align}
I_{\rm m}\to  I_{\rm m}^{\rm pp}=-\sum_A\int \tilde m_A(\phi)d\tilde s_A,\label{eq:JFaction2}
\end{align}
with $d\tilde s_A=\sqrt{-\tilde g_{\mu\nu}dx_A^\mu dx_A^\nu}$.
By contrast, we adopt here the Einstein-frame formulation of ST theories:
\begin{align}
I_{\rm ST} =  \int \frac{d^{4}x  \sqrt{-g}}{16 \pi}  \left(R - 2(\partial\varphi)^2\right)+I_{\rm m}[\Psi,\mathcal A^2g_{\mu\nu}],\label{eq:EFaction1}
\end{align}
where $(\partial\varphi)^2=g^{\mu\nu} \partial_{\mu}\varphi \partial_{\nu}\varphi$, and we account for compact bodies by the substitution
\begin{align}
I_{\rm m}\to I_{\rm m}^{\rm pp}=-\sum_A\int m_A(\varphi)ds_A,\label{eq:EFaction2}
\end{align}
with $ds_A=\sqrt{-g_{\mu\nu}dx_A^\mu dx_A^\nu}$.
The actions \eqref{eq:JFaction1}-\eqref{eq:JFaction2} and \eqref{eq:EFaction1}-\eqref{eq:EFaction2} are the same, modulo boundary terms and the following redefinitions:
\begin{subequations}
\begin{align}
\tilde g_{\mu\nu}&=\mathcal A^2 g_{\mu\nu},\\
3+2\omega(\phi)&=\left(\frac{d\ln\mathcal A}{d\varphi}\right)^{-2},\\
m_A(\varphi)&=\mathcal A(\varphi) \tilde m_A(\varphi),\label{eq:conversionJfEfmasses}%
\end{align}\label{eq:conversionJfEf}%
\end{subequations}
where $\varphi(\phi)$ is obtained by inverting $\mathcal A(\varphi)=1/\sqrt{\phi}$.
We also introduce the notation:
\begin{subequations}
\begin{align}
\alpha_0&=\frac{d\ln \mathcal A}{d\varphi}(\varphi_0),\label{eq:sensisUniversellesAlpha}\\
\beta_0&=\frac{d\alpha}{d\varphi}(\varphi_0),\\
{\beta'}_0&=\frac{d\beta}{d\varphi}(\varphi_0),\\
{\beta''}_0&=\frac{d{\beta'}}{d\varphi}(\varphi_0),
\end{align}\label{eq:sensisUniverselles}%
\end{subequations}
where the subscript $0$ denotes a quantity evaluated at infinity, $\varphi(\phi_0)=\varphi_0$.
The quantities above can be obtained by inserting Eq.~\eqref{eq:conversionJfEfmasses} into Eqs.~\eqref{eq:sensis} and taking the limit $\tilde m_A(\phi)=\rm const.$, in which case body $A$ is said to have negligible self-gravity.

We can now translate the parameters of Refs.~\cite{Mirshekari:2013vb,Bernard:2018hta,Bernard:2018ivi} into our conventions using Eqs.~\eqref{eq:conversionJfEf} and below. The results are gathered in Table~\ref{table:JFvsEFparameters}.

\begin{table*}
\centering
\begin{tabular}{|c |l|}
  \hline
    Refs.~\cite{Mirshekari:2013vb,Sennett:2016klh,Bernard:2018hta,Bernard:2018ivi,Bernard:2022noq} & 
    Refs.~\cite{Damour:1992we,Damour:1995kt,Julie:2017pkb,Julie:2017ucp,Julie:2019sab,Julie:2022qux} and this paper \\
      \hline \hline
Theory dependent &  \\
  $\tilde G$ & $\mathcal A_0^2 (1+\alpha_0^2)$\\
    $\zeta$ & $\frac{\alpha_0^2}{1+\alpha_0^2}$ \\ 
  $\lambda_1$ & $\frac{\beta_0}{2(1+\alpha_0^2)}$ \\
  $\lambda_2$ & $\frac{1}{4(1+\alpha_0^2)^2}(-\beta'_0\alpha_0-2\beta_0\alpha_0^2+4\beta_0^2)$ \\ 
  $\lambda_3$ & $\frac{1}{8(1+\alpha_0^2)^3}(6\beta'_0\alpha_0^3-24\beta_0^2\alpha_0^2+8\beta_0\alpha_0^4-13\beta'_0\beta_0\alpha_0+24\beta_0^3+{\beta''}_0\alpha_0^2)$ \\ 
    \hline
      Body dependent &  \\
  $m_A$ & $m_A^0/\mathcal A_0$ \\
  $s_A$ & $\frac{1}{2}(1-\alpha_A^0/\alpha_0)$ \\
  $s_A'$ & $\frac{1}{4\alpha_0}(\beta_A^0/\alpha_0-\alpha_A^0\beta_0/\alpha_0^2)$ \\
  $s_A''$ & $\frac{1}{8\alpha_0^5}(3\beta_A^0\alpha_0\beta_0-3\alpha_A^0\beta_0^2-{\beta'}_A^0\alpha_0^2+\alpha_A^0\beta'_0\alpha_0)$ \\
  $s_A'''$ & $\frac{1}{16\alpha_0^7}(15\beta_A^0\alpha_0\beta_0^2-15\alpha_A^0\beta_0^3-5{\beta'}_A^0\beta_0\alpha_0^2+10\alpha_A^0\beta_0'\beta_0\alpha_0-{\beta'}_A^0\alpha_0^2\beta_0$
  $-4\beta_A^0\alpha_0^2\beta_0'+{\beta''}_A^0\alpha_0^3-\alpha_A^0{\beta''}_A^0\alpha_0^2)$\\
  \hline
   0PN &  \\
  $\tilde G\tilde\alpha$ & $\mathcal A_0^2G_{AB}=\mathcal A_0^2(1+\alpha_A^0\alpha_B^0) $\\
  \hline   1PN &  \\
  $\bar\gamma$ & $\bar\gamma_{AB}=\frac{-2\alpha_A^0\alpha_B^0}{1+\alpha_A^0\alpha_B^0}$ \\
  $\bar\beta_A$ & $\bar\beta_A=\frac{\beta_A^0(\alpha_B^0)^2}{2(1+\alpha_A^0\alpha_B^0)^2}$ \\
  \hline
   2PN &  \\
  $\bar\delta_A$ & $\delta_A=\frac{(\alpha_A^0)^2}{(1+\alpha_A^0\alpha_B^0)^2}$ \\
  $-4\,\bar\chi_A$ & $\epsilon_A=\frac{{\beta'}_A^0(\alpha_B^0)^3}{(1+\alpha_A^0\alpha_B^0)^3}$ \\
    $-8\,\bar\beta_A\bar\beta_B/\bar\gamma$ & $\zeta_{AB}=\frac{{\beta}_A^0{\beta}_B^0\alpha_A^0\alpha_B^0}{(1+\alpha_A^0\alpha_B^0)^3}$ \\
  \hline
   3PN &  \\
  $\bar\kappa_A$ & $\kappa_A=\frac{(\alpha_B^0)^4{\beta''_A}^0}{8(1+\alpha_A^0\alpha_B^0)^4}$ \\
  $-4\,\bar\beta_A\bar\delta_A/\bar\gamma$ & $\psi_A=\frac{\alpha_A^0\alpha_B^0\beta_A^0}{(1+\alpha_A^0\alpha_B^0)^3}$ \\
  $16\,\bar\beta_A\bar\chi_B/\bar\gamma$ & $\xi_A=\frac{(\alpha_A^0)^2\alpha_B^0\beta_A^0{\beta'}_B^0}{(1+\alpha_A^0\alpha_B^0)^4}$ \\
$32\,\bar\beta_A(\bar\beta_B)^2/\bar\gamma^2$   & $\omega_A=\frac{(\alpha_A^0)^2\beta_A^0(\beta_B^0)^2}{(1+\alpha_A^0\alpha_B^0)^4}$ \\
  \hline
Other &  \\
$\sqrt{\zeta}\,\mathcal S_\pm$ & $\frac{\alpha_\pm}{2\sqrt{1+\alpha_A^0\alpha_B^0}}$ \\
  \hline
\end{tabular}
\caption{Translation of the parameters from Refs.~\cite{Mirshekari:2013vb,Sennett:2016klh,Bernard:2018hta,Bernard:2018ivi,Bernard:2022noq}. We rename their $\alpha$ as $\tilde\alpha$ here to avoid confusing it with ours.}
\label{table:JFvsEFparameters}
\end{table*}

\section{ESGB corrections to the EOB potentials\label{sec:ST_ESGB_HeffReminder}}

The coefficients
$(\delta \bar a_{3},\, \delta \bar a_{4},\, \bar a_{4,\ln})$,
$(\delta \bar d_{2},\, \delta \bar d_3)$
and $(\delta\bar q_{1},\, \bar q_{2})$
entering Eqs.~\eqref{eq:finalEOBpotentials} were calculated in Refs.~\cite{Julie:2017pkb,Julie:2022qux}. They read:
\begin{widetext}
\begin{subequations}
\begin{align}
\delta\bar a_3&=\frac{1 }{12} \Big[\nu(-24 \zeta _{A B}-36\bar{\beta }_++4   \bar{\gamma }_{A B}^2+40 \bar{\gamma }_{A B}+8\delta _++4  \epsilon _+)-24\langle\bar\beta\rangle\left(1-2 \bar{\gamma }_{A B}\right)-35 \bar{\gamma }_{A B}^2  -20 \bar{\gamma }_{A B}+4\langle\delta\rangle-4\langle\epsilon\rangle \Big],\\
\delta\bar a_4&=2 k_{\rm ESGB}-4k_{\rm tail} (\gamma _E+\ln 2)+\frac{\nu  \bar{\gamma }_{AB} \big(11 \left(\bar{\gamma }_{AB}+2\right)^2-4 \langle\delta\rangle\big)}{4 \tilde\alpha  \left(\bar{\gamma }_{AB}+2\right)}\nonumber\\
&+\frac{1}{12}\Big[2 \langle\bar\beta\rangle \left(47 \bar{\gamma }_{AB}^2-28 \bar{\gamma }_{AB}+6 \bar{\beta }_++4 \delta _+-12 \langle\delta\rangle-28\right)-60 \bar{\gamma }_{AB}^3-78 \bar{\gamma }_{AB}^2-24 \bar{\gamma }_{AB}+\langle\delta\rangle \left(16 \bar{\gamma }_{AB}+8\right)-12 \langle\epsilon\rangle \bar{\gamma }_{AB}\nonumber\\
&-4 \delta _- \bar{\beta }_--4 \delta _+ \bar{\beta }_++3 \bar{\beta }_-^2-3 \bar{\beta }_+^2+8 \langle\delta\rangle \bar{\beta }_++60 \langle\bar\beta\rangle^2+8 \langle\kappa\rangle\Big]\nonumber\\
&+\frac{\nu}{1152}\Big[-288 \bar{\beta }_+ \left(40 \bar{\gamma }_{AB}-3\right)+192 \langle\bar\beta\rangle \left(-4 \bar{\gamma }_{AB}^2-64 \bar{\gamma }_{AB}+30 \bar{\beta }_+-8 \delta _+-19\right)\nonumber\\
&+\delta _+ \left(126 \pi ^2 \bar{\gamma }_{AB}+768 \bar{\gamma }_{AB}+252 \pi ^2-6080\right)-6912 \zeta _{AB} \bar{\gamma }_{AB}+1152 \epsilon _+ \bar{\gamma }_{AB}+63 \pi ^2 \bar{\gamma }_{AB}^3-432 \bar{\gamma }_{AB}^3-180 \pi ^2 \bar{\gamma }_{AB}^2\nonumber\\
&+15296 \bar{\gamma }_{AB}^2-1350 \pi ^2 \bar{\gamma }_{AB}+37184 \bar{\gamma }_{AB}+192 \langle\delta\rangle\left(3 \bar{\gamma }_{AB}+2\right)+1536 \delta _- \bar{\beta }_--864 \bar{\beta }_-^2-288 \bar{\beta }_+^2+3456 \zeta _{AB}\nonumber\\
&+1152 \langle\bar\beta\rangle^2+1152 \langle\epsilon\rangle-768 \kappa _+-768 \langle\kappa\rangle-1536 \langle\psi\rangle+1152 \langle w\rangle-768 \psi _++1152 \langle\xi\rangle\Big],\label{eq:da4}\\
 \bar a_{4,\ln}&=-2k_{\rm tail},
\end{align}%
\end{subequations}

\begin{subequations}
\begin{align}
\delta\bar d_2&=\frac{1}{4} \Big[-3 \bar{\gamma }_{A B}^2-12 \bar{\gamma }_{A B}+4 \langle\delta \rangle-24 \langle\bar{\beta }\rangle+8 \nu(2 \bar{\gamma }_{A B} -\langle\bar{\beta }\rangle) \Big],\nonumber\\
\delta \bar d_3&=k_{\rm tail}(21-32 \ln 2 )+\frac{1}{12} \Big[4 (\langle\delta\rangle-6 \langle\bar\beta\rangle) \left(3 \bar{\gamma }_{AB}+8\right)-9 \bar{\gamma }_{AB}^3-52 \bar{\gamma }_{AB}^2-64 \bar{\gamma }_{AB} +16 \langle\epsilon\rangle\Big]\nonumber\\
&+\frac{\nu}{12} \Big[4 \left(-\delta _+ \left(6 \bar{\gamma }_{AB}+11\right)+9 \bar{\gamma }_{AB}^3+77 \bar{\gamma }_{AB}^2+173 \bar{\gamma }_{AB}+27 \zeta _{AB}-6 \langle\delta\rangle+3 \langle\epsilon\rangle-4 \epsilon _+\right)\nonumber\\
&+3 \bar{\beta }_+ \left(4 \bar{\gamma }_{AB}+69\right)-18 \langle\bar\beta\rangle \left(8 \bar{\gamma }_{AB}+11\right)\Big]+\nu^2\Big[-\bar{\gamma }_{AB}^2-10 \bar{\gamma }_{AB}+9 \bar{\beta }_++6 \zeta _{AB}-2 \delta _+-\epsilon _+\Big],
\end{align}
\end{subequations}
and, finally,
\begin{subequations}
\begin{align}
\delta\bar q_1&=\frac{1}{6}k_{\rm tail}(93+1753 \ln 2-729 \ln 6)
+\frac{\nu}{6}\Big[\bar{\gamma }_{AB} \left(15 \bar{\gamma }_{AB}+52\right)+4( \langle\bar\beta\rangle- \langle\bar\delta\rangle)\Big]
+2 \nu ^2 \Big[\langle\bar\beta\rangle-2 \bar{\gamma }_{AB}\Big],\\
\bar q_2&=\frac{3}{10}k_{\rm tail}(37-5707 \ln 2+2187 \ln 6).
\end{align}
\end{subequations}
\end{widetext}

\section{Factors $\hat H_{\ell m}$ of the metric modes\label{App:metricModes}}

Here, we give the explicit factors $\hat H_{\ell m}$ that are nonzero at 2PN, with $m\geq 0$.
Note that $\hat H_{20}$ originates from memory effects~\cite{Sennett:2016klh}.
It does not contribute to the metric flux on circular orbits \eqref{RRforce} because $m=0$, but it enters the metric waveform.
\begin{widetext}
\begin{subequations}
\begin{align}
\hat H_{22}&=1+x\,\left[ -\frac{2}{3} (\bar{\gamma }_{AB}+\bar{\beta }_+)-\frac{107}{42}+\frac{2m_-}{3}  \bar{\beta }_-+\frac{55 \nu }{42}\right]
+\frac{x^{3/2}}{1+\alpha^0_A \alpha^0_B}\left[2 \pi+i\, \left(\frac{1}{12} \left(\alpha _+^2-\alpha _-^2\right)-\frac{3  \nu }{8}\alpha _-^2\right)\right]\nonumber\\
&+x^2\Bigg[\frac{1}{1512}\left(630 \bar{\gamma }_{AB}^2-72 \bar{\gamma }_{AB}-504(\bar{\beta }_+^2+ \bar{\beta }_-^2)+1356 \bar{\beta }_++252 (\delta _++ \epsilon _+)-2173\right)+\frac{m_-}{126}  \left(84 \bar{\beta }_+\bar{\beta }_--113\bar{\beta }_-+21 \left(\delta _--\epsilon _-\right)\right)\nonumber\\
&\quad\ \ +\nu\left(\frac{2}{3} \zeta _{AB} \bar{\gamma }_{AB}-\frac{\bar{\gamma }_{AB}^2}{3}-\frac{74 \bar{\gamma }_{AB}}{21}+\frac{4 \bar{\beta }_+^2}{3}+\frac{23 \bar{\beta }_+}{14}+2 \zeta _{AB}-\frac{2 \delta _+}{3}-\frac{\epsilon _+}{3}-\frac{1069}{216}+\frac{19m_-}{14}  \bar{\beta }_-\right)+\frac{2047 \nu ^2}{1512}\Bigg],\\
\hat H_{21}&=\frac{im_-x^{1/2}}{3}
+i x^{3/2} \left[\frac{5 \nu  m_-}{21}-\frac{m_-}{84}  \left(-14 \bar{\gamma }_{AB}+28 \bar{\beta }_++17\right)+\frac{m_-^2}{3}  \bar{\beta }_-\right]\nonumber\\
&+\frac{x^2}{1+\alpha^0_A \alpha^0_B} \left[\frac{\nu}{9}\alpha _-   \left(\alpha _+ + m_-\alpha _-\right)+\frac{m_-}{6}  \left(1+2 i \pi +4 \ln 2\right)\right],\\
\hat H_{33}&=-\frac{3im_-x^{1/2}}{4}\sqrt{\frac{15}{14}}
+\frac{3im_- x^{3/2} }{4} \sqrt{\frac{15}{14}} \left[4+\bar{\gamma }_{AB}-2 \nu +\bar{\beta }_+-m_- \bar{\beta }_-\right]\nonumber\\
&+\frac{x^2}{1+\alpha^0_A \alpha^0_B}\frac{1}{\sqrt{210}}\left[\frac{5\nu}{2} \alpha _- \left(\alpha _+   -m_-\alpha _-\right)   +\frac{27 m_-}{32} \left(\alpha _+^2-\alpha _-^2  -40 i \pi -56+80 \ln (3/2)\right)\right],\\
\hat H_{32}&=\frac{ x}{3} \sqrt{\frac{5}{7}} (1-3 \nu )
+\frac{x^2}{54\sqrt{35}}\left[-193+120 (3 \nu -1) \left(\bar{\beta }_+-m_- \bar{\beta }_-\right)+725 \nu -365 \nu ^2\right],\\
\hat H_{31}&=\frac{i m_- x^{1/2}}{12 \sqrt{14}}
+\frac{i x^{3/2} }{\sqrt{14}}\left[-\frac{\nu  m_-}{18}-\frac{m_-}{36}  \left(3 \bar{\gamma }_{AB}+3 \bar{\beta }_++8\right)+\frac{m_-^2}{12} \bar{\beta }_-\right]\nonumber\\
&+\frac{x^2}{1+\alpha^0 _A \alpha^0_B}\frac{1}{\sqrt{14}} \left[-\frac{\nu}{18}  \alpha _-\left(\alpha _+-5 \alpha _- m_-\right)+\frac{m_- }{480} \left(40 i \pi+\alpha _-^2-\alpha _+^2 +56+80 \ln 2\right)\right],\\
\hat H_{44}&=\frac{8x}{9} \sqrt{\frac{5}{7}}(3 \nu -1) 
+\frac{4 x^2}{297 \sqrt{35}}\left[1779+440 \left(\bar{\gamma }_{AB}+\bar{\beta }_+- m_- \bar{\beta }_-\right)-1320\nu  \left( \frac{1273}{264}+\bar{\gamma }_{AB}+ \bar{\beta }_+- m_- \bar{\beta }_-\right)+2625 \nu ^2\right],\\
\hat H_{43}&=\frac{9 i m_- x^{3/2}}{4\sqrt{70}}(2 \nu -1),\\
\hat H_{42}&=\frac{x\sqrt{5}}{63}  (1-3 \nu )
+\frac{x^2}{4158 \sqrt{5}}\left[-1311-440 \left(\bar{\gamma }_{AB}+ \bar{\beta }_+- m_- \bar{\beta }_-\right)+\nu  \left( 4025+1320\left( \bar{\gamma }_{AB}+ \bar{\beta }_+- m_- \bar{\beta }_-\right)\right)-285 \nu ^2\right],\\
\hat H_{41}&=\frac{i m_- x^{3/2}}{84 \sqrt{10}}(1-2 \nu ),\\
\hat H_{55}&=\frac{625 i  m_- x^{3/2}}{96 \sqrt{66}}(1-2 \nu ),\\
\hat H_{54}&=-\frac{32  x^2}{9 \sqrt{165}}\left(1-5 \nu +5 \nu ^2\right),\\
\hat H_{53}&=-\frac{9im_- x^{3/2}}{32} \sqrt{\frac{3}{110}} (1-2 \nu ) ,\\
\hat H_{52}&=\frac{2  x^2}{27 \sqrt{55}}\left(1-5\nu+5 \nu ^2\right),\\
\hat H_{51}&=\frac{i m_- x^{3/2}}{288 \sqrt{385}}(1-2 \nu),\\
\hat H_{66}&=\frac{54  x^2}{5 \sqrt{143}}\left(1-5\nu+5 \nu ^2\right),\\
\hat H_{64}&=-\frac{128x^2}{495}\sqrt{\frac{2}{39}} \left(1-5\nu+5 \nu ^2\right) ,\\
\hat H_{62}&=\frac{2 x^2}{297 \sqrt{65}}\left(1-5\nu+5 \nu ^2\right) ,\\
\hat H_{20}&=\frac{1}{4 \sqrt{6}}.
\end{align}
\end{subequations}
\end{widetext}

\section{Coefficients of the scalar flux\label{App:scalarFlux}}

The coefficients
entering the scalar flux~\eqref{eq:scalar_flux} read:
\begin{widetext}
\begin{subequations}
\begin{align}
  f_{-1\rm{PN}}&=\frac{\alpha _-^2}{3},\\
  f_{0\rm{PN}}&=-\frac{2}{45}  \left[3 \bar{\gamma }_{AB}\left(\alpha _+^2+4\right) +\alpha _-^2 \left(2 \bar{\gamma }_{AB}+10 \langle\bar\beta\rangle+15\right)
  -\frac{15 \alpha _- }{\bar{\gamma }_{AB}}\bigg(\alpha _+ \left(\bar{\beta }_--m_- \bar{\beta }_+\right)+2 \alpha _- \langle\bar\beta\rangle\bigg)
  \right]
  -\frac{4\nu}{9} \alpha _-^2,\\
  f_{0.5\rm{PN}}&=\frac{1}{3} \pi  \alpha _-^2 \left(\bar{\gamma }_{AB}+2\right),\\
  f_{1\rm{PN}}&=\frac{1}{630}\left[
  \frac{1}{4} \bigg(70 \alpha _-^4 \left(\bar{\gamma }_{AB}+2\right)^2+
  \left(-\alpha _-^2+\alpha _+^2+4\right) \left(448\langle\bar\beta\rangle \left(2 \bar{\gamma }_{AB}-3\right)+448\bar{\gamma }_{AB}^2+1455 \bar{\gamma }_{AB}\right)+70 \alpha _+ \alpha _-^3 m_- \left(\bar{\gamma }_{AB}+2\right)^2\nonumber\right.\\
  &\qquad\ \ \left.-140 \alpha _-^2 \left(-2 \left(\bar{\gamma }_{AB}-3\right) \bar{\gamma }_{AB}-4 \bar{\beta }_+^2
  +8 \langle\bar\beta\rangle^2+12 \langle\bar\beta\rangle -2 \langle\epsilon\rangle+27\right)
  -140 \alpha _+ \alpha _- \left(4 \bar{\beta }_--m_- \left(18 \bar{\gamma }_{AB}+4 \bar{\beta }_++33\right)\right)\bigg)\right.\nonumber\\
  &\qquad\ \ \left. +\frac{35}{\bar{\gamma }_{AB}}\bigg(-3 \left(\alpha _+^2+4\right) 
  \left(\bar{\beta }_-^2-\bar{\beta }_+^2
  +4\bar{\beta }_+\langle\bar\beta\rangle\right)
  +\alpha _-^2 \left(9 \bar{\beta }_-^2-9 \bar{\beta }_+^2+36 \bar{\beta }_+\langle\bar\beta\rangle
  -40\langle\bar\beta\rangle^2-24\langle\bar\beta\rangle-6 \langle\epsilon\rangle\right)\right.\nonumber\\
  &\qquad\ \ \left.+\alpha _+ \alpha _- \left(-4 \left(2 \bar{\beta }_++3\right) \bar{\beta }_-+4 m_- \bar{\beta }_-^2+4 m_- \bar{\beta }_+ \left(\bar{\beta }_++3\right)+3 m_- \epsilon _+-3 \epsilon _-\right)\bigg)\right.\nonumber\\
  &\qquad\ \ \left. -\frac{210 \alpha _-}{\bar{\gamma }_{AB}^2}\bigg(-4 \alpha _+ \bar{\beta }_- \bar{\beta }_++\alpha _- \left(\bar{\beta }_-^2-\bar{\beta }_+^2+4 \langle\bar\beta\rangle \bar{\beta }_+-12 \langle\bar\beta\rangle^2\right)+2 \alpha _+ m_- \left(\bar{\beta }_-^2+\bar{\beta }_+^2\right)\bigg)
  \right]\nonumber\\
  &+\frac{ \nu }{18} \left[
  7 \left(\alpha _+^2+4\right) \bar{\gamma }_{AB}+\alpha _-^2 \left(-11 \bar{\gamma }_{AB}+24 \bar{\beta }_++32 \langle\bar\beta\rangle-2 \epsilon _++41\right)-\alpha _-^4 \left(\bar{\gamma }_{AB}+2\right)^2
  +\frac{24 }{\bar{\gamma }_{AB}^2}\alpha _- \bar{\beta }_+ \left(\alpha _- \bar{\beta }_+-2 \alpha _+ \bar{\beta }_-\right)\right.\nonumber\\
  &\qquad\left.+\frac{2 }{\bar{\gamma }_{AB}}\bigg(6 \left(\alpha _+^2+4\right) \bar{\beta }_+^2-\alpha _-^2 \left(18 \bar{\beta }_+ \left(\bar{\beta }_++1\right)+20 \langle\bar\beta\rangle-3 \epsilon _+\right)+\alpha _+ \alpha _- \left(4 \bar{\beta }_- \left(2 \bar{\beta }_+-7\right)+10 m_- \bar{\beta }_++3 \epsilon _-\right)\bigg)
  \right]+\frac{2}{9} \alpha _-^2 \nu ^2,\\
  f_{1.5\rm{PN}}&=-\frac{2 \pi}{45}
  \Bigg[
  6\bar{\gamma }_{AB}  \left(\bar{\gamma }_{AB}+2\right)\left(\alpha _+^2+4\right)+\alpha _-^2 \left(10 \langle\bar\beta\rangle \left(\bar{\gamma }_{AB}-1\right)-\bar{\gamma }_{AB}^2+\bar{\gamma }_{AB}+6\right)+3 \alpha _+ \alpha _- \left(m_- \left(4 \bar{\gamma }_{AB}+5 \bar{\beta }_++8\right)-5 \bar{\beta }_-\right)\nonumber\\
  &\qquad\ \ -\frac{30 \alpha _-}{\bar{\gamma }_{AB}}\bigg(\alpha _+ \left(\bar{\beta }_--m_- \bar{\beta }_+\right)+2 \alpha _- \langle\bar\beta\rangle\bigg)
  \Bigg]
  -\frac{151\nu}{90}\pi  \alpha _-^2   \left(\bar{\gamma }_{AB}+2\right).
\end{align}
\end{subequations}
\end{widetext}

\section{Decomposition of the scalar waveform on spherical harmonics\label{App:scalarModes}}

The scalar waveform~\eqref{def:scalarWaveform} can be decomposed in a similar fashion as its metric counterpart over spherical harmonics (of weight zero),
\begin{align}
\delta\varphi=\sum_{\ell= 0}^\infty\sum_{m=-\ell}^\ell Y_{\ell m} (\Theta, \Phi) \,\delta\varphi_{\ell m},\label{eq:sphericalHamonicsScalarDecomposition}
\end{align}
where the modes $\delta\varphi_{\ell m}$ were calculated in Ref.~\cite{Bernard:2022noq} at 1.5PN relative to the dipolar order.
We translate them following the steps detailed in Sec.~\ref{sec:scalar_flux} .
In the radiative coordinate system~\eqref{def:radiativeTime}, we find:
\begin{align}
\delta\varphi_{\ell m}(T,R)=-\frac{i M\nu \sqrt{x}\,\alpha_-}{R}\sqrt{\frac{2\pi}{3}}\delta\hat \varphi_{\ell m}e^{-im\psi}\,,\label{def:scalarModes}
\end{align}
where $x$ and $\psi$ were defined in Eqs.~\eqref{def:x} and~\eqref{def:effectivePhase}, respectively.
The nonzero, dimensionless factors $\delta\hat\varphi_{\ell m}$ then read, for $m\geq 0$:
\begin{widetext}
\begin{subequations}
\begin{align}
\delta\hat\varphi_{00}&=-\frac{i  \left(\alpha _++\alpha _- m_-\right)}{\alpha _- \nu  \sqrt{x}}\sqrt{\frac{3}{2}}-\frac{i  \sqrt{x}}{2 \alpha _- \bar{\gamma }_{AB}}\sqrt{\frac{3}{2}}\left[8 \alpha _- \bar{\beta }_-+8 \alpha _+ \bar{\beta }_+-5 \alpha _+ \bar{\gamma }_{AB}+\alpha _- m_- \bar{\gamma }_{AB}\right]\nonumber\\
&+\frac{i x^{3/2}}{8 \sqrt{6} \alpha _- \bar{\gamma }_{AB}}\Bigg[\nu\,\bigg(-16 \alpha _- \bar{\beta }_--16 \alpha _+ \bar{\beta }_++32 \alpha _- \bar{\beta }_- \bar{\gamma }_{AB}+7 \alpha _+ \bar{\gamma }_{AB}-5 \alpha _- m_- \bar{\gamma }_{AB}\bigg)+\frac{1}{\bar{\gamma }_{AB}}\bigg(\alpha_-\left(-4 \bar{\beta }_- \bar{\gamma }_{AB} \left(7 \bar{\gamma }_{AB}+8 \bar{\beta }_+-12\right)\right.\nonumber\\
&\qquad\qquad\quad\left.+12 \epsilon _- \bar{\gamma }_{AB}-12 \epsilon _+ \bar{\gamma }_{AB}-4 m_- \bar{\beta }_+ \bar{\gamma }_{AB}^2+32 m_- \bar{\beta }_-^2 \left(\bar{\gamma }_{AB}+3\right)+8 m_- \bar{\gamma }_{AB}^3+9 m_- \bar{\gamma }_{AB}^2-96 m_- \bar{\beta }_+^2\right)
+\alpha_+\left(4 \bar{\beta }_- \bar{\gamma }_{AB} \left(7 \bar{\gamma }_{AB}+8 \bar{\beta }_+\right)\right.\nonumber\\
&\qquad\qquad\quad\left.-32 \bar{\beta }_+^2 \left(\bar{\gamma }_{AB}+3\right)+12 \bar{\beta }_+ \bar{\gamma }_{AB} \left(4-5 \bar{\gamma }_{AB}\right)+\bar{\gamma }_{AB} \left(7 \bar{\gamma }_{AB} \left(8 \bar{\gamma }_{AB}+9\right)-12 \epsilon _-+12 \epsilon _+\right)+96 \bar{\beta }_-^2\right)\bigg)\Bigg],\\
\delta\hat\varphi_{11}&=1+\frac{x}{15 \alpha _- \bar{\gamma }_{AB}}\bigg[3 \alpha _+ \left(4 m_- \bar{\gamma }_{AB}+5 \bar{\beta }_--5 m_- \bar{\beta }_+\right)-\alpha _- \left(5 \bar{\beta }_+ \left(\bar{\gamma }_{AB}-3\right)+5 \bar{\gamma }_{AB}^2+27 \bar{\gamma }_{AB}-5 m_- \bar{\beta }_- \bar{\gamma }_{AB}+15 m_- \bar{\beta }_-\right)+14 \alpha _- \nu  \bar{\gamma }_{AB}\bigg]\nonumber\\
&-\frac{i x^{3/2}}{12 \left(1+\alpha _A \alpha _B\right)}\bigg[(1+6 i \pi +12 \ln 2) \left(\bar{\gamma }_{AB}+2\right) \left(1+\alpha _A \alpha _B\right) +\alpha _-^2+\alpha _+^2+2 \alpha _+ \alpha _- m_-+4 \alpha _-^2 \nu\bigg],\\
\delta\hat\varphi_{22}&=\frac{i \sqrt{x} \left(\alpha _+-\alpha _- m_-\right)}{\sqrt{5} \alpha _-}\nonumber\\
&+\frac{i x^{3/2}}{42 \sqrt{5} \alpha _- \bar{\gamma }_{AB}}\bigg[
-\alpha _+ \bigg(28 \bar{\beta }_+ \left(\bar{\gamma }_{AB}-3\right)+28 \bar{\gamma }_{AB}^2+159 \bar{\gamma }_{AB}-28 m_- \bar{\beta }_- \bar{\gamma }_{AB}+84 m_- \bar{\beta }_-\bigg)+\alpha _- \bigg(-28 \bar{\beta }_- \left(\bar{\gamma }_{AB}-3\right)+28 m_- \bar{\beta }_+ \left(\bar{\gamma }_{AB}-3\right)\nonumber\\
&+m_- \bar{\gamma }_{AB} \left(28 \bar{\gamma }_{AB}+159\right)\bigg)
+\nu \, \bigg(56 \alpha _- \bar{\beta }_- \left(2 \bar{\gamma }_{AB}-3\right)+\alpha _+ \left(211 \bar{\gamma }_{AB}-168 \bar{\beta }_+\right)-61 \alpha _- m_- \bar{\gamma }_{AB}\bigg)
\bigg],\\
\delta\hat \varphi_{33}&=\frac{9 x}{4 \alpha _-}\sqrt{\frac{3}{70}}\left( \alpha _+ m_--\alpha _-+2 \alpha _- \nu\right),\\
\delta\hat\varphi_{31}&=-\frac{x}{20 \sqrt{14} \alpha _-}\left(\alpha _+ m_- -\alpha _-+2 \alpha _- \nu\right),\\
\delta\hat\varphi_{44}&=-\frac{16 i x^{3/2}}{3 \sqrt{105} \alpha _-}\Big[\alpha _+-m_-\alpha _-+ \nu  \left(\alpha _- m_--3 \alpha _+\right)\Big],\\
\delta\hat\varphi_{42}&=\frac{2 i x^{3/2}}{21 \sqrt{15} \alpha _-}\Big[\alpha _+-m_-\alpha _- +\nu  \left(\alpha _- m_--3 \alpha _+\right)\Big].
\end{align}
\end{subequations}
\end{widetext}
The monopolar contribution $\delta\hat \varphi_{00}$ does not contribute to the scalar flux~\eqref{def:scalarFlux} on circular orbits, because $m=0$.
Thus, the scalar flux is dominated by the dipolar mode $\ell=1$.
Note that the orbit being planar, we have by symmetry~\cite{Faye:2012we}
\begin{align}
\delta\varphi_{\ell,-m}=(-1)^\ell (\delta\varphi_{\ell m})^*.
\end{align}

\bibliographystyle{JHEP}
\bibliography{../references}

\end{document}